\documentclass[%
 reprint,
 amsmath,amssymb,
 aps,pra
]{revtex4-1}
\usepackage{graphicx}
\usepackage{dcolumn}
\usepackage{bm}

\begin{document}

\title{Cold Bose Atoms Around the Crossing of Quantum Waveguides.}
\author{A. Markowsky and N. Schopohl }
\email{corresponding author: nils.schopohl@uni-tuebingen.de}
\affiliation{Institut f\"ur Theoretische Physik, CQ Center for Collective Quantum Phenomena and Their Applications in $LISA^{+}$, Eberhard Karls-Universit\"at T\"ubingen, Auf
der Morgenstelle 14, D-72076 T\"ubingen, Germany\\
}
\date{\today}

\begin{abstract}
We show that massive low energy particles traversing a branching zone or a
crossing of quantum waveguides may experience a non standard trapping force
that cannot be derived from a potential. For interacting cold Bose atoms we
report on the formation of a localised Hartree ground state for three
prototype waveguide geometries with broken translational symmetry: a cranked 
$L$-shaped waveguide $\mathcal{L}$, a $T$-shaped waveguide $\mathcal{T}$ ,
and the crossing $\mathcal{C}$ of two quantum waveguides. The phenomenon is
kinetic energy driven and cannot be described within the Thomas-Fermi
approximation. Depending on the ratio $\kappa ^{\left( \Gamma \right) }$ of
joining lateral tube diameters of the respective waveguides $\Gamma \in
\left\{ \mathcal{C},\mathcal{L},\mathcal{T}\right\} $ delocalisation
commences when the particle number $N$ approaches a critical value $%
N_{c}^{\left( \Gamma \right) }$. For the case of a binary mixture of two
different Bose atom species $A$ and $B$ we observe non standard trapping of
both atom species for subcritical particle numbers. A sudden demixing
quantum transition takes place as the total particle number $N=N_{A}+N_{B}$
is increased at fixed mixing ratio $N_{A}/N_{B}$. Depending on the mass
ratio $m_{A}/m_{B}$ the heavier atom species delocalises first for a wide
range of interaction parameters. The numerical calculations are based on a
splitting scheme involving an analytic approximation to the short time
asymptotics of the imaginary time quantum propagator of a single particle
obeying to Dirichlet boundary conditions at the walls inside the respective
waveguides.

\begin{description}
\item[PACS numbers] 03.75.Hh, 67.85.-d, 67.85.Hj, 05.30.Jp, 03.75.Be,
42.25.-p, 03.75.-b
\end{description}
\end{abstract}

\maketitle

\section{Introduction}

\label{Introduction}

Elementary quantum mechanics predicts, that the dispersion relation $E_{p}=%
\frac{p^{2}}{2m}$ of a massive particle in free space is modified, when the
particle is moving slowly inside a hollow micron-sized capillary tube with a
transverse size $w$ comparable to the thermal de Broglie wavelength $\lambda
_{th}$ of that particle. This is because boundary conditions at the hard
walls of such a tube eliminate an infinite number of solutions to the Schr%
\"{o}dinger equation in free space, and the ones remaining are the guided
matter waves. In full analogy to $TE$- and $TM$- modes used for transmitting
electromagnetic signals along waveguides, also matter waves propagating
along the axis of a hollow tube sustain a discrete set of guided modes. For
example, guided waves of ultracold neutrons have been observed in metallic
thin film waveguides \cite{Feng},\cite{Pogossian}.

More recently guided matter wave experiments with cold atoms, using various
optical techniques for atom confinement in hollow-core dielectric fibers,
have been carried out successfully by several groups \cite{Renn I}, \cite%
{Ito et al}, \cite{Mueller I}, \cite{Christensen I}, \cite{vorrath}, \cite%
{Pechkis}. Also the trapping and guiding of atoms in the evanescent light
field surrounding a thin subwavelength-diameter fiber \cite{Barnett et al}, 
\cite{K+B+H} has been observed recently \cite{Nayak et al}, \cite{Vetsch et
al}. A new type of atomic-cladding waveguide with a dimension on the
sub-micro-meter scale \cite{Stern} opens further new possibilities for
experiments with guided atoms. Also cylindrically blue-tuned dark hollow
light beams are capable to transport atoms along their dark core \cite%
{Schiffer et al}, \cite{jc0}.

With the emergence of guided matter wave experiments the question then
arises, what happens if ultra cold particles were carried along \emph{curved}
waveguides, or were transported across the \emph{branchin}g zone or the 
\emph{crossing} of two waveguides.

Theoretical studies of the motion of particles confined in branching planar
stripes \cite{srw} or curved quantum wires have been the subject of intense
theoretical research already for many years \cite{ez0},\cite{ez1},\cite{D+E}%
, \cite{Br+Es}, \cite{Avishai}, \cite{sad}, \cite{Nazarov I}, \cite{Dauge}.
It is well known, that inside an infinitely extended straight waveguide the
propagation of a stationary mode along the tube axis is enabled only if the
energy $E$ of that mode is above a certain excitation threshold $\varepsilon
_{xt}>0$, the precise value of $\varepsilon _{xt}$ depending on the
geometric shape of the cross section of that waveguide. However, as was
shown by Goldstone and Jaffe \cite{gj0}, even a slight deviation from being
exactly straight may then give rise to the formation of localised states,
i.e. there exist stationary eigenstates of the kinetic energy Hamiltonian
with an eigenvalue $E_{0}$ below the excitation threshold $\varepsilon _{xt}$%
. Localised states also exist at a crossing of two waveguides \cite{srw}.
Since such bound states originate from effects of interference, they are
absent within a classical point mechanics approach. The trapping force
confining the particles by this mechanism is non standard and cannot be
derived from a potential. It is based on the rapid variation of kinetic
energy of a quantum particle that traverses a crossing or branching region
of otherwise translational invariant waveguides.

\begin{figure}[tbp]
\begin{centering}
\includegraphics[scale=0.9]{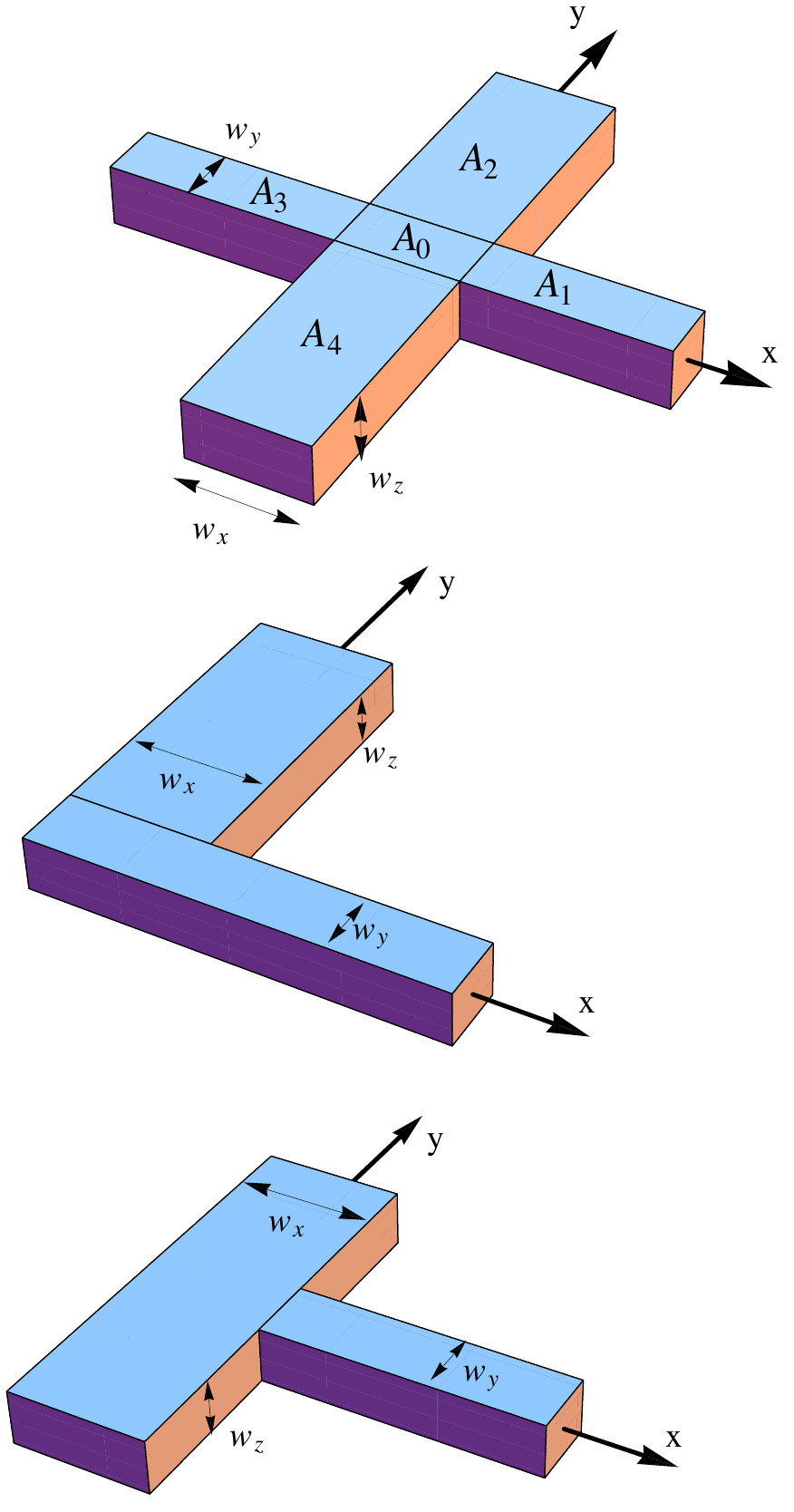}
\par\end{centering}
\caption{Three prototypes of waveguides with broken translational symmetry.
1) Cross shaped waveguide $\mathcal{C}$ as generated by two intersecting
tubes of rectangular cross-section. 2) $L$-shaped waveguide $\mathcal{L}$.
3) $T$-shaped waveguide $\mathcal{T}$ . The respective tube diameters are
denoted as $w_{x}$, $w_{y}$, and $w_{z}$. \newline
}
\label{fig:waveguides}
\end{figure}

A long hollow tube with hard walls and constant cross-section along the tube
axis will be referred to as a \emph{quantum waveguide} (QW), if the thermal
de Broglie wavelength $\lambda _{th}$ of a particle moving inside is
comparable to the transversal size $w$ of the tubes forming that QW. We
consider in the following three prototypes of QW geometries with broken
translational symmetry. The first consists of two intersecting orthogonal
tubes with rectangular cross-section, comprising four arms $\mathcal{A}_{1}$,%
$...$,$\mathcal{A}_{4}$ and a central zone $\mathcal{A}_{0}$, altogether
forming an open three-dimensional waveguide geometry in the guise of a swiss
cross $\mathcal{C}$ with boundary surface $\partial \mathcal{C}$ as
displayed schematically in Fig.\ref{fig:waveguides}. The second, in the
following referred to as $\mathcal{L}$, consists of a cranked tube that is $%
L $-shaped, the third, in the following referred to as $\mathcal{T}$ ,
consists of a $T$-shaped branching joining three tubes, see Fig.\ref%
{fig:waveguides}.

It appears then natural to ask if a QW with a bulge or bent like $\mathcal{L}
$, or with a branching like $\mathcal{T}$ , or a crossing of two waveguides
like $\mathcal{C}$, could be used as a particle trap for ultra cold
particles. With a repulsive interaction present, the number of Bose
particles that may occupy these bound states is limited to a critical
maximum value $N_{c}$ \cite{lp0}.

In the ensuing discussion we investigate localised ground states of
interacting cold Bose atoms inside the waveguides $\Gamma \in \left\{ 
\mathcal{C},\mathcal{L},\mathcal{T}\right\} $ for various cross section
areas and various particle numbers $N$. We determine the critical number $%
N_{c}^{\left( \Gamma \right) }$ of particles that can be trapped around the
respective crossing or branching regions. We show for cold Bose atoms
confined in such non classical traps that their kinetic energy is not
negligible (even for huge particle numbers), and that the Thomas-Fermi
approximation does not apply, a characteristic difference to the well known
BEC-atom traps with a parabolic potential.

Restricting to a mean field description of ultracold interacting Bose atoms
with mass $m$ we are interested in the Hartree ground state 
\begin{equation}
\Psi _{G}^{\left( \Gamma \right) }(\mathbf{r}_{1},\mathbf{r}_{2},...,\mathbf{%
r}_{N})=\psi ^{\left( \Gamma \right) }(\mathbf{r}_{1})\psi ^{\left( \Gamma
\right) }(\mathbf{r}_{2})\cdot \cdot \psi ^{\left( \Gamma \right) }(\mathbf{r%
}_{N})  \label{ground state BEC}
\end{equation}%
that forms inside the respective QW's subject to the constraint 
\begin{equation}
\int_{\Gamma }d^{3}r|\psi ^{\left( \Gamma \right) }(\mathbf{r})|^{2}=1
\label{normalization GP orbital}
\end{equation}%
The task is then to find the optimal one-particle orbital $\psi ^{\left(
\Gamma \right) }(\mathbf{r})$ that minimizes the energy of the interacting
Bose gas subject to a normalisation constraint ensuring conservation of
particle number $N$. Introducing a Lagrange parameter $\mu _{N}^{\left(
\Gamma \right) }$ for this constraint the optimal orbital $\psi ^{\left(
\Gamma \right) }(\mathbf{r})$ is then found solving the Gross-Pitaevskii
equation

\begin{widetext}

\begin{equation}
\left( -\frac{\hbar ^{2}}{2m}\nabla ^{2}+V_{T}^{\left( \Gamma \right) }(%
\mathbf{r})+\left( N-1\right) \frac{4\pi \hbar ^{2}a_{s}}{m}\ |\psi ^{\left(
\Gamma \right) }(\mathbf{r})|^{2}\right) \psi ^{\left( \Gamma \right) }(%
\mathbf{r})=\mu _{N}^{\left( \Gamma \right) }\psi ^{\left( \Gamma \right) }(%
\mathbf{r})  \label{Gross-Pitaevskii I}
\end{equation}%

\end{widetext}

Here, $a_{S}$ denotes the $s$-wave scattering length characterizing the
repulsive two-particle contact interaction. In order that such a mean field
description applies all tube size parameters $w_{a}^{\left( \Gamma \right) }$%
should be large compared to $a_{s}$. For hollow tubes $\Gamma \in \left\{ 
\mathcal{C},\mathcal{L},\mathcal{T}\right\} $ with hard walls $\partial
\Gamma $ the effect of the trap potential 
\begin{equation}
V_{T}^{\left( \Gamma \right) }(\mathbf{r})=\left\{ 
\begin{array}{c}
0\text{ for }\mathbf{r}\in \Gamma \\ 
\infty \text{ for }\mathbf{r}\in \Gamma%
\end{array}%
\right.
\end{equation}
will be taken into account in the following posing Dirichlet boundary value
conditions at these walls: 
\begin{equation}
\psi ^{\left( \Gamma \right) }(\mathbf{r})|_{\mathbf{r}\in \partial \Gamma
}=0  \label{boundary value condition GP-orbital}
\end{equation}

In the numerical calculations we use scaled units: $r_{a}\rightarrow r_{a}/L$%
, $\,\mu _{N}^{\left( \Gamma \right) }\rightarrow \mu _{N}^{\left( \Gamma
\right) }/\varepsilon _{L}$ , where $\varepsilon _{L}=\frac{\hbar ^{2}}{%
2mL^{2}}$ defines the units of energy and $L$ defines the units of length.
In particular $a_{S}\rightarrow a_{S}/L\ $and $w_{a}^{\left( \Gamma \right)
}\rightarrow w_{a}^{\left( \Gamma \right) }/L$ for $a\in \left\{
x,y,z\right\} $. In these units then$\frac{4\pi \hslash ^{2}}{m}a_{s}=\frac{%
8\pi a_{s}}{L}\times \left[ \varepsilon _{L}L^{3}\right] $ $\rightarrow 8\pi
a_{s}$.

\subsection{Two-Dimensional or Three-Dimensional Laplace Operator\ ?}

As a first step, we consider the case $N=1$. So we look for a solution of
the Schr\"{o}dinger eigenvalue problem%
\begin{equation}
H_{kin}\psi ^{\left( \Gamma \right) }(\mathbf{r})=E\psi ^{\left( \Gamma
\right) }(\mathbf{r})  \label{one particel Schroedinger equation}
\end{equation}%
describing the stationary modes of a single particle of mass $m$ moving
inside a waveguide $\Gamma \in \left\{ \mathcal{C},\mathcal{L},\mathcal{T}%
\right\} $ , see Fig. \ref{fig:waveguides}. For a large tube height $%
w_{z}^{\left( \Gamma \right) }\gg \max (w_{x}^{\left( \mathcal{C}\right)
},w_{y}^{\left( \mathcal{C}\right) })$ the transversal part $-\frac{\hslash
^{2}}{2m}\frac{\partial ^{2}}{\partial z^{2}}$ of the kinetic energy
operator $H_{kin}=-\frac{\hslash ^{2}}{2m}\left( \frac{\partial ^{2}}{%
\partial x^{2}}+\frac{\partial ^{2}}{\partial y^{2}}+\frac{\partial ^{^{2}}}{%
\partial z^{2}}\right) $ has a vanishing contribution in the ground state,
so that the Laplace operator becomes effectively two-dimensional.
Theoretical studies on guided matter waves in branching waveguides often
assume explicitely such a planar geometry connected to a two-dimensional
Laplace operator $\frac{\partial ^{2}}{\partial x^{2}}+\frac{\partial ^{2}}{%
\partial y^{2}}$, for example \cite{ez0}, \cite{ez1}, \cite{D+E}, \cite%
{Br+Es}, \cite{Avishai}, \cite{sad}, \cite{Dauge}, \cite{ez3}. But a
realistic thin film geometry cannot be described assuming a large thickness
parameter $w_{z}^{\left( \Gamma \right) }$.

To elucidate this paradox consider a simple model of a thin film, say a flat
box with lateral sizes $w_{x}$, $w_{y}$ and thickness (height) $w_{z}$. For
a single particle with mass $m$ moving inside such a box, and obeying to
Dirichlet boundary conditions at the walls, the energy levels are well known:%
\begin{eqnarray}
E_{n_{x},n_{y},n_{z}}^{\left( 3\right) } &=&\frac{\hslash ^{2}}{2m}\left[
\left( \frac{n_{x}\pi }{w_{x}}\right) ^{2}+\left( \frac{n_{y}\pi }{w_{y}}%
\right) ^{2}+\left( \frac{n_{z}\pi }{w_{z}}\right) ^{2}\right] \\
\text{ }n_{x},n_{y},n_{z} &\in &\left\{ 1,2,3,...\right\}  \nonumber
\end{eqnarray}%
In a realistic thin film there holds $w_{z}\ll \min \left(
w_{x},w_{y}\right) $. If then the kinetic energy (respectively the
temperature $k_{B}T$) of a particle is small compared to the level distance $%
E_{n_{x},n_{y},n_{z}=2}^{\left( 3\right) }-E_{n_{x},n_{y},n_{z}=1}^{\left(
3\right) }$ , the motion of the particle in the low-energy subspace $n_{z}=1$
may be considered effectively as two-dimensional, the associated energy
eigenvalue of the particle being%
\begin{equation}
E_{n_{x},n_{y},n_{z}=1}^{\left( 3\right) }=E_{n_{x},n_{y}\ }^{\left(
2\right) }+\frac{\hslash ^{2}}{2m}\left( \frac{\pi }{w_{z}}\right) ^{2}
\label{E_3 =E_2+shift}
\end{equation}%
Here $E_{n_{x},n_{y}\ }^{\left( 2\right) }$ denotes an eigenvalue of the
two-dimensional kinetic energy operator corresponding to a \emph{planar}
geometry ($w_{z}\rightarrow \infty $):%
\begin{equation}
E_{n_{x},n_{y}\ }^{\left( 2\right) }=\frac{\hslash ^{2}}{2m}\left[ \left( 
\frac{n_{x}\pi }{w_{x}}\right) ^{2}+\left( \frac{n_{y}\pi }{w_{y}}\right)
^{2}\right]
\end{equation}%
With increasing thickness of the film, i.e. for $w_{z}\rightarrow \infty $ ,
there holds then $E_{n_{x},n_{y},n_{z}=1}^{\left( 3\right) }\rightarrow
E_{n_{x},n_{y}\ }^{\left( 2\right) }$.

The relation (\ref{E_3 =E_2+shift}) applies for a single ultra cold
particle, $N=1$. The GP-equation (\ref{Gross-Pitaevskii I}) being nonlinear
for $N>1$, the value obtained for $\mu _{N}^{\left( \Gamma \right) }$ (the
chemical potential of the $N$-particle ground state of a BEC) in the limit
of a planar geometry ($w_{z}\rightarrow \infty $) cannot be related to the
value obtained for a finite thickness $w_{z}$ by a simple shift like in (\ref%
{E_3 =E_2+shift}). For this reason we treat in what follows the full
three-dimensional problem.

\section{ Localised Single Particle Ground States Around Branching Zones in 
\newline
$\mathcal{C}$, $\mathcal{L}$ and $\mathcal{T}$.}

\label{Localised Single Particle Eigenmodes}

Because the arms $\mathcal{A}_{j}$ of the waveguides $\Gamma \in \left\{ 
\mathcal{C},\mathcal{L},\mathcal{T}\right\} $ all have a rectangular
cross-section, see Fig.\ref{fig:waveguides}, the excitation threshold $%
\varepsilon _{xt}^{\left( \Gamma \right) }$ of a massive particle moving
inside those arms is readily identified: 
\begin{eqnarray}
\frac{\varepsilon _{xt}^{\left( \Gamma \right) }}{\varepsilon _{L}} &=&\left[
\frac{\pi L}{\max \left( w_{x}^{\left( \Gamma \right) },w_{y}^{\left( \Gamma
\right) }\right) }\right] ^{2}+\left( \frac{\pi L}{w_{z}^{\left( \Gamma
\right) }}\right) ^{2}  \label{excitation threshold  C} \\
\varepsilon _{L} &=&\frac{\hslash ^{2}}{2mL^{2}}  \nonumber
\end{eqnarray}%
The excitation threshold of a \emph{planar} wave guide geometry $\mathcal{C}$%
, as considered by Schult et al. \cite{srw}, corresponds to the limit $%
w_{z}^{\left( \mathcal{C}\right) }\rightarrow \infty $.

Generally speaking, the spectrum of the kinetic energy operator $H_{kin}$,
when it acts on wave functions with a support identical to the cross shaped
domain $\mathcal{C}$, consists of two parts, the continuous spectrum, with
associated propagating modes of \emph{infinite} $L_{2}$-norm that obey to
the boundary conditions (\ref{boundary value condition GP-orbital}), but are
extended over the entire QW, and the discrete (point-like) spectrum, with at
least one localised eigenfunctions $\psi _{0,\gamma }^{\left( \mathcal{C}%
\right) }(\mathbf{r})$ of \emph{finite} $L_{2}$-norm (\ref{normalization GP
orbital}). A quantum particle with energy equal to the eigenvalue $%
E_{0,\gamma }^{\left( \mathcal{C}\right) }$ of a localised eigenmode $\psi
_{0,\gamma }^{\left( \mathcal{C}\right) }(\mathbf{r})$ is trapped in the
localisation region around the crossing zone $\mathcal{A}_{0}$, so it cannot
propagate along the arms $\mathcal{A}_{1}$,$...$,$\mathcal{A}_{4}$ of the
domain $\mathcal{C}$.

The ground state mode $\psi _{0}^{\left( \mathcal{C}\right) }(\mathbf{r})$
of $H_{kin}$ not only solves (\ref{one particel Schroedinger equation}), but
obeys to the normalization constraint (\ref{normalization GP orbital}) and
fulfills the hard wall boundary condition (\ref{boundary value condition
GP-orbital}). The associated eigenvalue $E_{0}^{\left( \mathcal{C}\right) }$
of the ground state mode is below the excitation threshold $\varepsilon
_{xt}^{\left( \mathcal{C}\right) }$ of the QW:%
\begin{equation}
0<E_{0}^{\left( \mathcal{C}\right) }<\varepsilon _{xt}^{\left( \mathcal{C}%
\right) }
\end{equation}

For the cross shaped waveguide $\mathcal{C}$ with its infinitely extended
arms $\mathcal{A}_{1}$,...,$\mathcal{A}_{4}$ the normalization condition (%
\ref{normalization GP orbital}) cannot be fulfilled if the energy eigenvalue 
$E_{0}^{\left( \mathcal{C}\right) }$ of the particle was above the
excitation threshold $\varepsilon _{xt}^{\left( \mathcal{C}\right) }$.
Remarkably, the continuous spectrum of the kinetic energy operator $H_{kin}$
may also contain embedded \emph{discrete} eigenvalues $E_{0,\gamma }^{\left( 
\mathcal{C}\right) }>\varepsilon _{xt}^{\left( \mathcal{C}\right) }$, with
associated eigenfunctions $\psi _{0,\gamma }^{\left( \mathcal{C}\right) }(%
\mathbf{r})$ that are localised \cite{srw}, but display characteristic nodes
along various symmetry planes of the domain $\mathcal{C}$. Below we identify
some of these embedded eigenstates $\psi _{0,\gamma }^{\left( \mathcal{C}%
\right) }(\mathbf{r})$ of the Hamiltonian $H_{kin}$ as new ground states
associated with the action of $H_{kin}$ being restricted to wavefunctions
with a support equal to the waveguides $\mathcal{L}$ and $\mathcal{T}$.

Starting at initial time $\tau _{0}$ from an intial configuration $\psi _{%
\mathcal{A}_{l}}\left( \mathbf{r};\tau _{0}\right) $ prescribed in the
subdomains $\mathcal{A}_{l}$ $\subset \mathcal{C}$ , we now calculate for $%
j\in \left\{ 0,1,2,3,4\right\} $ intermediate configurations $\psi _{%
\mathcal{A}_{j}}\left( \mathbf{r};\tau _{n}\right) $ from the recursion \cite%
{Supplementary Material}%
\begin{eqnarray}
\tau _{n+1} &=&\tau _{n}+\Delta \tau \text{ for }n=0,1,2,3,...
\label{Chapman-Kolmogorov II} \\
\psi _{\mathcal{A}_{j}}\left( \mathbf{r};\tau _{n+1}\right)
&=&\sum_{l=0}^{4}\int_{\mathcal{A}_{l}}d^{3}r^{\prime }\mathcal{K}_{\mathcal{%
A}_{j},\mathcal{A}_{l}}\left( \mathbf{r},\mathbf{r}^{\prime };\Delta \tau
\right) \psi _{\mathcal{A}_{l}}\left( \mathbf{r}^{\prime };\tau _{n}\right) 
\nonumber
\end{eqnarray}%
According to what has been stated in \cite{Supplementary Material}, a
normalized ground state mode $\psi _{0}^{\left( \mathcal{C}\right) }(\mathbf{%
r})$ with energy eigenvalue $E_{0}^{\left( \mathcal{C}\right) }$ is then
determined by the limit $n\rightarrow \infty $ of this process:%
\begin{equation}
\psi _{0}^{\left( \mathcal{C}\right) }(\mathbf{r})=\lim_{n\rightarrow \infty
}\frac{\psi (\mathbf{r};\tau _{n})}{\sqrt{\int_{\mathcal{C}}d^{3}r^{\prime
}\ \left\vert \psi (\mathbf{r}^{\prime };\tau _{n})\right\vert ^{2}}}
\label{ground state mode N=1}
\end{equation}%
Here, for $\mathbf{r\in }\mathcal{A}_{j}$ and $\mathbf{r}^{\prime }\mathbf{%
\in }\mathcal{A}_{l}$ , the kernel functions $\mathcal{K}_{\mathcal{A}_{j},%
\mathcal{A}_{l}}\left( \mathbf{r},\mathbf{r}^{\prime };\Delta \tau \right) =%
\left[ \left\langle \mathbf{r}\right\vert e^{-\Delta \tau H_{kin}}\left\vert 
\mathbf{r}^{\prime }\right\rangle \right] _{\mathbf{r}\in \mathcal{A}_{j},%
\mathbf{r}^{\prime }\in \mathcal{A}_{l}}$ represent the various pieces of
the \emph{short-time} expansion of the imaginary time quantum propagator
obeying to Dirichlet boundary value conditions at the walls $\partial 
\mathcal{C}$ of our waveguide. Explicit expressions for the kernel functions 
$\mathcal{K}_{\mathcal{A}_{j},\mathcal{A}_{l}}\left( \mathbf{r},\mathbf{r}%
^{\prime };\Delta \tau \right) $ for small $\Delta \tau $ are listed in
appendix \ref{Appendix B}.

The short-time asymptotics of the associated quantum propagator at real time 
$\left[ \left\langle \mathbf{r}\right\vert e^{-i\Delta tH_{kin}}\left\vert 
\mathbf{r}^{\prime }\right\rangle \right] _{\mathbf{r}\in \mathcal{A}_{j},%
\mathbf{r}^{\prime }\in \mathcal{A}_{l}}$actually describes an isotropic
source of particles that emanate during time $\Delta t$ from the location $%
\mathbf{r}^{\prime }$ of the source at intial time $t=0$ along \emph{%
classical} trajectories to the endpoint $\mathbf{r}$, possibly undergoing
mirror reflection at the hard walls $\partial \mathcal{C}$. The full quantum
mechanics at later times $t_{n}=n\Delta t$ is recovered then by the
superposition principle as represented by the Chapman-Kolmogorov identity (%
\ref{Chapman-Kolmogorov II}). For a thorough discussion why quantum motion
of a massive particle for short times $\Delta t $ may indeed be considered
as classical see \cite{Gosson&Hiley}.

In the numerical calculations we represent the functions $\psi _{\mathcal{A}%
_{l}}\left( \mathbf{r}^{\prime };\tau _{n}\right) $ by the method of
barycentric interpolation \cite{Trefethen II}, restricting the points $%
\mathbf{r}\in $ $\mathcal{A}_{j}$ and $\mathbf{r}^{\prime }\in \mathcal{A}%
_{l}$ to a (non equidistant) Chebyshev tensor grid. The number of grid
points, say in the arm $\mathcal{A}_{1}$, we chose $N_{x}\times N_{y}\times
N_{z}=40\times 20\times 20$. As time step we chose $\Delta \tau =0.01\times %
\left[ \frac{\hbar }{\varepsilon _{L}}\right] $. The integrals with the
kernel functions need then to be evaluated (with high accuracy) for fixed $%
\Delta \tau $ and a fixed geometry with tube diameters $w_{a}^{\left( 
\mathcal{C}\right) }=2L_{a}$ just once, at the start of the iteration.
Details of the analytical and numerical calculations can be found in \cite%
{Supplementary Material}.

\subparagraph{Symmetry Classification.}

The group of discrete symmetry operations leaving the cross shaped domain $%
\mathcal{C}$ invariant is the well known (abelian) point group $D_{2h}$. It
consists of eight discrete symmetry operations, namely the identity $id$ and
the inversion operation $I$, the rotations $C_{2}\left( x\right) $, $%
C_{2}\left( y\right) $, $C_{2}\left( z\right) $ around the axes $\mathbf{e}%
_{x},\mathbf{e}_{y},\mathbf{e}_{z}\ $by an angle $\pi $, and the reflections 
$\sigma \left( xy\right) $ , $\sigma \left( xz\right) $, $\sigma \left(
yz\right) $ at the respective $xy$-, $xz$- and $yz$-symmetry planes.
Therefore, because the kernel $K(\mathbf{r},\mathbf{r}^{\prime };\Delta \tau
)$ is invariant under all operations of the point group $D_{2h}$ (applied
simultaneously to $\mathbf{r}$ and $\mathbf{r}^{\prime }$), choosing an
intial wave function $\psi _{\gamma }^{\left( in\right) }\left( \mathbf{r}%
\right) $ that is a representation of $D_{2h}$, all the iterated functions $%
\psi _{\gamma }(\mathbf{r};\tau _{n})$ will preserve the parity $\pm 1$ of
the intial wave function $\psi _{\gamma }^{\left( in\right) }\left( \mathbf{r%
}\right) $ under these eight symmetry operations. Here the label $\gamma \in
\left\{ A_{g},B_{1g},B_{2g},B_{3g},A_{u},B_{1u},B_{2u},B_{3u}\right\} $
specifies the possible (irreducible) representations of $D_{2h}$.

Being interested mainly in the ground state of the kinetic energy operator $%
H_{kin}$, when the latter is restricted to operate on wave functions with a
support equal to the domain $\mathcal{C}$ and obeying to Dirichlet boundary
conditions at the walls $\partial \mathcal{C}$, we restrict in the following
to a subspace of eigenmodes that \emph{all} are even under reflection at the
symmetry plane $z=0$, thus prohibiting for $\gamma $ any other option but $%
\gamma \in \left\{ A_{g},B_{1g},B_{2u},B_{3u}\right\} $. Also let us assume
(without loss of generality) a restriction for the lateral tube diameters, $%
w_{y}^{\left( \mathcal{C}\right) }\leq w_{x}^{\left( \mathcal{C}\right) }$.

\subparagraph{The Ground State Modes in $\mathcal{C}$ , $\mathcal{L}$ and $%
\mathcal{T}$.}

Depending on the choice of symmetry of the intial wave function $\psi
_{\gamma }^{\left( in\right) }\left( \mathbf{r}\right) $ at the start, we
find employing the iteration explained in \cite{Supplementary Material},
besides the ground state $\psi _{0,A_{g}}^{\left( \mathcal{C}\right) }(%
\mathbf{r})$ with eigenvalue $E_{0,A_{g}}^{\left( \mathcal{C}\right)
}<\varepsilon _{xt}^{\left( \mathcal{C}\right) }$, for symmetries $\gamma
\in \left\{ B_{1g},B_{2u},B_{3u}\right\} $ further localised modes $\psi
_{0,\gamma }^{\left( \mathcal{C}\right) }(\mathbf{r})$ with a corresponding
eigenvalue $E_{0,\gamma }^{\left( \mathcal{C}\right) }>\varepsilon
_{xt}^{\left( \mathcal{C}\right) }$. It is a feature of such eigenmodes $%
\psi _{0,\gamma }^{\left( \mathcal{C}\right) }(\mathbf{r})$ that on one hand
the corresponding eigenvalue $E_{0,\gamma }^{\left( \mathcal{C}\right) }$
belongs to the point spectrum of $H_{kin}$, on the other hand it is embedded
into the continuous spectrum of $H_{kin}$ comprising the stationary modes
with infinite $L_{2}$-norm propagating along the infinitely extended arms $%
\mathcal{A}_{j}$ of the domain $\mathcal{C}$.

In the case of $A_{g}$-symmetry the localised mode $\psi _{0,A_{g}}^{\left( 
\mathcal{C}\right) }(\mathbf{r})$ stays invariant under all symmetry
operations of the group $D_{2h}$. The corresponding eigenvalue $%
E_{0,A_{g}}^{\left( \mathcal{C}\right) }$ of $\psi _{0,A_{g}}^{\left( 
\mathcal{C}\right) }(\mathbf{r})$ is below the excitation threshold $%
\varepsilon _{xt}^{\left( \mathcal{C}\right) }$of the waveguide $\mathcal{C}$
, so that $0<$ $E_{0,A_{g}}^{\left( \mathcal{C}\right) }<\varepsilon
_{xt}^{\left( \mathcal{C}\right) }$. The mode $\psi _{0,A_{g}}^{\left( 
\mathcal{C}\right) }(\mathbf{r})$ is nodeless inside $\mathcal{C}$, and it
remains for arbitrary tube widths $w_{y}^{\left( \mathcal{C}\right) }$ and $%
w_{x}^{\left( \mathcal{C}\right) }$ localised around the crossing zone $%
\mathcal{A}_{0}\subset \mathcal{C}$. The mode $\psi _{0}^{\left( \mathcal{C}%
\right) }(\mathbf{r})\equiv \psi _{0,A_{g}}^{\left( \mathcal{C}\right) }(%
\mathbf{r})$ represents the highly symmetric ground state of a particle
moving inside $\mathcal{C}$.

In the case of $B_{1g}$-symmetry the localised eigenmode $\psi
_{0,B_{1g}}^{\left( \mathcal{C}\right) }(\mathbf{r})$ has odd parity under
the reflections $\sigma \left( xz\right) $, $\sigma \left( yz\right) $: 
\begin{eqnarray}
\psi _{0,B_{1g}}^{\left( \mathcal{C}\right) }(x,y,z) &=&-\psi
_{0,B_{1g}}^{\left( \mathcal{C}\right) }(-x,y,z) \\
\psi _{0,B_{1g}}^{\left( \mathcal{C}\right) }(x,y,z) &=&-\psi
_{0,B_{1g}}^{\left( \mathcal{C}\right) }(x,-y,z)  \nonumber
\end{eqnarray}%
Clearly, the mode $\psi _{0,B_{1g}}^{\left( \mathcal{C}\right) }(\mathbf{r})$
displays inside the domain $\mathcal{C}$ two nodal surfaces coinciding with
the symmetry planes $x=0$ and $y=0$. In the limit of a large tube height $%
w_{z}^{\left( \mathcal{C}\right) }\gg L$ and assuming tube widths $%
w_{x}^{\left( \mathcal{C}\right) }=w_{y}^{\left( \mathcal{C}\right) }=2L$
the localised eigenmode $\psi _{0,B_{1g}}^{\left( \mathcal{C}\right) }(%
\mathbf{r})$ was first obtained in \cite{srw}. However, a localised embedded
mode $\psi _{0,B_{1g}}^{\left( \mathcal{C}\right) }(\mathbf{r})$ ceases to
exist if the tube widths ratio $\kappa ^{\left( \mathcal{C}\right) }=\frac{%
w_{y}^{\left( \mathcal{C}\right) }}{w_{x}^{\left( \mathcal{C}\right) }}\leq
1 $ is too small. A localised mode $\psi _{0,B_{1g}}^{\left( \mathcal{C}%
\right) }(\mathbf{r})$ only exists if $\kappa _{c,B_{1g}}^{\left( \mathcal{C}%
\right) }<\kappa ^{\left( \mathcal{C}\right) }\leq 1$, where according to
our calculations the lower bound is $\kappa _{c,B_{1g}}^{\left( \mathcal{C}%
\right) }\simeq 0.89$ , independent on $w_{z}^{\left( \mathcal{C}\right) }$.

In the case of $B_{3u}$-symmetry the localised eigenmode $\psi
_{0,B_{3u}}^{\left( \mathcal{C}\right) }(\mathbf{r})$ has odd parity under
the reflection $\sigma \left( yz\right) $, but has even parity under the
reflection $\sigma \left( xz\right) $: 
\begin{eqnarray}
\psi _{0,B_{3u}}^{\left( \mathcal{C}\right) }(x,y,z) &=&-\psi
_{0,B_{3u}}^{\left( \mathcal{C}\right) }(-x,y,z) \\
\psi _{0,B_{3u}}^{\left( \mathcal{C}\right) }(x,y,z) &=&\psi
_{0,B_{3u}}^{\left( \mathcal{C}\right) }(x,-y,z)  \nonumber
\end{eqnarray}%
The mode $\psi _{0,B_{3u}}^{\left( \mathcal{C}\right) }(\mathbf{r})$ reveals
inside the domain $\mathcal{C}$ a nodal surface coinciding with the plane $%
x=0$. Assuming a symmetrical choice of tube widths $w_{x}^{\left( \mathcal{C}%
\right) }=w_{y}^{\left( \mathcal{C}\right) }\ $ no localised embedded
eigenmode $\psi _{0,B_{3u}}^{\left( \mathcal{C}\right) }(\mathbf{r})$ exists
for any box height $w_{z}^{\left( \mathcal{C}\right) }$. But a localised
embedded mode $\psi _{0,B_{3u}}^{\left( \mathcal{C}\right) }(\mathbf{r})$
indeed exists for $\kappa ^{\left( C\right) }<\kappa _{c,B_{3u}}^{\left( 
\mathcal{C}\right) }$, where according to our calculations the upper bound
is $\kappa _{c,B_{3u}}^{\left( \mathcal{C}\right) }\simeq 0.63$ ,
independent on $w_{z}^{\left( \mathcal{C}\right) }$. The case of $B_{2u}$%
-symmetry is very similar to the case of $B_{3u}$-symmetry. Corresponding to
a transposition of coordinate labels $x$ and $y$ it needs here no separate
discussion.

Next we consider two subdomains of $\mathcal{C}$, the $T$-shaped subdomain $%
\mathcal{T}$ , see Fig.\ref{fig:waveguides}, and the $L$-shaped subdomain $%
\mathcal{L}$, see Fig.\ref{fig:waveguides}: 
\begin{eqnarray}
\mathcal{T} &=&\left\{ \left( x,y,z\right) \in \mathcal{C\;}|\;x\geq 0\text{ 
}\right\}  \label{definition L- and T-QW} \\
\mathcal{L} &=&\left\{ \left( x,y,z\right) \in \mathcal{C\;}|\;\left( x\geq
0\right) \wedge \left( y\geq 0\right) \text{ }\right\}  \nonumber
\end{eqnarray}%
There holds $\mathcal{L}\subset \mathcal{T\subset C}$, the domain $\mathcal{L%
}$ forming a quarter and the domain $\mathcal{T}$ forming a half of the
original cross shaped domain $\mathcal{C}$. The respective tube diameters $%
w_{a}^{\left( \Gamma \right) }$ are then connected: 
\begin{eqnarray}
w_{z}^{\left( \mathcal{L}\right) } &=&w_{z}^{\left( \mathcal{T}\right)
}=w_{z}^{\left( \mathcal{C}\right) }  \label{tube size parameters C L T} \\
2w_{y}^{\left( \mathcal{L}\right) } &=&w_{y}^{\left( \mathcal{T}\right)
}=w_{y}^{\left( \mathcal{C}\right) }  \nonumber \\
2w_{x}^{\left( \mathcal{L}\right) } &=&2w_{x}^{\left( \mathcal{T}\right)
}=w_{x}^{\left( \mathcal{C}\right) }  \nonumber
\end{eqnarray}%
This implies for the excitation thresholds $\varepsilon _{xt}^{\left( 
\mathcal{L}\right) }$and $\varepsilon _{xt}^{\left( \mathcal{T}\right) }$ of
the waveguides $\mathcal{L}$ and $\mathcal{T}$ according to (\ref{excitation
threshold C}) the property 
\begin{equation}
\varepsilon _{xt}^{\left( \mathcal{C}\right) }=\,\varepsilon _{xt}^{\left( 
\mathcal{L}\right) }\leq \varepsilon _{xt}^{\left( \mathcal{T}\right) }
\end{equation}

Incidentally, if the localised embedded mode $\psi _{0,B_{1g}}^{\left( 
\mathcal{C}\right) }(\mathbf{r})$ is restricted to the $L$-shaped subdomain $%
\mathcal{L\subset C}$, it coincides with the ground state mode $\psi _{0\
}^{\left( \mathcal{L}\right) }(\mathbf{r})$ of the Hamiltonian $H_{kin}$ ,
granted the action of the operator $H_{kin}$ is restricted solely to wave
functions with a support identical to $\mathcal{L}$. By construction, the
function 
\begin{equation}
\psi _{0\ }^{\left( \mathcal{L}\right) }(\mathbf{r})=\psi
_{0,B_{1g}}^{\left( \mathcal{C}\right) }(\mathbf{r})|_{\mathbf{r\in }%
\mathcal{L}}
\end{equation}%
obeys at the walls $\partial \mathcal{L}$ of $\mathcal{L}$ to Dirichlet
boundary value conditions, because both nodal planes of the mode $\psi
_{0,B_{1g}}^{\left( \mathcal{C}\right) }(\mathbf{r})$, namely $x=0$ and $y=0$%
, now also belong to the boundary $\partial \mathcal{L}$ of $\mathcal{L}$,
see Fig.\ref{fig:waveguides}. Inside $\mathcal{L}$ the mode $\psi _{0\
}^{\left( \mathcal{L}\right) }(\mathbf{r})$ is nodeless. For the
corresponding eigenvalue $E_{0}^{\left( \mathcal{L}\right) }\equiv
E_{0,B_{1g}}^{\left( \mathcal{C}\right) }$ there holds $E_{0}^{\left( 
\mathcal{L}\right) }<\,\varepsilon _{xt}^{\left( \mathcal{L}\right) }$.

Similarly, if the localised solution $\psi _{0,B_{3u}}^{\left( \mathcal{C}%
\right) }(\mathbf{r})$ is restricted to the $T$-shaped subdomain $\mathcal{%
T\subset C}$, it coincides with the ground state $\psi _{0\ }^{\left( 
\mathcal{T}\right) }(\mathbf{r})$ of the Hamiltonian $H_{kin}$, granted the
action of the operator $H_{kin}$ is restricted solely to wave functions with
a support identical to $\mathcal{T}$. The function 
\begin{equation}
\psi _{0\ }^{\left( \mathcal{T}\right) }(\mathbf{r})=\psi
_{0,B_{3u}}^{\left( \mathcal{C}\right) }(\mathbf{r})|_{\mathbf{r\in }%
\mathcal{T}}
\end{equation}%
obeys at the walls $\partial \mathcal{T}$ of $\mathcal{T}$ to Dirichlet
boundary value conditions, the nodal plane $x=0$ now also belonging to the
boundary $\partial \mathcal{T}$ , see Fig.\ref{fig:waveguides}. Inside $%
\mathcal{T}$ the mode $\psi _{0\ }^{\left( \mathcal{T}\right) }(\mathbf{r})$
is nodeless. For the corresponding eigenvalue $E_{0}^{\left( \mathcal{T}%
\right) }\equiv E_{0,B_{3u}}^{\left( \mathcal{C}\right) }$ there holds $%
E_{0}^{\left( \mathcal{T}\right) }<\,\varepsilon _{xt}^{\left( \mathcal{T}%
\right) }$.

\begin{figure}[tbp]
\begin{centering}
\includegraphics[scale=0.9]{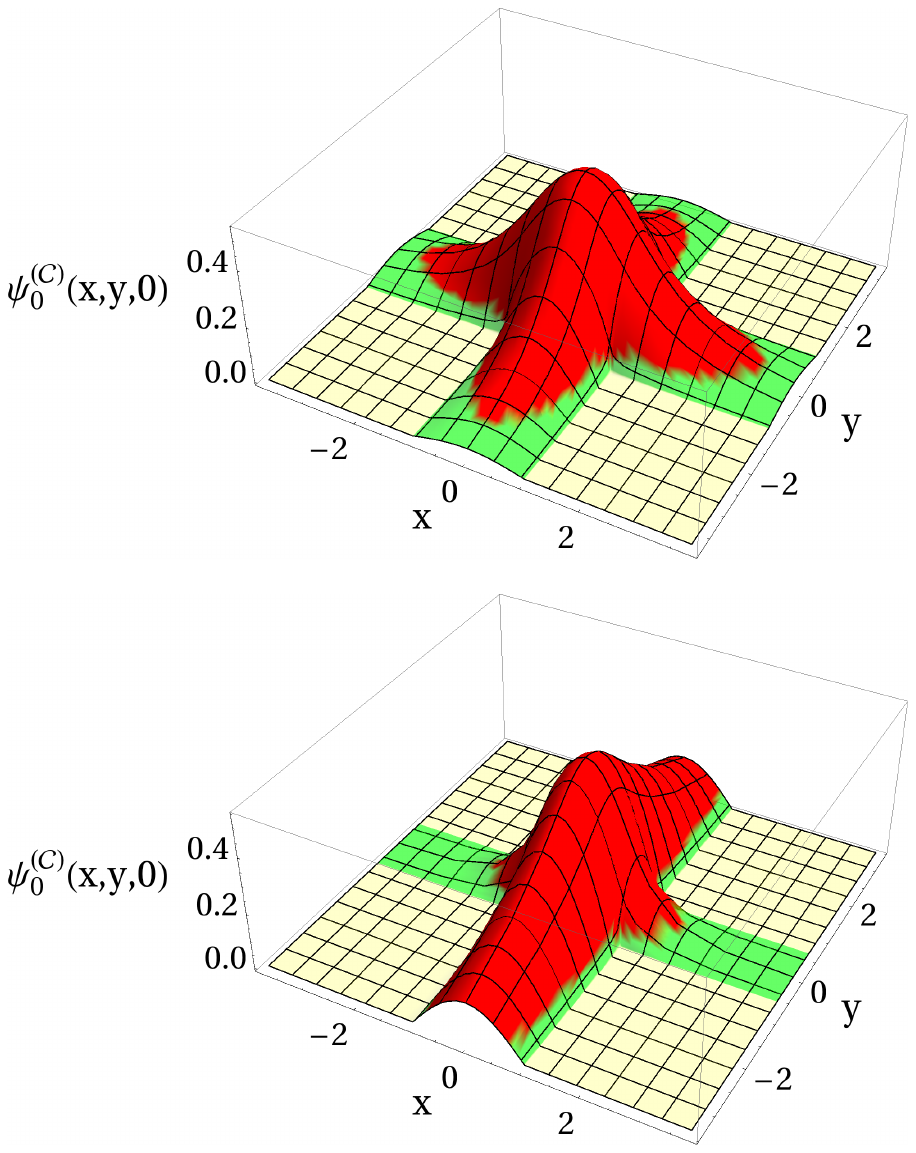}
\par\end{centering}
\par
\centering{}
\caption{The highly symmetric groundstate $\protect\psi _{0}^{\left( 
\mathcal{C}\right) }\left( \mathbf{r}\right) $ localised around the crossing
zone of a wave-guide $\mathcal{C}$ for different choice of tube widths.
Upper plot $w_{x}^{\left( \mathcal{C}\right) }=w_{y}^{\left( \mathcal{C}%
\right) }=w_{z}^{\left( \mathcal{C}\right) }=2L$ , lower plot $w_{x}^{\left( 
\mathcal{C}\right) }=w_{z}^{\left( \mathcal{C}\right) }=2L$ and $%
w_{y}^{\left( \mathcal{C}\right) }=0.6w_{x}^{\left( \mathcal{C}\right) }$.
Both plots restrict to the plane $z=0$. Length measured in units of $L$.}
\label{fig:psi3Dplot}
\end{figure}

In Fig.\ref{fig:psi3Dplot} we display the highly symmetric localised ground
state $\psi _{0}^{\left( \mathcal{C}\right) }(\mathbf{r})$ of $H_{kin}$ for
a particle moving inside $\mathcal{C}$, for two sets of tube diameters $%
w_{a}^{\left( \mathcal{C}\right) }$, restricting to the plane $z=0$. The
wave function $\psi _{0}^{\left( \mathcal{C}\right) }(\mathbf{r})$ takes on
its maximum value at the center $\mathbf{r}_{M}=(0,0,0)$ of the crossing
zone $\mathcal{A}_{0}$, while it vanishes everywhere at the hard walls $%
\partial \mathcal{C}$, and it decays exponentially along the axes of the
arms $\mathcal{A}_{1}$,$...$,$\mathcal{A}_{4}$.

\begin{figure}[tbp]
\begin{centering}
\includegraphics[scale=0.9]{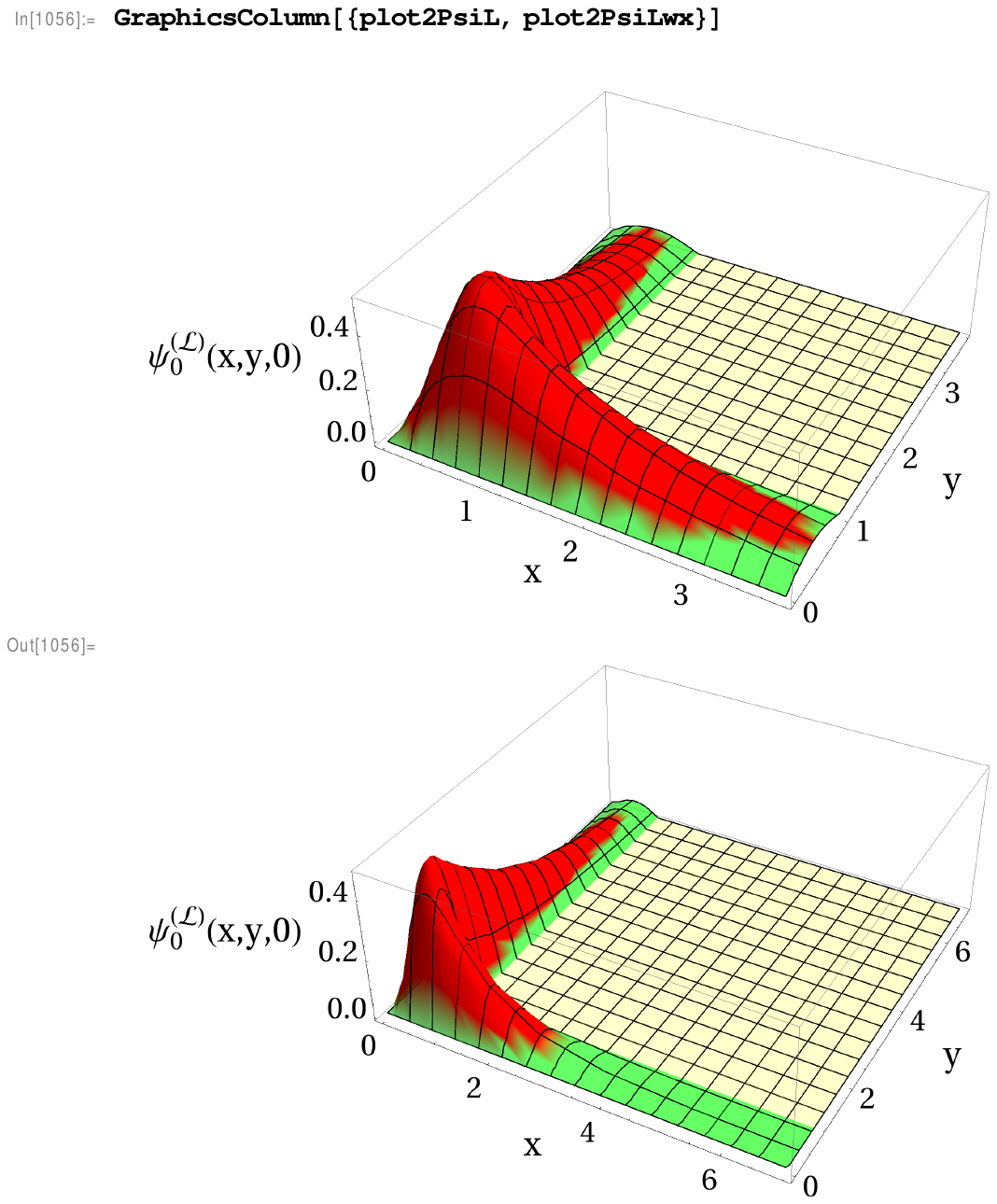}
\par\end{centering}
\par
\centering{}
\caption{The localised groundstate $\protect\psi _{0}^{\left( \mathcal{L}%
\right) }\left( \mathbf{r}\right) $ around the corner zone of an $L$-shaped
waveguide $\mathcal{L}$ for different choice of tube widths. Upper plot $%
w_{x}^{\left( \mathcal{L}\right) }=w_{y}^{\left( \mathcal{L}\right) }=L$ and 
$w_{z}^{\left( \mathcal{L}\right) }=2L$, lower plot $w_{x}^{\left( \mathcal{L%
}\right) }=L$, $w_{z}^{\left( \mathcal{L}\right) }=2L$ and $w_{y}^{\left( 
\mathcal{L}\right) }=0.95w_{x}^{\left( \mathcal{L}\right) }$. Both plots
restrict to the plane $z=0$. Length measured in units of $L$.}
\label{fig:psidxycontourL}
\end{figure}

In Fig.\ref{fig:psidxycontourL} we display the localised ground state $\psi
_{0}^{\left( \mathcal{L}\right) }(\mathbf{r})$ of $H_{kin}$ for a particle
moving inside $\mathcal{L}$, for two sets of tube diameters $w_{a}^{\left( 
\mathcal{L}\right) }$, restricting to the plane $z=0$. The wave function $%
\psi _{0}^{\left( \mathcal{L}\right) }(\mathbf{r})$ takes on its maximum
value at the center of the corner zone of $\mathcal{L}$, while it vanishes
everywhere at the hard walls $\partial \mathcal{L}$, and it decays
exponentially along the axes directions $\mathbf{e}_{x}$ and $\mathbf{e}_{y}$%
.

\begin{figure}[tbph]
\begin{centering}
\includegraphics[scale=0.9]{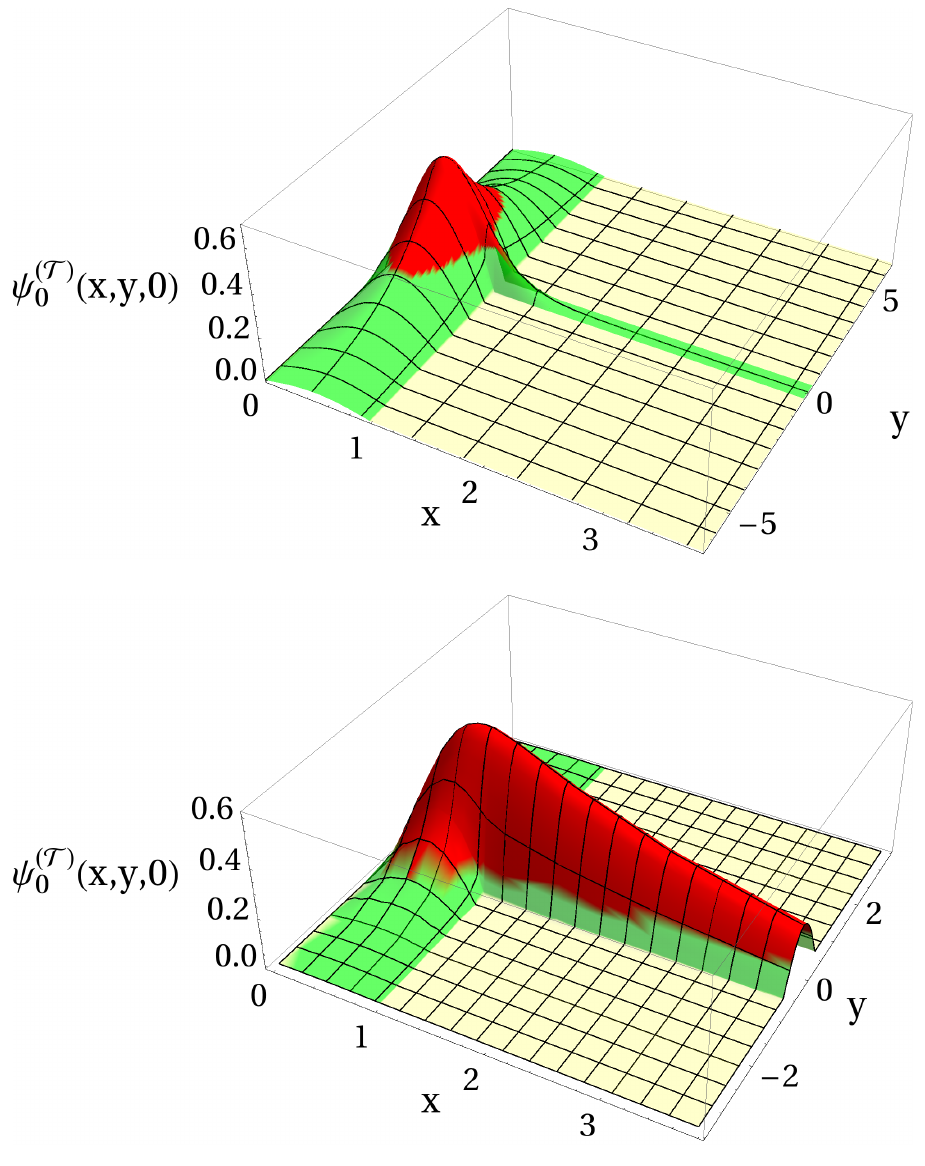}
\par\end{centering}
\caption{The localised groundstate $\protect\psi _{0}^{\left( \mathcal{T}%
\right) }\left( \mathbf{r}\right) $ around the branching zone of the $T$%
-shaped waveguide $\mathcal{T}$ for different choice of tube widths. Upper
plot $w_{x}^{\left( \mathcal{T}\right) }=L$, $w_{y}^{\left( \mathcal{T}%
\right) }=0.6L$ and $w_{z}^{\left( \mathcal{T}\right) }=2L$ , lower plot $%
w_{y}^{\left( \mathcal{T}\right) }=1.2L$, $w_{x}^{\left( \mathcal{T}\right)
}=L$ and $w_{z}^{\left( \mathcal{T}\right) }=2L$. Both plots restrict to the
plane $z=0$. Length measured in units of $L$.}
\label{fig:psicontourT}
\end{figure}

In Fig.\ref{fig:psicontourT} we display the localised ground state $\psi
_{0}^{\left( \mathcal{T}\right) }(\mathbf{r})$ of $H_{kin}$ for a particle
moving inside $\mathcal{T}$, for two sets of tube diameters $w_{a}^{\left( 
\mathcal{T}\right) }$, restricting to the plane $z=0$. The wave function $%
\psi _{0}^{\left( \mathcal{T}\right) }(\mathbf{r})$ takes on its maximum
value at the center of the branching zone of $\mathcal{T}$, while it
vanishes everywhere at the hard walls $\partial \mathcal{T}$, and it decays
exponentially along the axes directions $\mathbf{e}_{x}$ and $\mathbf{e}_{y}$%
.

\begin{figure}[tbp]
\begin{centering}
\includegraphics[scale=1.2]{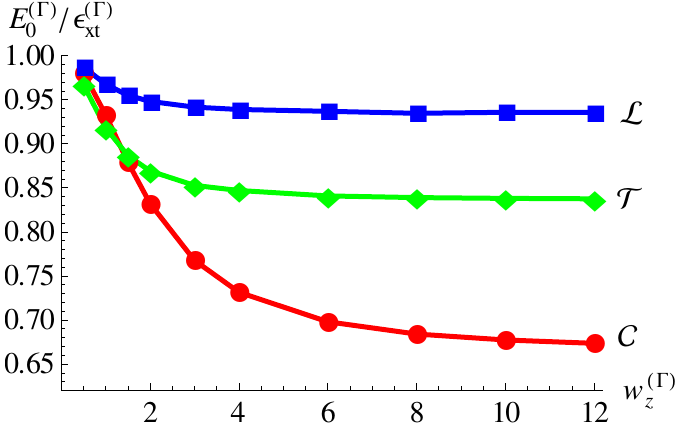}
\par\end{centering}
\caption{The ratio $E_{0}^{\left( \Gamma \right) }/\protect\varepsilon %
_{xt}^{\left( \Gamma \right) }$ of the eigenvalue $E_{0}^{\left( \Gamma
\right) }$ to the threshold energy $\protect\varepsilon _{xt\ }^{\left(
\Gamma \right) }$ corresponding to the localised groundstate mode $\protect%
\psi _{0}^{\left( \Gamma \right) }\left( \mathbf{r}\right) $ in the
respective waveguide geometries $\Gamma \in \left\{ \mathcal{C},\mathcal{T},%
\mathcal{L}\right\} $ as a function of the tube height $w_{z}^{\left( \Gamma
\right) }$, assuming fixed tube widths $w_{x}^{\left( \Gamma \right)
}=w_{y}^{\left( \Gamma \right) }$. Length measured in units of $L$.}
\label{fig:energyratio}
\end{figure}

In Fig.\ref{fig:energyratio} we display for all three waveguides $\Gamma \in
\left\{ \mathcal{C},\mathcal{T},\mathcal{L}\right\} $ the eigenvalues $E_{0\
}^{\left( \Gamma \right) }$ of the associated ground state eigenmode $\psi
_{0\ }^{\left( \Gamma \right) }(\mathbf{r})$ , plotting the ratios $E_{0\
}^{\left( \Gamma \right) }/\varepsilon _{xt\ }^{\left( \Gamma \right) }$ as
a function of the thickness parameter $w_{z}^{\left( \Gamma \right) }$ of
the respective waveguides, restricting to a \emph{symmetric} choice of tube
widths, $\kappa ^{\left( \Gamma \right) }\equiv w_{y}^{\left( \Gamma \right)
}/w_{x}^{\left( \Gamma \right) }=1$. In the limit of a \textit{thin }layer, $%
w_{z}^{\left( \Gamma \right) }\rightarrow 0$ , there holds $E_{0\ }^{\left(
\Gamma \right) }\rightarrow \varepsilon _{xt\ }^{\left( \Gamma \right) }$.
For \textit{thick} layers (not a 'thin' film) corresponding to the planar
limit $w_{z}^{\left( \Gamma \right) }\rightarrow \infty $ (two-dimensional
Laplace operator), calculations based on our heat kernel method confirm the
eigenvalue $E_{0\ }^{\left( \mathcal{C}\right) }=0.659$ $\times \varepsilon
_{xt\ }^{\left( \mathcal{C}\right) }$ for the ground state $\psi _{0\
}^{\left( \mathcal{C}\right) }(\mathbf{r})$ for a symmetric crossing $%
\mathcal{C}$ with $\kappa ^{\left( \mathcal{C}\right) }=1$ , and the
eigenvalue $E_{0\ }^{\left( \mathcal{L}\right) }=0.929$ $\times \varepsilon
_{xt\ }^{\left( \mathcal{L}\right) }$for the ground state $\psi _{0\
}^{\left( \mathcal{L}\right) }(\mathbf{r})$ for a symmetric cranked
waveguide $\mathcal{L}$ with $\kappa ^{\left( \mathcal{L}\right) }=1$, in
complete agreement with previous calculations \cite{srw}, \cite{ez2}, \cite%
{Amore}, \cite{Avishai}, \cite{Trefethen I} based on solving the
two-dimensional Schr\"{o}dinger eigenvalue problem with a variational
collocation ansatz.

\begin{figure}[tbph]
\begin{centering}
\includegraphics[scale=1.2]{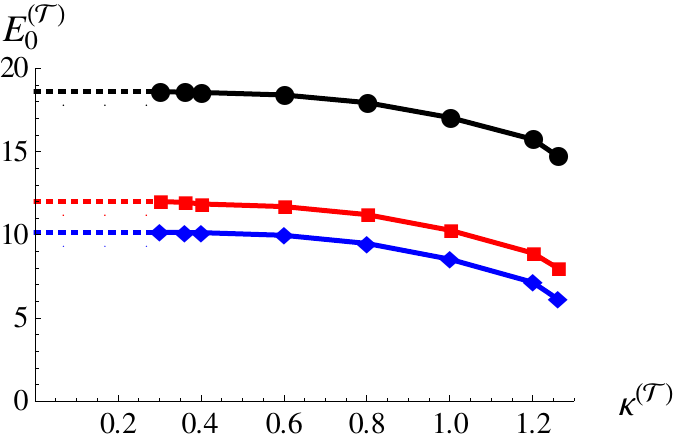}
\par\end{centering}
\caption{The eigenvalue $E_{0}^{\left( \mathcal{T}\right) }$ of the
localised groundstate mode $\protect\psi _{0}^{\left( \mathcal{T}\right)
}\left( \mathbf{r}\right) $ in the $T$-shaped waveguide geometry vs. the
lateral tube widths ratio $\protect\kappa ^{\left( \mathcal{T}\right)
}=w_{y}^{\left( \mathcal{T}\right) }/w_{x}^{\left( \mathcal{T}\right) }$
assuming different tube heights: $w_{z}^{\left( \mathcal{T}\right) }=L$
(black), $w_{z}^{\left( \mathcal{T}\right) }=2L$ (red), $w_{z}^{\left( 
\mathcal{T}\right) }=4L$ (blue). Energy measured in units of $\protect%
\varepsilon _{L}$.}
\label{fig:EnVsLxTShape}
\end{figure}

We find that the eigenvalue $E_{0\ }^{\left( \Gamma \right) }$ of the ground
state modes $\psi _{0\ }^{\left( \Gamma \right) }(\mathbf{r})$ of a massive
particle moving inside a realistic thin film or \emph{three-dimensional} QW
depends indeed strongly on the thickness parameter $w_{z}^{\left( \Gamma
\right) }$, as can be seen from the results displayed in Fig.\ref%
{fig:energyratio}, Fig.\ref{fig:EnVsLxTShape}. While the eigenvalues $E_{0\
}^{\left( \Gamma \right) }$certainly depend on $w_{z}^{\left( \Gamma \right)
}$, the localisation lengths $\lambda _{x}^{\left( \Gamma \right) }$ and $%
\lambda _{y}^{\left( \Gamma \right) }$ of the eigenmodes $\psi _{0\
}^{\left( \Gamma \right) }(\mathbf{r})$ along the respective tube axes $%
\mathbf{e}_{x}$ and $\mathbf{e}_{y}$ of the waveguides $\Gamma \in \left\{ 
\mathcal{C},\mathcal{T},\mathcal{L}\right\} $ are independent on the
tickness parameter $w_{z}^{\left( \Gamma \right) }$, because at a large
distance to the respective branching zones of $\Gamma $ the Schr\"{o}dinger
eigenvalue problem (\ref{one particel Schroedinger equation}) is completely
separable.

For an \emph{asymmetric} crossing of two waveguides with different tube
widths, assuming $w_{y}^{\left( \mathcal{C}\right) }<w_{x}^{\left( \mathcal{C%
}\right) }$, see Fig.\ref{fig:waveguides}, the localised ground state $\psi
_{0\ }^{\left( \mathcal{C}\right) }(\mathbf{r})$ then decays exponentially
along the axes $\mathbf{e}_{x}$ and $\mathbf{e}_{y}$ of $\mathcal{C}$,
displaying a smaller decay length $\lambda _{x}^{\left( \mathcal{C}\right) }$
along the tube axes $\pm \mathbf{e}_{x}$ of the arms with shorter lateral
size $w_{y}^{\left( \mathcal{C}\right) }$, and a larger decay length $%
\lambda _{y}^{\left( \mathcal{C}\right) }$ along the tube axes $\pm \mathbf{e%
}_{y}$ of the arms with wider lateral size $w_{x}^{\left( \mathcal{C}\right)
}$.

\begin{figure}[tbph]
\begin{centering}
\includegraphics[scale=1]{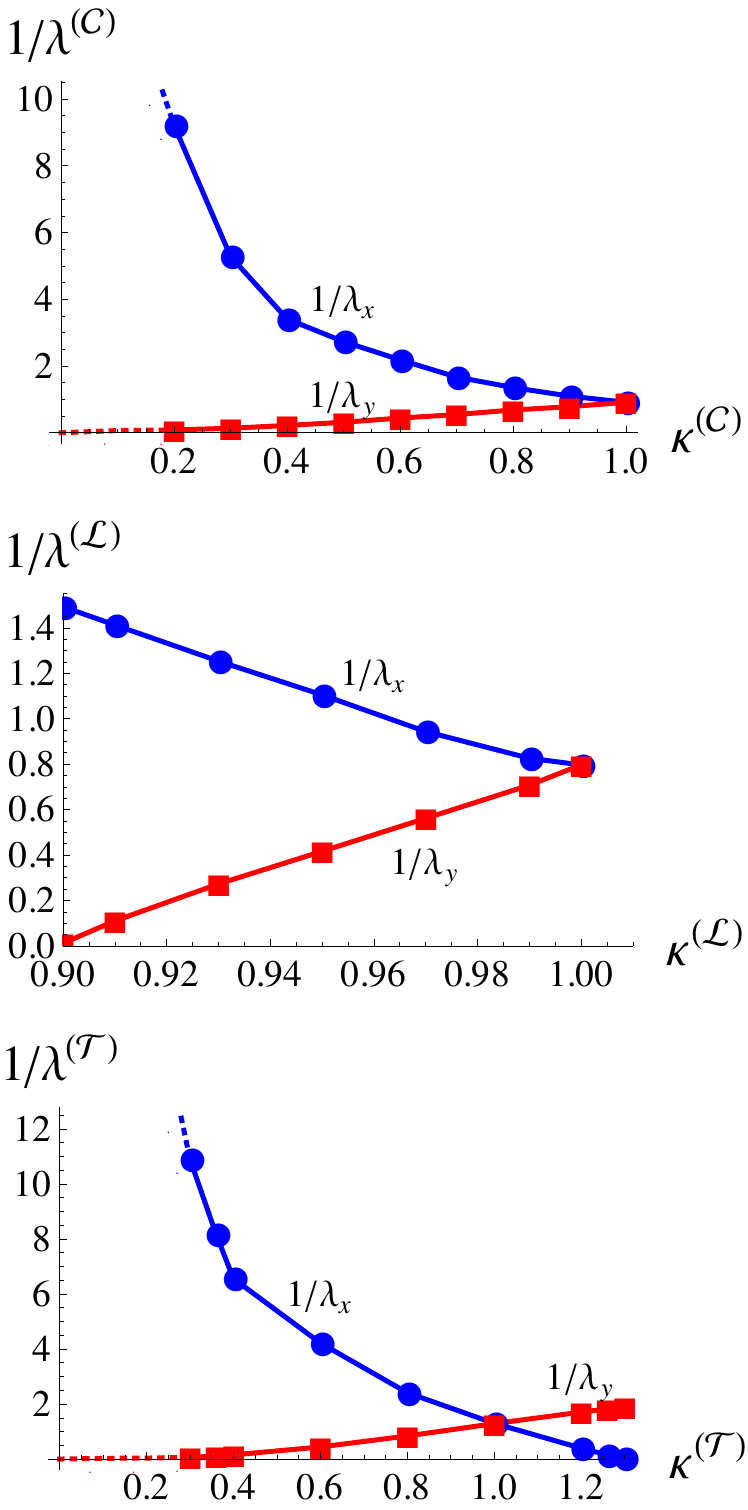}
\par\end{centering}
\caption{ Inverse localisation lengths $1/\protect\lambda _{y}^{\left(
\Gamma \right) }$ along tube axis $\mathbf{e}_{y}$ (red line) and $1/\protect%
\lambda _{x}^{\left( \Gamma \right) }$ along tube axis $\mathbf{e}_{x}$
(blue line) of prototype waveguides $\Gamma \in \left\{ \mathcal{C},\mathcal{%
L},\mathcal{T}\right\} $ as function of respective tube widths ratio $%
\protect\kappa ^{\left( \Gamma \right) }=w_{y}^{\left( \Gamma \right)
}/w_{x}^{\left( \Gamma \right) }$ for constant tube height $w_{z}^{\left(
\Gamma \right) }=2L$. Upper plot $w_{x}^{\left( \mathcal{C}\right) }=2L$,
middle plot $w_{x}^{\left( \mathcal{L}\right) }=L$, lower plot $%
w_{x}^{\left( \mathcal{T}\right) }=L$. If $\protect\kappa ^{\left( \mathcal{L%
}\right) }<\protect\kappa _{c}^{\left( \mathcal{L}\right) }\simeq 0.89$ no
localised ground state $\protect\psi _{0}^{\left( \mathcal{L}\right) }\left( 
\mathbf{r}\right) $ exists, if $\protect\kappa ^{\left( \mathcal{T}\right) }>%
\protect\kappa _{c}^{\left( \mathcal{T}\right) }\simeq 1.26$ no localised
ground state $\protect\psi _{0}^{\left( \mathcal{T}\right) }\left( \mathbf{r}%
\right) $ exists. Length measured in units of $L$.}
\label{fig:LambdaKappaNN1}
\end{figure}

While there always exists a localised ground state $\psi _{0\ }^{\left( 
\mathcal{C}\right) }(\mathbf{r})$ around the crossing zone $\mathcal{A}%
_{0}\subset $ $\mathcal{C}$ for any choice of tube widths $w_{y}^{\left( 
\mathcal{C}\right) }\ $and $w_{x}^{\left( \mathcal{C}\right) }$, see Fig.\ref%
{fig:LambdaKappaNN1}, a localised ground state $\psi _{0\ }^{\left( \mathcal{%
L}\right) }(\mathbf{r})$ around the corner of the cranked $L$-shaped
waveguide $\mathcal{L}$ only exists if the tube widths ratio $\kappa
^{\left( \mathcal{L}\right) }=\frac{w_{y}^{\left( \mathcal{L}\right) }}{%
w_{x}^{\left( \mathcal{L}\right) }}$ is not too small, i.e. a localised
ground state exists provided $\kappa _{c}^{\left( \mathcal{L}\right)
}<\kappa ^{\left( \mathcal{L}\right) }\ $, with $\kappa _{c}^{\left( 
\mathcal{L}\right) }$ denoting a characteristic lower bound of tube widths
ratios. Choosing $w_{z}^{\left( \mathcal{L}\right) }=2L=w_{x}^{\left( 
\mathcal{L}\right) }$ we obtain from our three-dimensional numerical
calculations a value around $\kappa _{c}^{\left( \mathcal{L}\right) }\simeq
\allowbreak 0.89$ , see Fig.\ref{fig:LambdaKappaNN1}.

On the other hand, for an \emph{asymmetric }wave guide $\mathcal{T}$ a
localised ground state $\psi _{0\ }^{\left( \mathcal{T}\right) }(\mathbf{r})$
around the branching zone of $\mathcal{T}$ only exists, if the ratio $\kappa
^{\left( \mathcal{T}\right) }=\frac{w_{y}^{\left( \mathcal{T}\right) }}{%
w_{x}^{\left( \mathcal{T}\right) }}$ is not too big, i.e. a localised ground
state exists provided $0<\kappa ^{\left( \mathcal{T}\right) }<\kappa
_{c}^{\left( \mathcal{T}\right) }$. Choosing $w_{z}^{\left( \mathcal{T}%
\right) }=L=w_{x}^{\left( \mathcal{T}\right) }$ we find from our
three-dimensional numerical calculations a value around $\kappa _{c}^{\left( 
\mathcal{T}\right) }\simeq 1.26$ , see Fig.\ref{fig:LambdaKappaNN1}.

Similar (equivalent) results for asymmetric (but planar) waveguides were
recently obtained by Nazarov \cite{Nazarov II} , and independently by Amore
et al. \cite{Amore} using precise numerical collocation (using many grid
points). Coupled waveguide geometries of finite extension may also display a
high sensitivity of the localisation of the ground state mode to slight
changes of the geometrical shape \cite{dng}, \cite{dng1}.

A possible physical explanation why for a single particle a localised ground
state ceases to exist around the corner zone in $\mathcal{L}$ for $\kappa
^{\left( \mathcal{L}\right) }\leq \kappa _{c}^{\left( \mathcal{L}\right) }$,
and likewise ceases to exist around the branching zone in $\mathcal{T}$ for $%
\kappa _{c}^{\left( \mathcal{T}\right) }\leq \kappa ^{\left( \mathcal{T}%
\right) }$, but always exists around the crossing zone $\mathcal{A}%
_{0}\subset \mathcal{C}$ for any $\kappa ^{\left( \mathcal{C}\right) }>0$,
we discuss in the next section \ref{Kinetic Energy Induced Confinement at a
Crossing.}

\section{Reason for Non Standard Trapping Around Branching Zones in Quantum
Waveguides}

\label{Kinetic Energy Induced Confinement at a Crossing.}

Besides bouncing back and forth from the hard walls a \emph{classical}
particle senses no extra force when it moves, say along the tube axis $%
\mathbf{e}_{y}$, inside the cross shaped waveguide $\mathcal{C}$. The
question is then, why in quantum mechanics the ground state $\psi
_{0}^{\left( \mathcal{C}\right) }(\mathbf{r})$ of a particle moving inside $%
\mathcal{C}$ is always localised around the crossing zone $\mathcal{A}%
_{0}\subset \mathcal{C}$ with an eigenvalue $E_{0}^{\left( \mathcal{C}%
\right) }$ below the excitation threshold $\varepsilon _{xt}^{\left( 
\mathcal{C}\right) }$. The key observation to answer this question is, that
far from the crossing zone $\mathcal{A}_{0}$, say deep inside the arms $%
\mathcal{A}_{2}$ and $\mathcal{A}_{4}$, the Schr\"{o}dinger eigenvalue
problem is separable, so that $\psi _{0}^{\left( \mathcal{C}\right) }\left( 
\mathbf{r}\right) =\psi _{\perp }^{\left( \mathcal{C}\right) }\left( \mathbf{%
r}_{\perp }\right) \phi _{0}^{\left( \mathcal{C}\right) }(y)$. Introducing
the function 
\begin{equation}
\phi _{0}^{\left( \mathcal{C}\right) }(y)=\int_{-\infty }^{\infty
}dx\int_{-L_{z}}^{L_{z}}dz\ \psi _{\perp }^{\left( \mathcal{C}\right)
}\left( \mathbf{r}\right) \psi _{0}^{\left( \mathcal{C}\right) }\left( 
\mathbf{r}\right)
\end{equation}%
we see that (\ref{one particel Schroedinger equation}) is equivalent to a 
\emph{one-dimensional} Schr\"{o}dinger eigenvalue problem 
\begin{equation}
\left[ -\partial _{y}^{2}+V_{\perp }^{\left( \mathcal{C}\right) }(y)\right]
\phi _{0}^{\left( \mathcal{C}\right) }(y)=E_{0}^{\left( \mathcal{C}\right)
}\phi _{0}^{\left( \mathcal{C}\right) }(y)
\label{effective 1D-Schroedinger eigen value problem}
\end{equation}%
, but with an effective potential $V_{\perp }^{\left( \mathcal{C}\right)
}(y) $ generated by the \emph{transversal} kinetic energy, 
\begin{equation}
V_{\perp }^{\left( \mathcal{C}\right) }(y)=\frac{\int_{-\infty }^{\infty
}dx\int_{-L_{z}}^{L_{z}}dz\ \psi _{\perp }^{\left( \mathcal{C}\right)
}\left( \mathbf{r}\right) \left( -\partial _{x}^{2}-\partial _{z}^{2}\right)
\psi _{0}^{\left( \mathcal{C}\right) }\left( \mathbf{r}\right) }{%
\int_{-\infty }^{\infty }dx\int_{-L_{z}}^{L_{z}}dz\ \psi _{0}^{\left( 
\mathcal{C}\right) }\left( \mathbf{r}\right) \psi _{\perp }^{\left( \mathcal{%
C}\right) }\left( \mathbf{r}\right) }  \label{effective 1D-potential}
\end{equation}%
, as the coordinate $y$ runs along the tube axis $\mathbf{e}_{y}$.

\subsection{Non Standard Trapping in $\mathcal{C}$.}

For example, consider equal tube diameters $w_{x}^{\left( \mathcal{C}\right)
}=w_{y}^{\left( \mathcal{C}\right) }=w_{z}^{\left( \mathcal{C}\right) }=2L$.
Then a rather accurate fit to the spatial variation of the transversal part $%
\psi _{\perp }^{\left( \mathcal{C}\right) }\left( \mathbf{r}\right) $ of the
numerically calculated three-dimensional ground state wave function $\psi
_{0}^{\left( \mathcal{C}\right) }\left( \mathbf{r}\right) $ inside the
respective tube segments $\mathcal{A}_{j}$ is%
\[
\psi _{\perp }^{\left( \mathcal{C}\right) }\left( \mathbf{r}\right) =\left\{ 
\begin{array}{c}
a_{\perp }\cos \left( \frac{\pi }{2L}x\right) \cos \left( \frac{\pi }{2L_{z}}%
z\right) \text{ for }\mathbf{r\in }\mathcal{A}_{2}\cup \mathcal{A}_{4} \\ 
\frac{a_{\perp }}{\cosh ^{2}\left( \frac{x}{2\lambda }\right) }\cos \left( 
\frac{\pi }{2L_{z}}z\right) \text{for }\mathbf{r\in }\mathcal{A}_{0}\cup 
\mathcal{A}_{1}\cup \mathcal{A}_{3} \\ 
0\text{ for }\mathbf{r\notin }\mathcal{C}%
\end{array}%
\right. 
\]%
Here, the length $\lambda $ is equal to the numerically determined
localisation length of the three-dimensional ground state wave function $%
\psi _{0}^{\left( \mathcal{C}\right) }\left( \mathbf{r}\right) $ for a
single particle, see Fig.\ref{fig:psi3Dplot}.

As can be seen in Fig.\ref{fig:potentialwell}, the effective potential $%
V_{\perp }^{\left( \mathcal{C}\right) }(y)$ calculated from (\ref{effective
1D-potential}) takes on the form of a one-dimensional box-shaped potential
as one traverses the crossing zone $\mathcal{A}_{0}\subset \mathcal{C}$
along the tube axis $\mathbf{e}_{y}$:

\begin{figure}[tbp]
\begin{centering}
\includegraphics[scale=1.]{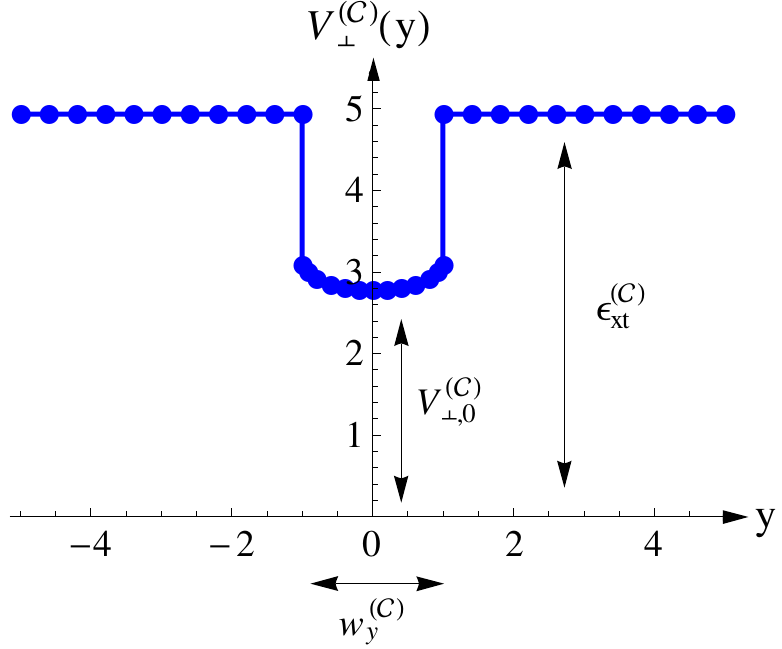}
\par\end{centering}
\caption{Effective one-dimensional potential $V_{\perp }^{\left( \mathcal{C}%
\right) }(y)\ $ vs. $y$ as sensed by a particle traversing the crossing
region of the waveguide $\mathcal{C}$. Length and energy measured in units
of $L$ and $\protect\varepsilon _{L}$, respectively.}
\label{fig:potentialwell}
\end{figure}

\begin{equation}
V_{\perp }^{\left( \mathcal{C}\right) }(y)=\left\{ 
\begin{array}{c}
V_{\perp ,0}^{\left( \mathcal{C}\right) }\text{\ for }\left\vert
y\right\vert <L \\ 
\varepsilon _{xt}^{\left( \mathcal{C}\right) }\text{ for }\left\vert
y\right\vert >L%
\end{array}%
\right.  \label{box-shaped  attractive 1D potential}
\end{equation}%
, with $\varepsilon _{xt}^{\left( \mathcal{C}\right) }=\lim_{\left\vert
y\right\vert \rightarrow \infty }V_{\perp }^{\left( \mathcal{C}\right)
}(y)>V_{\perp ,0}^{\left( \mathcal{C}\right) }\geq 0$ denoting here the 
\emph{excitation threshold} in the arms $\mathcal{A}_{2}$ , $\mathcal{A}_{4}$%
. Actually there holds $\varepsilon _{xt}^{\left( \mathcal{C}\right)
}>V_{\perp ,0}^{\left( \mathcal{C}\right) }$ for arbitrary cross shaped
waveguides $\mathcal{C}$, so the potential $V_{B}^{\left( \mathcal{C}\right)
}(y)$ is always attractive and has therefore a finite binding strength
(scaled units) 
\begin{equation}
b_{\perp \ }^{\left( \mathcal{C}\right) }=\sqrt{\varepsilon _{xt}^{\left( 
\mathcal{C}\right) }-V_{\perp ,0}^{\left( \mathcal{C}\right) }}
\end{equation}%
The existence of a bound state with \emph{even parity} localised inside the
box $\left\vert y\right\vert <L$ is then granted (see any standard text on
Quantum Mechanics, e.g. \cite{QM}). So it is the rapid change of the
transversal kinetic energy that occurs in our waveguide system around the
crossing zone $\mathcal{A}_{0}\subset \mathcal{C}$ , see Figure (\ref%
{fig:waveguides}), that provides the physical mechanism for trapping a
quantum particle of mass $m$ in that region. Via the excitation threshold $%
\varepsilon _{xt}^{\left( \mathcal{C}\right) }$ , see \ref{excitation
threshold C}, the strength of this unconventional trapping force is not only
dependent on the respective tube sizes $w_{a}^{\left( \mathcal{C}\right) }$
, but also dependent on mass, a lighter particle thus experiencing a
stronger trapping force than a heavier one!

Because for an \emph{attractive} one-dimensional box-shaped potential $%
V_{\perp }^{\left( \mathcal{C}\right) }(y)$ there always exists a localised
ground state with even parity for any value of the binding strength $%
b_{\perp \ }^{\left( \mathcal{C}\right) }>0$, one may further simplify the
problem. Being only interested in the asymptotic behaviour of the ground
state $\phi _{0}^{\left( \mathcal{C}\right) }(y)$ at a large distance $%
\left\vert y\right\vert >>L$ to the crossing zone $\mathcal{A}_{0}$ , we may
replace $V_{\perp }^{\left( \mathcal{C}\right) }(y)$ by an equivalent
attractive delta function potential (scaled units): 
\begin{equation}
V_{\perp }^{\left( \mathcal{C}\right) }(y)\rightarrow \widetilde{V}_{\perp
}^{\left( \mathcal{C}\right) }(y)=\varepsilon _{xt}^{\left( \mathcal{C}%
\right) }-\frac{2}{\lambda }\delta (y)
\end{equation}%
The associated normalised bound state wave function $\widetilde{\phi }%
_{0}^{\left( \mathcal{C}\right) }(y)$ is then (see any standard text on
Quantum Mechanics, for example \cite{QM}): 
\begin{equation}
\widetilde{\phi }_{0}^{\left( \mathcal{C}\right) }(y)=\sqrt{\frac{1}{\lambda 
}}\exp \left( -\frac{\left\vert y\right\vert }{\lambda }\right)
\label{bound state delta-function potential}
\end{equation}%
This formula provides for $\left\vert y\right\vert >>L$ the asymptotic
behaviour of the ground state $\phi _{0}^{\left( \mathcal{C}\right) }(y)$.
It follows at once that the eigenvalue $E_{0}^{\left( \mathcal{C}\right) }$
associated with $\phi _{0}^{\left( \mathcal{C}\right) }(y)$ is given by
(scaled units) 
\begin{equation}
E_{0}^{\left( \mathcal{C}\right) }=\varepsilon _{xt}^{\left( \mathcal{C}%
\right) }-\frac{1}{\lambda ^{2}}
\end{equation}%
In order that such a toy model actually makes sense it is mandatory that $%
\lambda >L$ , where $w_{y}^{\left( \mathcal{C}\right) }=2L$ measures the
lateral size of the crossing zone, see Fig.\ref{fig:potentialwell}. It turns
out that our results from the full three-dimensional numerical calculations
for the localised ground state $\psi _{0}^{\left( \mathcal{C}\right) }\left( 
\mathbf{r}\right) $ indeed fulfill this requirement.

\begin{figure}[tbph]
\begin{centering}
\includegraphics[scale=1.1]{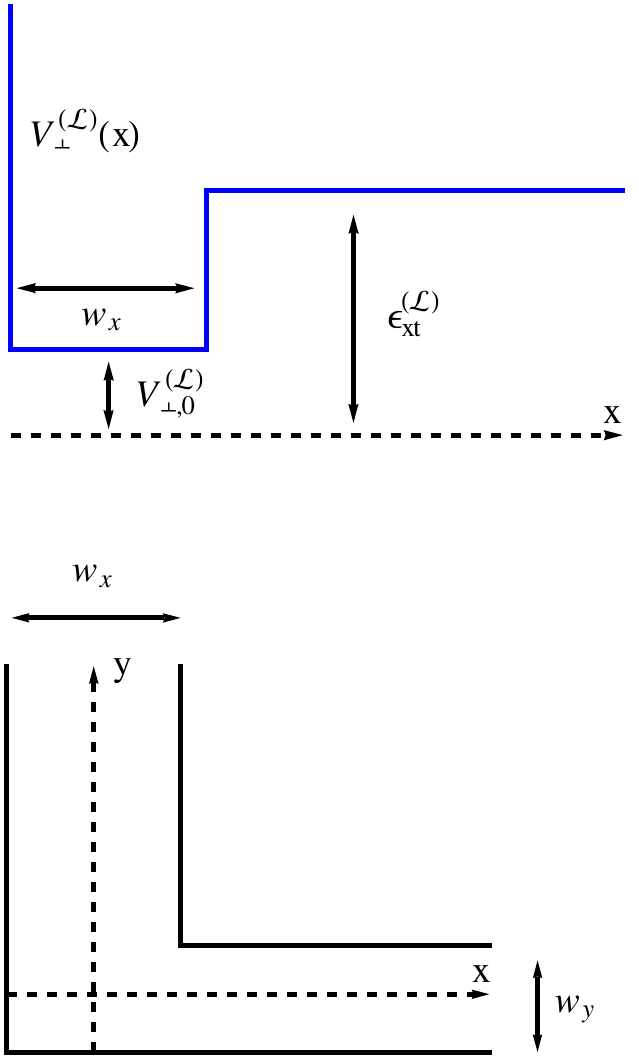}
\par\end{centering}
\caption{Effective one-dimensional attractive potential $V_{\perp }^{\left( 
\mathcal{L}\right) }(x)\ $ vs. $x$ as sensed by a particle approaching the
corner zone of the $L$-shaped waveguide $\mathcal{L}$. Length and energy
measured in units of $L$ and $\protect\varepsilon _{L}$, respectively.}
\label{fig:LGeoV0}
\end{figure}

\subsection{Non Standard Trapping in $\mathcal{L}$.}

The cranked $L$-shaped waveguide $\mathcal{L}$ displayed in Fig.\ref%
{fig:waveguides} may be considered as a subdomain (one quarter) of the
crossing geometry $\mathcal{C}$. To keep the notation compatible with the
one employed to describe the original waveguide $\mathcal{C}$, the height of
the tubes comprising $\mathcal{L}$ is denoted as $w_{z}^{\left( \mathcal{L}%
\right) }=2L_{z}$, while the lateral widths of the tubes with axis $\mathbf{e%
}_{y}$ and $\mathbf{e}_{x}$, respectively, are denoted as $w_{x}^{\left( 
\mathcal{L}\right) }=\frac{w_{x}^{\left( \mathcal{C}\right) }}{2}=L$ and $%
w_{y}^{\left( \mathcal{L}\right) }=\frac{w_{y}^{\left( \mathcal{C}\right) }}{%
2}=L_{y}$. Traversing the domain $\mathcal{L}$, say parallel to the tube
axis $\mathbf{e}_{x}$ , see Fig.\ref{fig:waveguides}, there results in
analogy to the previous consideration for the domain $\mathcal{C}$ an
effective one-dimensional Schr\"{o}dinger eigenvalue problem 
\begin{equation}
\left[ -\partial _{x}^{2}+V_{\perp }^{\left( \mathcal{L}\right) }(x)\right]
\phi _{0}^{\left( \mathcal{L}\right) }(x)=E_{0}^{\left( \mathcal{L}\right)
}\phi _{0}^{\left( \mathcal{L}\right) }(x)
\end{equation}%
Here, the support of $\phi _{0}^{\left( \mathcal{L}\right) }(x)$ is
restricted to the half line $x>0$ with the effective one-dimensional
potential, see (\ref{effective 1D-potential}), now describing the drop in
the transversal kinetic energy around the corner of $\mathcal{L}$ : 
\begin{equation}
V_{\perp }^{\left( \mathcal{L}\right) }(x)=\left\{ 
\begin{array}{c}
\infty \text{ for }x=0 \\ 
V_{\perp ,0}^{\left( \mathcal{L}\right) }\text{ for }0<x<L \\ 
\varepsilon _{xt}^{\left( \mathcal{L}\right) }\ \text{for }L<x%
\end{array}%
\right.  \label{L-shaped geometry effective 1D-potential}
\end{equation}%
The constant $V_{\perp ,0}^{\left( \mathcal{L}\right) }$ $<\varepsilon
_{xt}^{\left( \mathcal{L}\right) }$ describes the effect, that the
transversal kinetic energy may assume a finite value inside the central
region $\mathcal{A}_{0}$, possibly also depending on the tube size
parameters $w_{a}^{\left( \mathcal{L}\right) }$.

The problem to find the ground state for this potential $V_{\perp }^{\left( 
\mathcal{L}\right) }(x)$ on the half line $0<x<\infty $ is equivalent to
looking for the lowest lying eigenstate with \emph{odd} parity for an
attractive one-dimensional box-shaped potential $V_{\perp }^{\left( \mathcal{%
C}\right) }(x)$ of the type (\ref{box-shaped attractive 1D potential})
extended along the full real axis $-\infty <x<\infty $. As is well known,
the existence of a localised eigenstate with \emph{odd} parity for an
attractive box-shaped potential like $V_{\perp }^{\left( \mathcal{C}\right)
}(x)$ requires a sufficiently strong binding strength (scaled units) 
\begin{equation}
b_{\perp \ }^{\left( \mathcal{L}\right) }=\sqrt{\varepsilon _{xt}^{\left( 
\mathcal{L}\right) }-V_{\perp ,0}^{\left( \mathcal{L}\right) }}>\frac{\pi }{2%
}
\end{equation}%
(see any standard text on Quantum Mechanics, for example \cite{QM}). It is
clear from what has been said, that there exists no localised ground state
around the corner of a quantum waveguide $\mathcal{L}$ if its tube widths
ratio $\kappa ^{\left( \mathcal{L}\right) }=\frac{w_{y}^{\left( \mathcal{L}%
\right) }}{w_{x}^{\left( \mathcal{L}\right) }}$ is below a critical value $%
\kappa _{c}^{\left( \mathcal{L}\right) }$, in agreement with results
obtained from the full three-dimensional numerical calculations presented in
Fig.\ref{fig:LambdaKappaNN1}. As the ratio $\kappa ^{\left( \mathcal{L}%
\right) }$ approaches its lower bound $\kappa _{c}^{\left( \mathcal{L}%
\right) }$, the binding strength $b_{\perp \ }^{\left( \mathcal{L}\right) }$
approaches (from above) the critical value $\frac{\pi }{2}$, and the
localisation length $\lambda $ diverges.

\subsection{Non Standard Trapping in $\mathcal{T}$.}

Traversing the waveguide $\mathcal{T}$ along the tube axis $\mathbf{e}_{y}$,
see Fig.\ref{fig:waveguides}, the effective one-dimensional potential
associated with the drop in the transversal kinetic energy around the
braching zone may be described by a potential $V_{\perp }^{\left( \mathcal{T}%
\right) }\left( x=\frac{w_{x}^{\left( \mathcal{T}\right) }}{2},y\right) $
vs. $y$ similar to the one displayed in Fig.\ref{fig:potentialwell} for the
waveguide $\mathcal{C}$. But traversing $\mathcal{T}$ along the tube axis $%
\mathbf{e}_{x}$ the corresponding effective potential $V_{\perp }^{\left( 
\mathcal{T}\right) }\left( x,y=\frac{w_{y}^{\left( \mathcal{T}\right) }}{2}%
\right) $ vs. $x$ is similar to the one displayed in Fig.\ref{fig:LGeoV0}
for the domain $\mathcal{L}$. It follows from what has been said before,
that a localised ground state around the branching zone of a $T$-shaped
waveguide $\mathcal{T}$ exists only if the ratio of tube widths $\kappa
^{\left( \mathcal{T}\right) }=\frac{w_{y}^{\left( \mathcal{T}\right) }}{%
w_{x}^{\left( \mathcal{T}\right) }}$ is not too large, thus ensuring a large
enough binding strength $b_{\perp \ }^{\left( \mathcal{T}\right) }>\frac{\pi 
}{2}$ of the effective potential $V_{\perp }^{\left( \mathcal{T}\right)
}\left( x\right) $, in agreement with the results of the full
three-dimensional numerical calculations presented in Fig.\ref%
{fig:LambdaKappaNN1}.

\section{ Localised BEC Ground States Around Branching Zones in $\mathcal{C}$
, $\mathcal{L}$ and $\mathcal{T}$.}
\label{ Heat Kernel Method for Numerical Solution of Gross-Pitaevskii
Equation.}

To find the optimal GP-orbital $\psi ^{\left( \mathcal{C}\right) }(\mathbf{r}%
)$ determining the Hartree ground state (\ref{ground state BEC}) of an
interacting BEC confined around the crossing zone of the waveguide $\mathcal{%
C}$ we need to solve the Gross-Pitaevskii equation (\ref{Gross-Pitaevskii I}%
). To construct a suitable splitting scheme we consider an auxiliary
diffusion process:%
\begin{equation}
-\frac{\partial }{\partial \tau }\psi (\mathbf{r},\tau )=\left[
H_{kin}+U_{\psi }(\mathbf{r},\tau )\right] \psi (\mathbf{r},\tau )
\label{diffusion process GP}
\end{equation}%
However, because the amplitude of the auxiliary wave function $\psi (\mathbf{%
r},\tau )$ decays exponentially with diffusion time $\tau $ as the diffusion
process progresses the interaction term neeeds explicit normalization \cite%
{Krotscheck}: 
\begin{equation}
U_{\psi }(\mathbf{r},\tau )=\left( N-1\right) \frac{4\pi \hbar ^{2}a_{s}}{m}%
\ \frac{\left\vert \psi (\mathbf{r},\tau )\right\vert ^{2}}{\int_{\mathcal{C}%
}d^{3}r^{\prime }\ \left\vert \psi (\mathbf{r}^{\prime },\tau )\right\vert
^{2}}
\end{equation}%
Apparently, for large diffusion time $\tau $ then $U_{\psi }(\mathbf{r},\tau
)$ becomes independent on $\tau $. The seeked localised GP-orbital is thus
given by 
\begin{equation}
\psi ^{\left( \mathcal{C}\right) }(\mathbf{r})=\lim_{\tau \rightarrow \infty
}\frac{\psi (\mathbf{r},\tau )}{\sqrt{\int_{\mathcal{C}}d^{3}r^{\prime
}\left\vert \psi (\mathbf{r}^{\prime },\tau )\right\vert ^{2}}}
\end{equation}

In sharp contrast to the behaviour in a harmonic trap, now the kinetic
energy in the localised Hartree ground of a BEC, that is confined around the
crossing zone $\mathcal{A}_{0}$ of $\mathcal{C}$ by the described non
standard trapping force, dominates over the interaction energy even for a
large particle number $N$. To solve for a large particle number $N$ the
Gross-Pitaevskii equation (\ref{Gross-Pitaevskii I}) accurately, a specially
tailored splitting scheme is useful as described in the appendix \ref%
{Appendix A}. The update rule that determines $\psi (\mathbf{r},\tau _{n+1})$
from a given $\psi (\mathbf{r},\tau _{n})$ for a short diffusion time
interval $\Delta \tau $ consists of the following five steps:%
\begin{eqnarray}
\psi (\mathbf{r},\tau _{0}) &=&\psi ^{\left( in\right) }(\mathbf{r})
\label{splitting scheme GP-equation} \\
\tau _{n+1} &=&\tau _{n}+\Delta \tau \text{ for }n=0,1,2,...  \nonumber \\
\psi ^{\left( IV\right) }(\mathbf{r},\tau _{n}) &=&e^{-\frac{\Delta \tau }{6}%
U_{\psi }(\mathbf{r},\tau _{n})}\psi (\mathbf{r},\tau _{n})  \nonumber \\
\psi ^{\left( III\right) }(\mathbf{r},\tau _{n}) &=&\int_{\mathcal{C}}d^{3}%
\mathbf{r}^{\prime }K(\mathbf{r},\mathbf{r}^{\prime },\frac{\Delta \tau }{2}%
)\psi ^{\left( IV\right) }(\mathbf{r}^{\prime },\tau _{n})  \nonumber \\
\psi ^{\left( II\right) }(\mathbf{r},\tau _{n}) &=&e^{-\frac{2\Delta \tau }{3%
}U_{\psi }(\mathbf{r},\tau _{n})}\psi ^{\left( III\right) }(\mathbf{r},\tau
_{n})  \nonumber \\
\psi ^{\left( I\right) }(\mathbf{r},\tau _{n}) &=&\int_{\mathcal{C}}d^{3}%
\mathbf{r}^{\prime }K(\mathbf{r},\mathbf{r}^{\prime },\frac{\Delta \tau }{2}%
)\psi ^{\left( II\right) }(\mathbf{r}^{\prime },\tau _{n})  \nonumber \\
\psi (\mathbf{r},\tau _{n+1}) &=&e^{-\frac{\Delta \tau }{6}U_{\psi }(\mathbf{%
r},\tau _{n})}\psi ^{\left( I\right) }(\mathbf{r},\tau _{n})  \nonumber
\end{eqnarray}%
Here the kernel $K(\mathbf{r},\mathbf{r}^{\prime },\frac{\Delta \tau }{2})$
is associated with the kinetic energy operator $H_{kin}$ of a single
particle moving inside $\mathcal{C}$ and obeying to Dirichlet boundary value
conditions at the walls $\partial \mathcal{C}$ of that waveguide, see
appendix \ref{Appendix B}. It acts on the functions $\psi ^{\left( IV\right)
}(\mathbf{r}^{\prime },\tau _{n})$ and $\psi ^{\left( II\right) }(\mathbf{r}%
^{\prime },\tau _{n})$ as described in the previous section \ref{Localised
Single Particle Eigenmodes}.

In the numerical calculations with the proposed splitting scheme we chose $%
\Delta \tau =0.01\times \left[ \frac{\hbar }{\varepsilon _{L}}\right] $. The
obtained results clearly show, that the optimal GP-orbital is indeed
localised around the crossing zone $\mathcal{A}_{0}$, provided $1\leq N\leq
N_{c}^{\left( \mathcal{C}\right) }$, where $N_{c}^{\left( \mathcal{C}\right)
}$ denotes a critical particle number depending on the tube sizes $%
w_{a}^{\left( \mathcal{C}\right) }=2L_{a}$ and the $s$-wave scattering
length $a_{s}$ of the Bose atoms. Physically, $N_{c}^{\left( \mathcal{C}%
\right) }$ has the meaning of the maximal number of particles that can be
trapped in the localised Hartree ground state by the described non standard
confinement mechanism. In particular, like in the case $N=1$, there exist
localised GP-orbitals $\psi _{\gamma }\left( \mathbf{r}\right) $ displaying
different discrete symmetries $\gamma \in \left\{
A_{g},B_{1g},B_{2u},B_{3u}\right\} $.

\begin{figure}[tbp]
\begin{centering}
\includegraphics[scale=0.9]{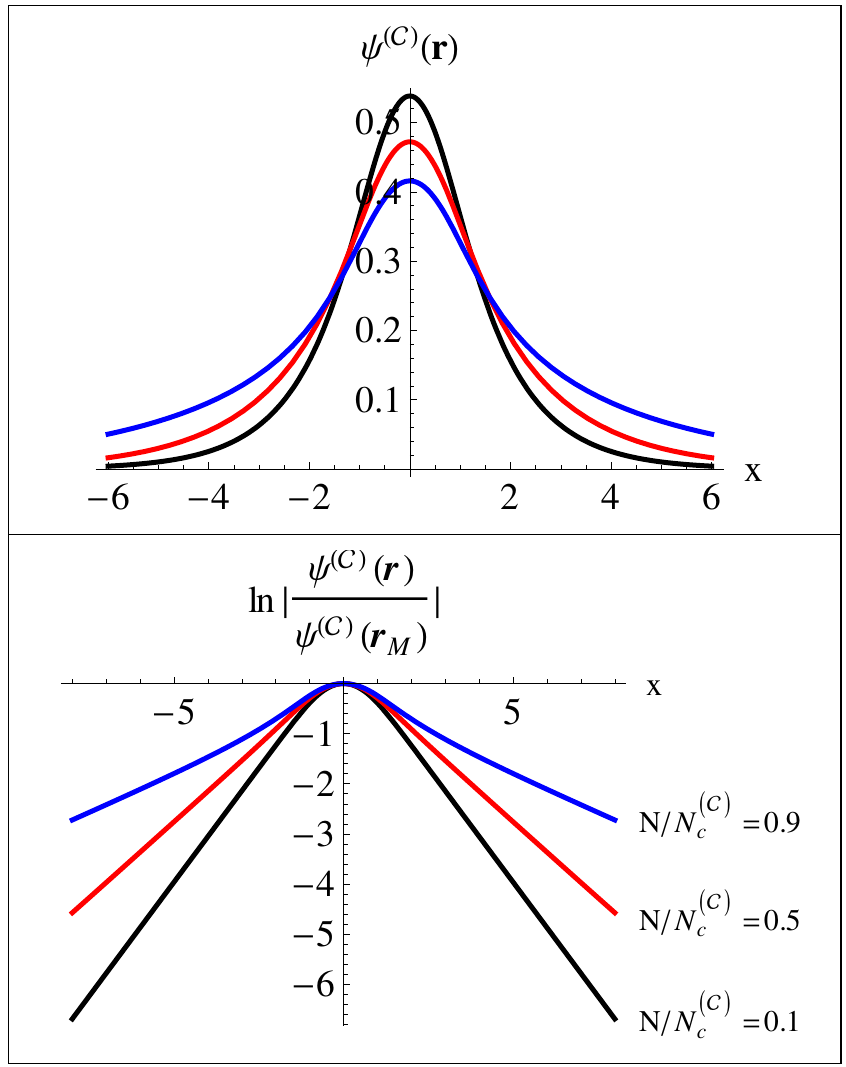}
\par\end{centering}
\caption{Profile of the GP-orbital $\protect\psi ^{(\mathcal{C})}\left( 
\mathbf{r}\right) $ comprising the Hartree ground state localised around the
crossing zone of the waveguide $\mathcal{C}$ for different particle number
ratios $N/N_{c}^{\left( \mathcal{C}\right) }$: black line $N/N_{c}^{\left( 
\mathcal{C}\right) }=0.1$ , red line $N/N_{c}^{\left( \mathcal{C}\right)
}=0.5$, blue line $N/N_{c}^{\left( \mathcal{C}\right) }=0.9$. Results shown
correspond to tube sizes $w_{x}^{\left( \mathcal{C}\right) }=w_{y}^{\left( 
\mathcal{C}\right) }=w_{z}^{\left( \mathcal{C}\right) }=2L$. Distance $x$
measured in units of $L$.}
\label{fig:PsiProfileC}
\end{figure}

Not unexpectedly, the Hartree ground state (\ref{ground state BEC}) with the
lowest energy in the waveguide system $\mathcal{C}$ is buildt from the
orbital $\psi ^{\left( \mathcal{C}\right) }(\mathbf{r})\equiv \psi
_{A_{g}}\left( \mathbf{r}\right) $ , which orbital is nodeless in $\mathcal{C%
}$. Like in the single particle case, the optimal GP-orbital $\psi ^{\left( 
\mathcal{C}\right) }(\mathbf{r})$ is localised around the origin $\mathbf{r}%
= $ $\mathbf{0}$ of the crossing zone $\mathcal{A}_{0}\subset \mathcal{C}$,
but the exponential decay of $\psi ^{\left( \mathcal{C}\right) }(\mathbf{r})$
with increasing distance to the crossing zone is slower for higher particle
numbers $N$, see Fig.\ref{fig:PsiProfileC}.

\begin{figure}[tbp]
\begin{centering}
\includegraphics[scale=0.9]{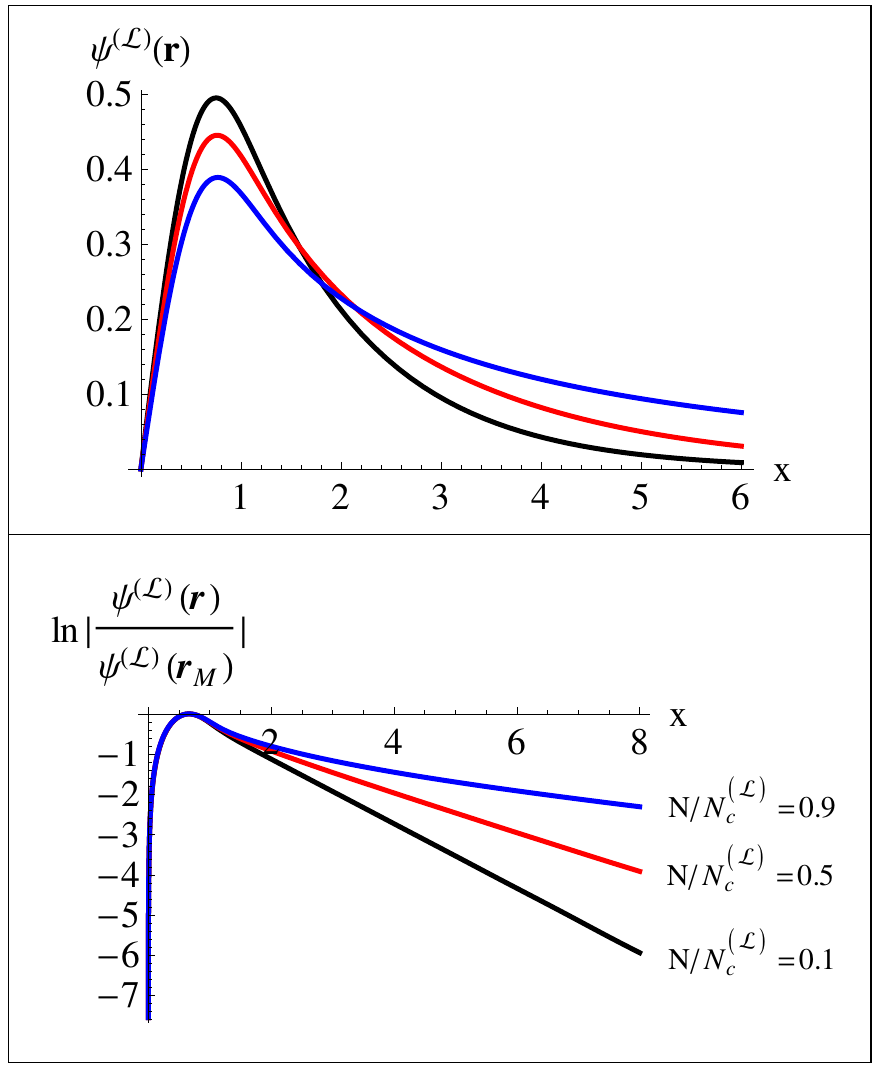}
\par\end{centering}
\caption{Profile of the GP-orbital $\protect\psi ^{(\mathcal{L})}\left( 
\mathbf{r}\right) $ comprising the Hartree ground state localised around the
corner zone of the waveguide $\mathcal{L}$ for different particle number
ratios $N/N_{c}^{\left( \mathcal{L}\right) }$: black line $N/N_{c}^{\left( 
\mathcal{L}\right) }=0.1$ , red line $N/N_{c}^{\left( \mathcal{L}\right)
}=0.5$, blue line $N/N_{c}^{\left( \mathcal{L}\right) }=0.9$. Results shown
correspond to tube sizes $w_{x}^{\left( \mathcal{L}\right) }=w_{y}^{\left( 
\mathcal{L}\right) }=L$, $w_{z}^{\left( \mathcal{L}\right) }=2L$. Distance $%
x $ measured in units of $L$.}
\label{fig:PsiProfileL}
\end{figure}

\begin{figure}[tbp]
\begin{centering}
\includegraphics[scale=0.9]{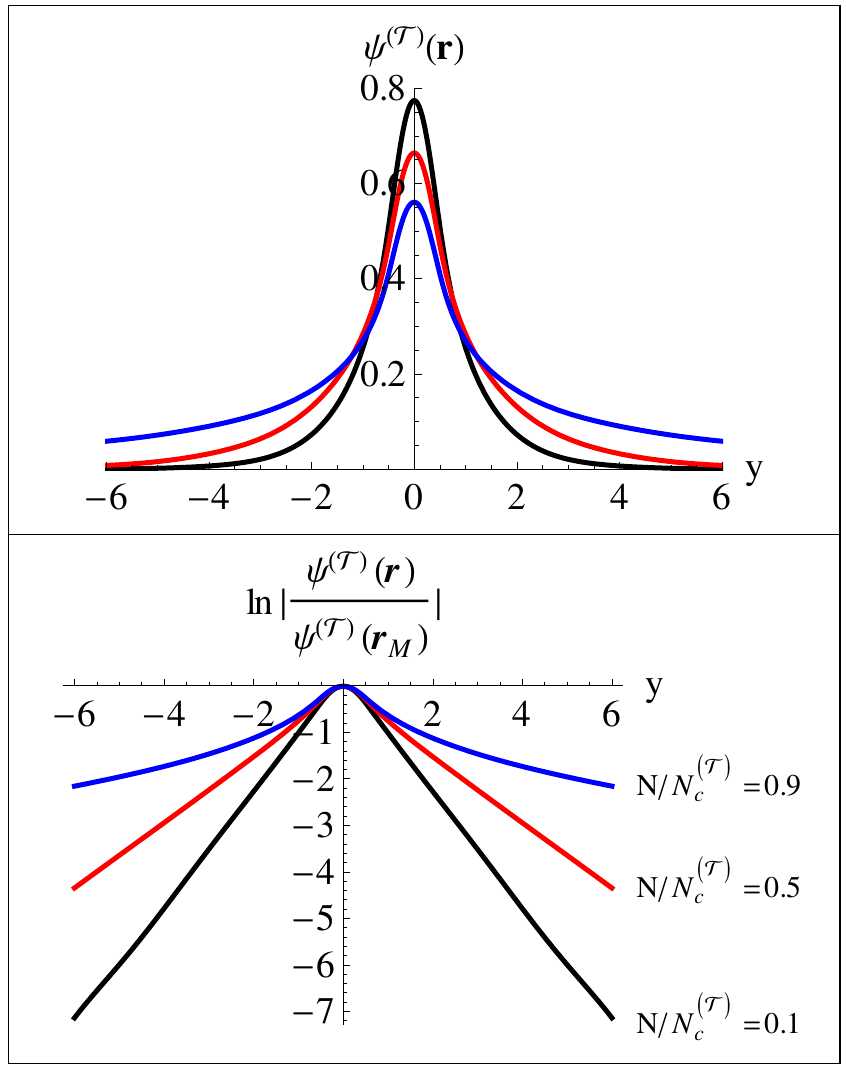}
\par\end{centering}
\caption{Profile along tube axis direction $\mathbf{e}_{y}$ of the
GP-orbital $\protect\psi ^{(\mathcal{T})}\left( \mathbf{r}\right) $
comprising the Hartree ground state localised around the branching zone of
the waveguide $\mathcal{T}$ for different particle number ratios $%
N/N_{c}^{\left( \mathcal{T}\right) }$: black line $N/N_{c}^{\left( \mathcal{T%
}\right) }=0.1$ , red line $N/N_{c}^{\left( \mathcal{T}\right) }=0.5$, blue
line $N/N_{c}^{\left( \mathcal{T}\right) }=0.9$. Results shown correspond to
tube sizes $w_{x}^{\left( \mathcal{T}\right) }=w_{y}^{\left( \mathcal{T}%
\right) }=L$, $w_{z}^{\left( \mathcal{T}\right) }=2L$. Distance $y$ measured
in units of $L$.}
\label{fig:PsiProfileTY}
\end{figure}

A similar behaviour is also found for the other localised GP-orbitals $\psi
_{\gamma }\left( \mathbf{r}\right) $ with symmetry representation $\gamma
\in \left\{ B_{1g},B_{3u}\right\} $, corresponding to the localised
GP-orbitals $\psi ^{\left( \mathcal{L}\right) }(\mathbf{r})$ and $\psi
^{\left( \mathcal{T}\right) }(\mathbf{r})$ comprising the localised BEC
ground states around the branching zones of the quantum waveguides $\mathcal{%
L}$ and $\mathcal{T}$ , respectively. In Fig.\ref{fig:PsiProfileL}, Fig.\ref%
{fig:PsiProfileTX}, Fig.\ref{fig:PsiProfileTY} we show for the waveguides $%
\mathcal{L}$ and $\mathcal{T}$ the profiles of the localised GP-orbitals $%
\psi ^{\left( \mathcal{L}\right) }(\mathbf{r})$ and $\psi ^{\left( \mathcal{T%
}\right) }(\mathbf{r})$ for different particle numbers $N$. For $N\geq
N_{c}^{\left( \Gamma \right) }$ , where $N_{c}^{\left( \Gamma \right) }$
denotes a critical particle number associated with the respective waveguide
geometries $\Gamma \in \left\{ \mathcal{C},\mathcal{L},\mathcal{T}\right\} $%
, a localised GP-orbital $\psi ^{\left( \Gamma \right) }(\mathbf{r})$ ceases
to exist.

\begin{figure}[tbp]
\begin{centering}
\includegraphics[scale=0.92]{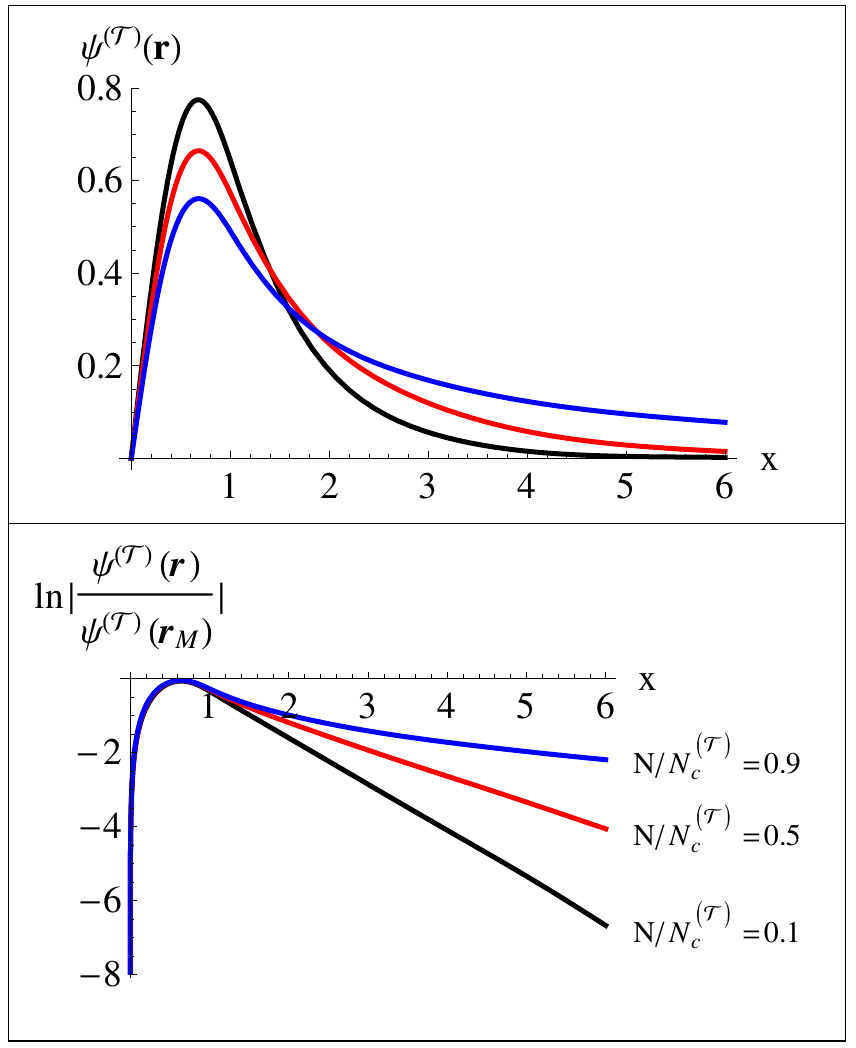}
\par\end{centering}
\caption{Profile along tube axis direction $\mathbf{e}_{x}$ of the
GP-orbital $\protect\psi ^{(\mathcal{T})}\left( \mathbf{r}\right) $
comprising the Hartree ground state localised around the branching zone of
the waveguide $\mathcal{T}$ for different particle number ratios $%
N/N_{c}^{\left( \mathcal{T}\right) }$: black line $N/N_{c}^{\left( \mathcal{T%
}\right) }=0.1$ , red line $N/N_{c}^{\left( \mathcal{T}\right) }=0.5$, blue
line $N/N_{c}^{\left( \mathcal{T}\right) }=0.9$. Results shown correspond to
tube sizes $w_{x}^{\left( \mathcal{T}\right) }=w_{y}^{\left( \mathcal{T}%
\right) }=L$, $w_{z}^{\left( \mathcal{T}\right) }=2L$. Distance $x$ measured
in units of $L$.}
\label{fig:PsiProfileTX}
\end{figure}

\begin{figure}[tbp]
\begin{centering}
\includegraphics[scale=1.2]{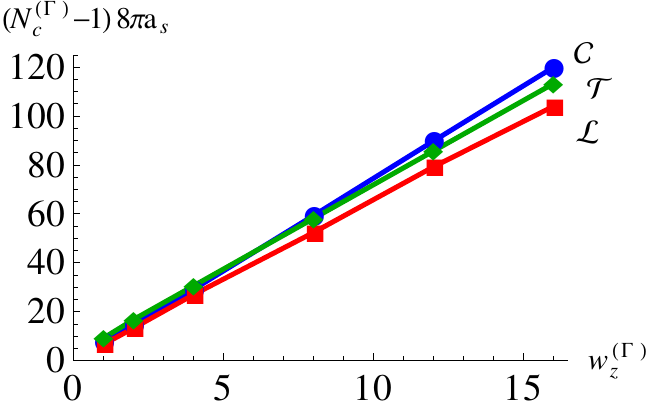}
\par\end{centering}
\caption{Plot of the critical particle number $N_{c}^{(\Gamma )}-1$ vs. box
height $w_{z}^{\left( \Gamma \right) }$ for the cross shaped geometry $%
\mathcal{C}$ (blue line), $T$-shaped geometry $\mathcal{T}$ (green line) and 
$L$-shaped geometry $\mathcal{L}$ (red line). Choice of tube width
parameters: $w_{x}^{\left( \mathcal{C}\right) }=w_{y}^{\left( \mathcal{C}%
\right) }=2L$ , $w_{x}^{\left( \mathcal{T}\right) }=w_{y}^{\left( \mathcal{T}%
\right) }=L$ and $w_{x}^{\left( \mathcal{L}\right) }=w_{y}^{\left( \mathcal{L%
}\right) }=L$. Tube height $w_{z}^{\left( \Gamma \right) }$ measured in
units of $L$.}
\label{fig:NcPlot}
\end{figure}

For the determination of the optimal orbital $\psi ^{\left( \Gamma \right)
}\left( \mathbf{r}\right) $ and the associated chemical potential $\mu
_{N}^{\left( \Gamma \right) }$ of the Hartree ground state of the BEC only
the effective interaction parameter $\left( N-1\right) $ $\frac{8\pi a_{s}}{L%
}\ $matters (scaled units). As displayed in Fig.\ref{fig:NcPlot}, the
critical particle number $N_{c}^{\left( \Gamma \right) }$ displays the
expected linear increase as the respective tube diameter $w_{z}^{\left(
\Gamma \right) }$ increases.

Once the (normalized!) optimal GP-orbital $\psi ^{\left( \Gamma \right) }(%
\mathbf{r})$ has been found for the respective waveguide geometries $\Gamma
\in \left\{ \mathcal{C},\mathcal{L},\mathcal{T}\right\} $, it follows
directly from (\ref{Gross-Pitaevskii I}) by taking a scalar product with the
adjoint orbital $\left[ \psi ^{\left( \Gamma \right) }(\mathbf{r})\right]
^{\dag }$ an explicit expression for the Lagrange parameter $\mu
_{N}^{\left( \Gamma \right) }$: 
\begin{equation}
\mu _{N}^{\left( \Gamma \right) }=\frac{E_{kin}^{\left( \Gamma \right)
}\left( N\right) +2E_{int}^{\left( \Gamma \right) }\left( N\right) }{N}
\label{chemical potential GP-ground state}
\end{equation}%
Here%
\begin{equation}
E_{int}^{\left( \Gamma \right) }\left( N\right) =\frac{N\left( N-1\right) }{2%
}\frac{4\pi \hbar ^{2}a_{s}}{m}\int_{\Gamma }d^{3}r|\psi ^{\left( \Gamma
\right) }(\mathbf{r})|^{4}  \label{interaction energy in BEC groundstate}
\end{equation}%
and%
\begin{equation}
E_{kin}^{\left( \Gamma \right) }\left( N\right) =N\int_{\Gamma }d^{3}r\left[
\psi ^{\left( \Gamma \right) }(\mathbf{r})\right] ^{\dag }H_{kin}\psi
^{\left( \Gamma \right) }(\mathbf{r})
\label{kinetic energy in BEC groundstate}
\end{equation}%
, respectively, denote the interaction energy and the kinetic energy of the $%
N$-particle Hartree ground state (\ref{ground state BEC}) associated with $%
\psi ^{\left( \Gamma \right) }(\mathbf{r})$. With 
\begin{equation}
E^{\left( \Gamma \right) }(N)=E_{kin}^{\left( \Gamma \right) }\left(
N\right) +E_{int}^{\left( \Gamma \right) }\left( N\right)
\end{equation}%
denoting the total energy of the respective $N$-particle Hartree ground
states (\ref{ground state BEC}) there holds as an identity 
\begin{equation}
\mu _{N}^{\left( \Gamma \right) }=E^{\left( \Gamma \right) }(N)-E^{\left(
\Gamma \right) }(N-1)
\end{equation}%
So it is manifest that the Lagrange parameter $\mu _{N}^{\left( \Gamma
\right) }$ has the physical meaning of the chemical potential in the ground
state of a BEC.

\begin{figure}[tbp]
\begin{centering}
\includegraphics[scale=0.9]{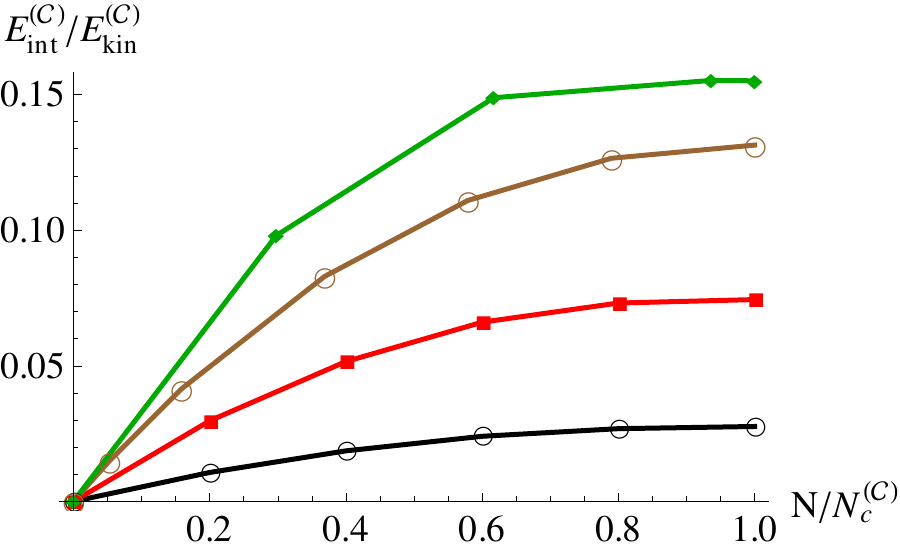}
\par\end{centering}
\caption{Ratio $E_{int}^{\left( \mathcal{C}\right) }/E_{kin}^{\left( 
\mathcal{C}\right) }$ as a function of normalized particle number $%
(N-1)/N_{c}^{\left( \mathcal{C}\right) }$ for the Hartree ground state
localised around the crossing zone of the waveguide $\mathcal{C}$. The
curves refer to at fixed choice of lateral tube diameters $w_{x}^{\left( 
\mathcal{C}\right) }=w_{y}^{\left( \mathcal{C}\right) }=2L$, but different
tube heights: $w_{z}^{\left( \mathcal{C}\right) }=L$ (open black circles), $%
w_{z}^{\left( \mathcal{C}\right) }=2L$ (red squares), $w_{z}^{\left( 
\mathcal{C}\right) }=4L$ (open brown circles), $w_{z}^{\left( \mathcal{C}%
\right) }=8L$ (green diamonds). Further increase of $w_{z}^{\left( \mathcal{C%
}\right) }$ gives for $E_{int}^{\left( \mathcal{C}\right) }/E_{kin}^{\left( 
\mathcal{C}\right) }$ results close to the results obtained for $%
w_{z}^{\left( \mathcal{C}\right) }=8L$.}
\label{fig: ratio E_int/E_kin}
\end{figure}

In (\ref{fig: ratio E_int/E_kin}) we plot for the Hartree ground state of a
BEC localised around the crossing zone inside $\mathcal{C}$ the ratio of the
interaction energy $E_{int}^{\left( \mathcal{C}\right) }$ to the kinetic $%
E_{kin}^{\left( \mathcal{C}\right) }$ energy vs. particle number $N$. The
plot clearly indicates, that such a BEC is dominated by its kinetic energy,
so that the profile of the particle density cannot be described by the
Thomas-Fermi approximation. A similar behaviour we find for the waveguides $%
\mathcal{L}$ and $\mathcal{T}$. This finding is in sharp contrast to an
interacting cold Bose gas confined in a harmonic trap, where the kinetic
energy compared to the interaction energy becomes negligible small for large 
$N$ proportional to $N\ ^{-\frac{4}{5}}$ .

The localisation lengths $\lambda _{N,x\ }^{\left( \Gamma \right) }$ and $%
\lambda _{N,y\ }^{\left( \Gamma \right) }$, which describes the exponential
decay of the GP-orbital $\psi ^{\left( \Gamma \right) }\left( \mathbf{r}%
\right) $ away from the localisation zone along the tube axes $\mathbf{e}%
_{x} $ and $\mathbf{e}_{y}$ of $\Gamma $, are given by 
\begin{eqnarray}
1/\lambda _{N,a}^{\left( \Gamma \right) } &=&-\lim_{s\rightarrow \infty }%
\frac{1}{\left\vert \mathbf{r}_{M}+s\mathbf{e}_{a}\right\vert }\ln
\left\vert \frac{\psi ^{\left( \Gamma \right) }\left( \mathbf{r}_{M}+s%
\mathbf{e}_{a}\right) }{\psi ^{\left( \Gamma \right) }\left( \mathbf{r}%
_{M}\right) }\right\vert  \label{localisation length GP orbital} \\
a &\in &\left\{ x,y\right\}  \nonumber
\end{eqnarray}%
, where $\mathbf{r}_{M}$ denotes a suitable reference position, say where
the modulus $\left\vert \psi ^{\left( \Gamma \right) }\left( \mathbf{r}%
\right) \right\vert $ attains its maximum, see also Fig.\ref{fig:PsiProfileC}%
, Fig.\ref{fig:PsiProfileL}, Fig.\ref{fig:PsiProfileTX}, Fig.\ref%
{fig:PsiProfileTY}. As a rule the arms of $\Gamma \in \left\{ \mathcal{C},%
\mathcal{L},\mathcal{T}\right\} $ with a narrower lateral diameter are
associated with a shorter localisation length.

\begin{figure}[tbp]
\begin{centering}
\includegraphics[scale=0.7]{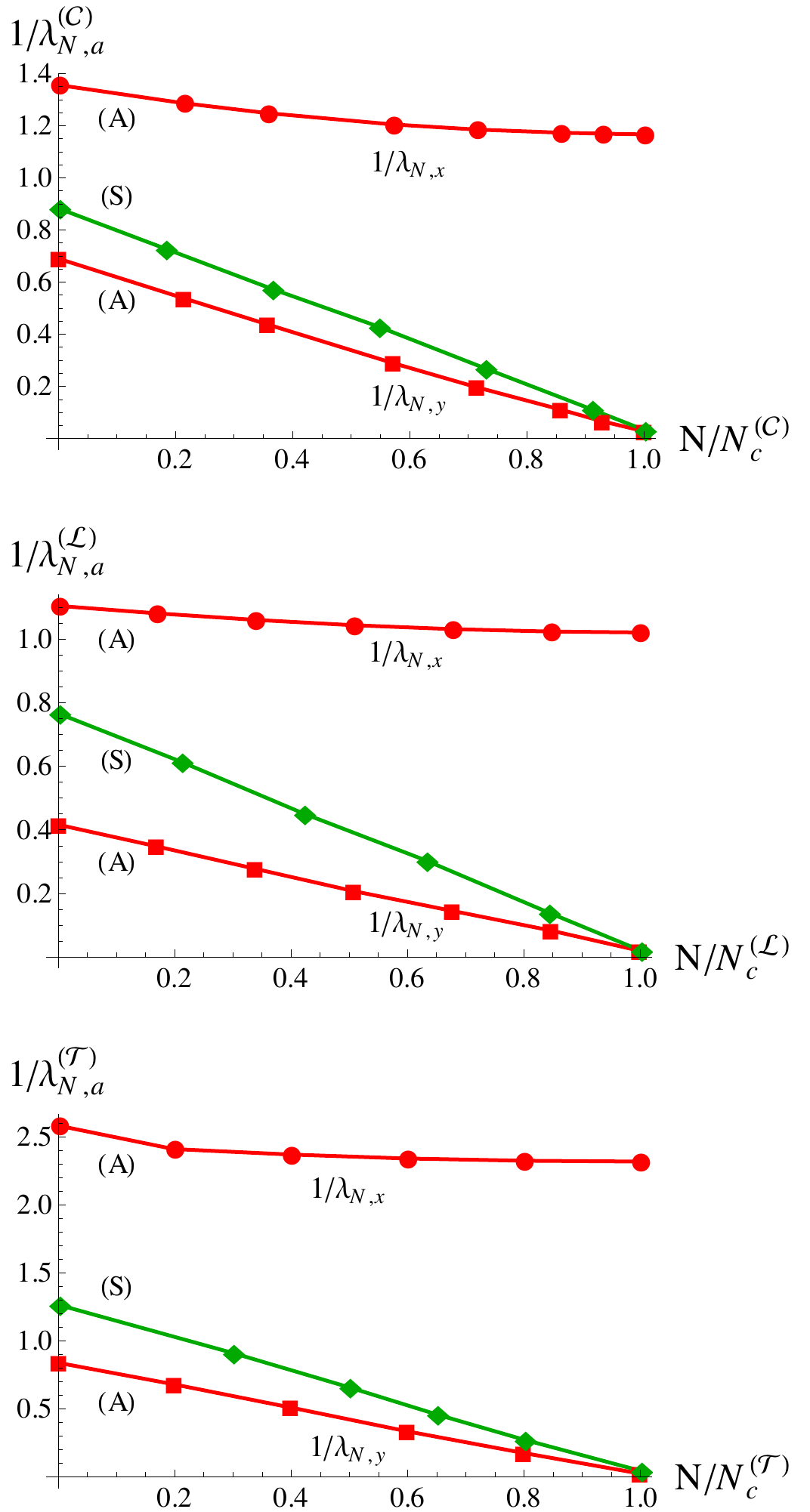}
\par\end{centering}
\caption{The inverse localisation lengths $1/\protect\lambda _{N,x}^{\left(
\Gamma \right) }$ and $1/\protect\lambda _{N,y}^{\left( \Gamma\right) }$
along the respective tube axes $\mathbf{e}_{x}$ and $\mathbf{e}_{y}$ vs.
particle number $N$ of the Hartree ground state of a BEC localised around
the crossing or branching zone of three prototype waveguides $\Gamma \in
\left\{ \mathcal{C},\ \mathcal{L},\mathcal{T}\right\} $. All plots refer to
a tube height $w_{z}^{\left( \Gamma \right) }=2L$. All green lines (S)
correspond to a symmetric choice of lateral tube diameters $w_{y}^{\left(
\Gamma \right) }=w_{x}^{\left( \Gamma \right) }$, all red lines (A)
correspond to an asymmetric choice $w_{y}^{\left( \Gamma \right)
}<w_{x}^{\left( \Gamma \right) }$. \protect\linebreak\ 1) waveguide $%
\mathcal{C}$ : $w_{x}^{\left( \mathcal{C}\right) }=2L$, $w_{y}^{\left( 
\mathcal{C}\right) }=0.8w_{x}^{\left( \mathcal{C}\right) }$. \protect%
\linebreak 2) waveguide $\mathcal{L}$ : $w_{x}^{\left( \mathcal{L}\right)
}=L $, $w_{y}^{\left( \mathcal{L}\right) }=0.95w_{x}^{\left( \mathcal{L}%
\right) } $. \protect\linebreak 3) waveguide $\mathcal{T}$ : $w_{x}^{\left( 
\mathcal{T}\right) }=L$, $w_{y}^{\left( \mathcal{T}\right)
}=0.8w_{x}^{\left( \mathcal{T}\right) }$. \protect\linebreak In all plots $L$
denotes the unit of length.}
\label{fig:LambdaCLTPlot}
\end{figure}

In Fig.\ref{fig:LambdaCLTPlot} the inverse localisation lengths $1/\lambda
_{N,x}^{\left( \Gamma \right) }$ and $1/\lambda _{N,y}^{\left( \Gamma
\right) }$ are plotted vs. particle number $N$ (using $L$ as unit of length)
for two sets of lateral tube diameters with ratio $\kappa ^{\left( \Gamma
\right) }=\frac{w_{y}^{\left( \Gamma \right) }}{w_{x}^{\left( \Gamma \right)
}}$. The green curves refer to a symmetric choice $\kappa ^{\left( \Gamma
\right) }=1$ assuming $w_{x}^{\left( \mathcal{C}\right) }=2L$, $%
w_{x}^{\left( \mathcal{L}\right) }=L$, $w_{x}^{\left( \mathcal{T}\right) }=L$
and tube heights $w_{z}^{\left( \Gamma \right) }=2L$. The blue and red
curves refer to an asymmetric choice of tube widths, $\kappa ^{\left( 
\mathcal{C}\right) }=0.8$ , $\kappa ^{\left( \mathcal{L}\right) }=0.95$, $%
\kappa ^{\left( \mathcal{T}\right) }=0.8$. The results of our full
three-dimensional numerical calculations clearly show that depending on the
choice of $\kappa ^{\left( \Gamma \right) }$ the localisation length along
the arms of wider lateral diameter commences to diverge when the particle
number $N$ approaches the critical particle number $N_{c}^{\left( \Gamma
\right) }=N_{c}^{\left( \Gamma \right) }\left( w_{x}^{\left( \Gamma \right)
},w_{y}^{\left( \Gamma \right) },w_{z}^{\left( \Gamma \right) }\right) $.
Apparently the inverse of the larger localisation length scales linearly
with particle number $N$ over the full range $1\leq N\leq N_{c}^{\left(
\Gamma \right) }$.

\begin{figure}[tbph]
\begin{centering}
\includegraphics[scale=0.7]{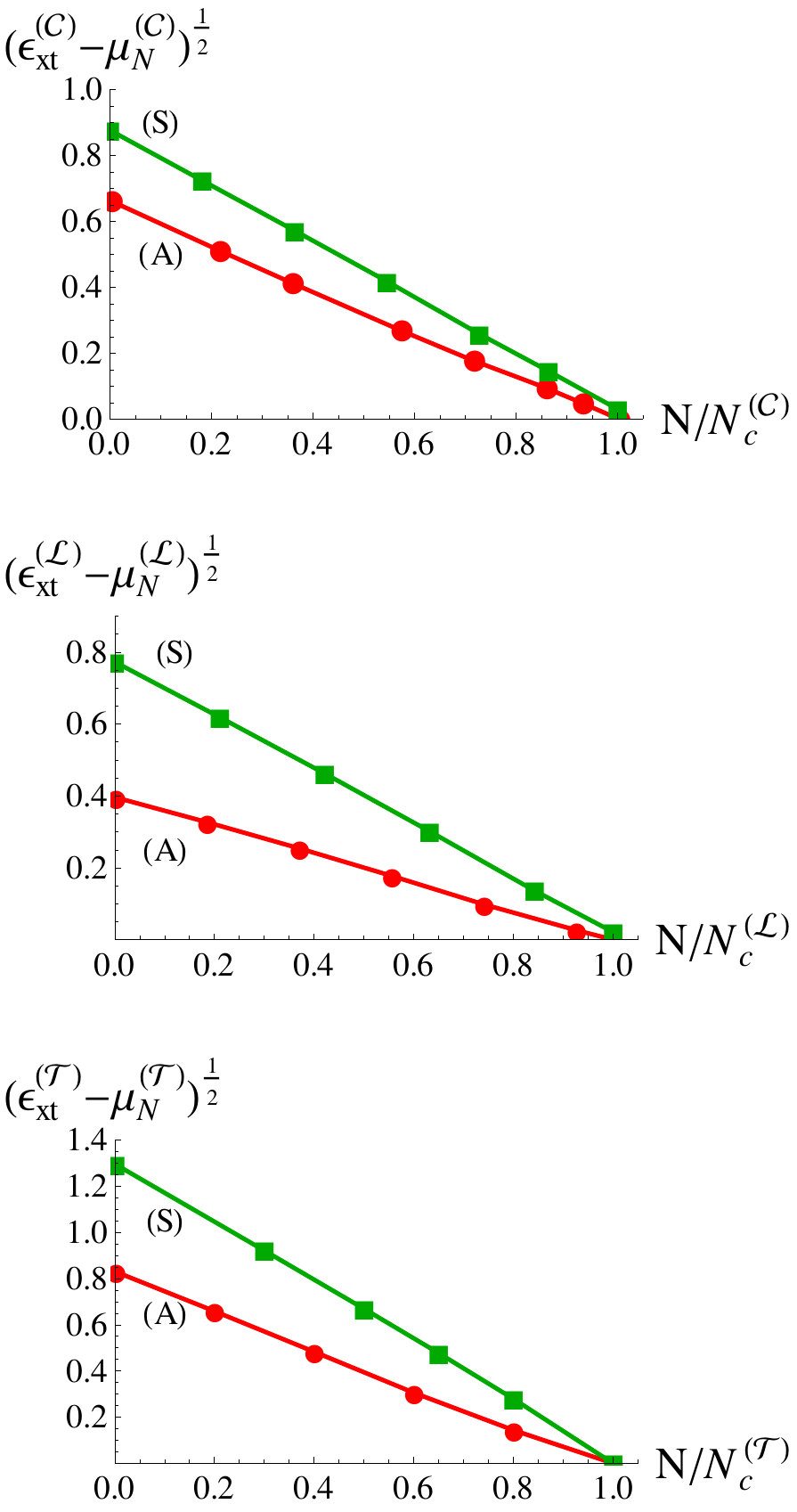}
\par\end{centering}
\caption{The chemical potential $\protect\mu _{N}^{\left( \Gamma \right) }$
vs. particle number $N$ of the Hartree ground state of a BEC localised
around the crossing or branching zone of three prototype waveguides $\Gamma
\in \left\{ \mathcal{C},\ \mathcal{L},\mathcal{T}\right\} $. All plots refer
to a tube height $w_{z}^{\left( \Gamma \right) }=2L$. All green lines (S)
correspond to a symmetric choice of lateral tube diameters $w_{y}^{\left(
\Gamma \right) }=w_{x}^{\left( \Gamma \right) }$, all red lines (A)
correspond to an asymmetric choice $w_{y}^{\left( \Gamma \right)
}<w_{x}^{\left( \Gamma \right) }$. \protect\linebreak\ 1) waveguide $%
\mathcal{C}$ : $w_{x}^{\left( \mathcal{C}\right) }=2L$, $w_{y}^{\left( 
\mathcal{C}\right) }=0.8w_{x}^{\left( \mathcal{C}\right) }$. \protect%
\linebreak 2) waveguide $\mathcal{L}$ : $w_{x}^{\left( \mathcal{L}\right)
}=L $, $w_{y}^{\left( \mathcal{L}\right) }=0.95w_{x}^{\left( \mathcal{L}%
\right) } $. \protect\linebreak 3) waveguide $\mathcal{T}$ : $w_{x}^{\left( 
\mathcal{T}\right) }=L$, $w_{y}^{\left( \mathcal{T}\right)
}=0.8w_{x}^{\left( \mathcal{T}\right) }$. \protect\linebreak In all plots $%
\protect\varepsilon _{L}$ denotes the unit of energy.}
\label{fig:muCLTPlot}
\end{figure}

For the chemical potential $\mu _{N}^{\left( \Gamma \right) }$ of the
localised Hartree ground state inside the respective waveguides $\Gamma \in
\left\{ \mathcal{C},\mathcal{T},\mathcal{L}\right\} $ there holds $0<\mu
_{N}^{\left( \Gamma \right) }<\varepsilon _{xt}^{\left( \Gamma \right) }$.
In Fig.\ref{fig:muCLTPlot} we plot the square root $\sqrt{\varepsilon
_{xt}^{\left( \Gamma \right) }-\mu _{N}^{\left( \Gamma \right) }}$ of the
difference of the excitation threshold $\varepsilon _{xt}^{\left( \Gamma
\right) }$ to the chemical potential $\mu _{N}^{\left( \Gamma \right) }$ vs.
particle number $N$ choosing the respective tube diameters like in Fig.\ref%
{fig:LambdaCLTPlot}. A \emph{linear} decrease with increasing particle
number $N$ of the function $\sqrt{\varepsilon _{xt}^{\left( \Gamma \right)
}-\mu _{N}^{\left( \Gamma \right) }}$ is clearly visible in all results of
our numerical calculations over the full range $1\leq N\leq N_{c}^{\left(
\Gamma \right) }$. For $N\ \rightarrow N_{c}^{\left( \Gamma \right) }$ with $%
N_{c}^{\left( \Gamma \right) }=N_{c}^{\left( \Gamma \right) }\left(
w_{x}^{\left( \Gamma \right) },w_{y}^{\left( \Gamma \right) },w_{z}^{\left(
\Gamma \right) }\right) $ the chemical potential $\mu _{N}^{\left( \Gamma
\right) }$ approaches the excitation threshold $\varepsilon _{xt}^{\left(
\Gamma \right) }$ for a single particle. We find excellent agreement of the
numerical results for $\mu _{N}^{\left( \Gamma \right) }$ with the following
scaling relation 
\begin{equation}
\frac{\mu _{N=N_{c}}^{\left( \Gamma \right) }-\mu _{N}^{\left( \Gamma
\right) }}{\mu _{N=N_{c}}^{\left( \Gamma \right) }-\mu _{N=1}^{\left( \Gamma
\right) }}=\left( 1-\frac{N-1}{N_{c}^{\left( \Gamma \right) }-1}\right) ^{2}
\label{scaling law chemical potential}
\end{equation}%
The displayed apparent linear scaling of $\sqrt{\mu _{N=N_{c}}^{\left(
\Gamma \right) }-\mu _{N}^{\left( \Gamma \right) }}$ vs. particle number $N$
is in sharp contrast to the well known scaling $\mu _{N}$ $\varpropto N^{%
\frac{2}{5}}$ of the chemical potential $\mu _{N}$ in a conventional
harmonic atom trap \cite{Pethick&Smith}. Though for (anharmonic) shallow
atom trap potentials there also exists a critical particle number $N_{c}$,
only small deviations to the scaling $\mu _{N}$ $\varpropto N^{\frac{2}{5}}$
have been reported \cite{Martikainen}. For comparison we show in Fig. \ref%
{fig:muComparisonHarmonicTrap} the function $\sqrt{\mu _{N=N_{c}}-\mu _{N}}$
as obtained for shallow conservative trap potentials \cite{Martikainen}, and
also for a harmonic trap of finite depth. It is clearly visible, that the
described non standard trapping around a crossing or branching zone of a QW
with regard to the dependence on particle number $N$ noticeably differs from
results obtained for conservative trap potentials.

\begin{figure}[tbp]
\begin{centering}
\includegraphics[scale=0.7]{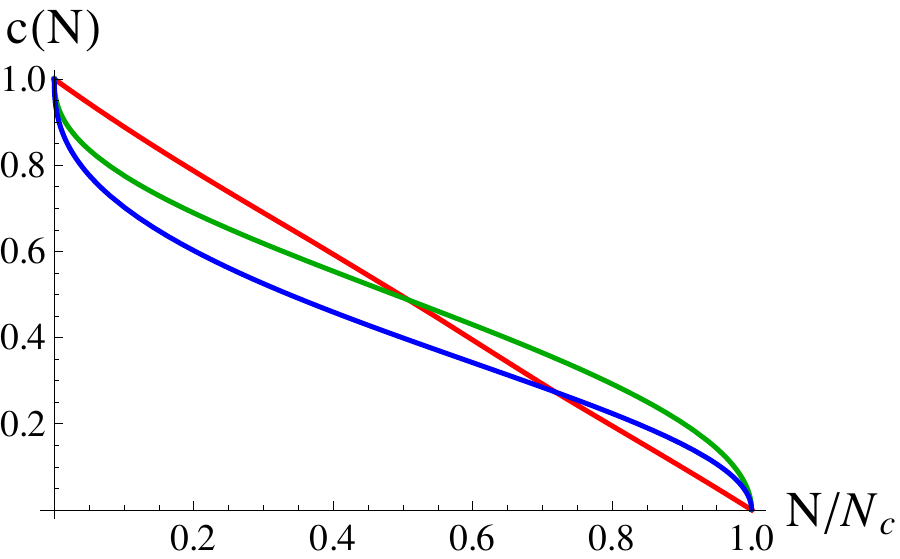}
\par\end{centering}
\caption{The function $c\left( N\right) =\protect\sqrt{\frac{\protect\mu %
_{N=N_{c}}-\protect\mu _{N}}{\protect\mu _{N=N_{c}}-\protect\mu _{N=1}}}$
vs. normalized particle numbers $N/N_{c}$ for chemical potential $\protect%
\mu _{N}$ corresponding to: (i) a shallow conservative trap potential as
considered in Ref.\protect\cite{Martikainen} (blue line), (ii) a harmonic
trap potential of finite depth (green line), (iii) the non standard trapping
around the crossing of a QW (red line). The particle number $N_{c}$ denotes
the maximum number of particles in the respective traps.}
\label{fig:muComparisonHarmonicTrap}
\end{figure}

\begin{figure}[tbph]
\begin{centering}
\includegraphics[scale=0.75]{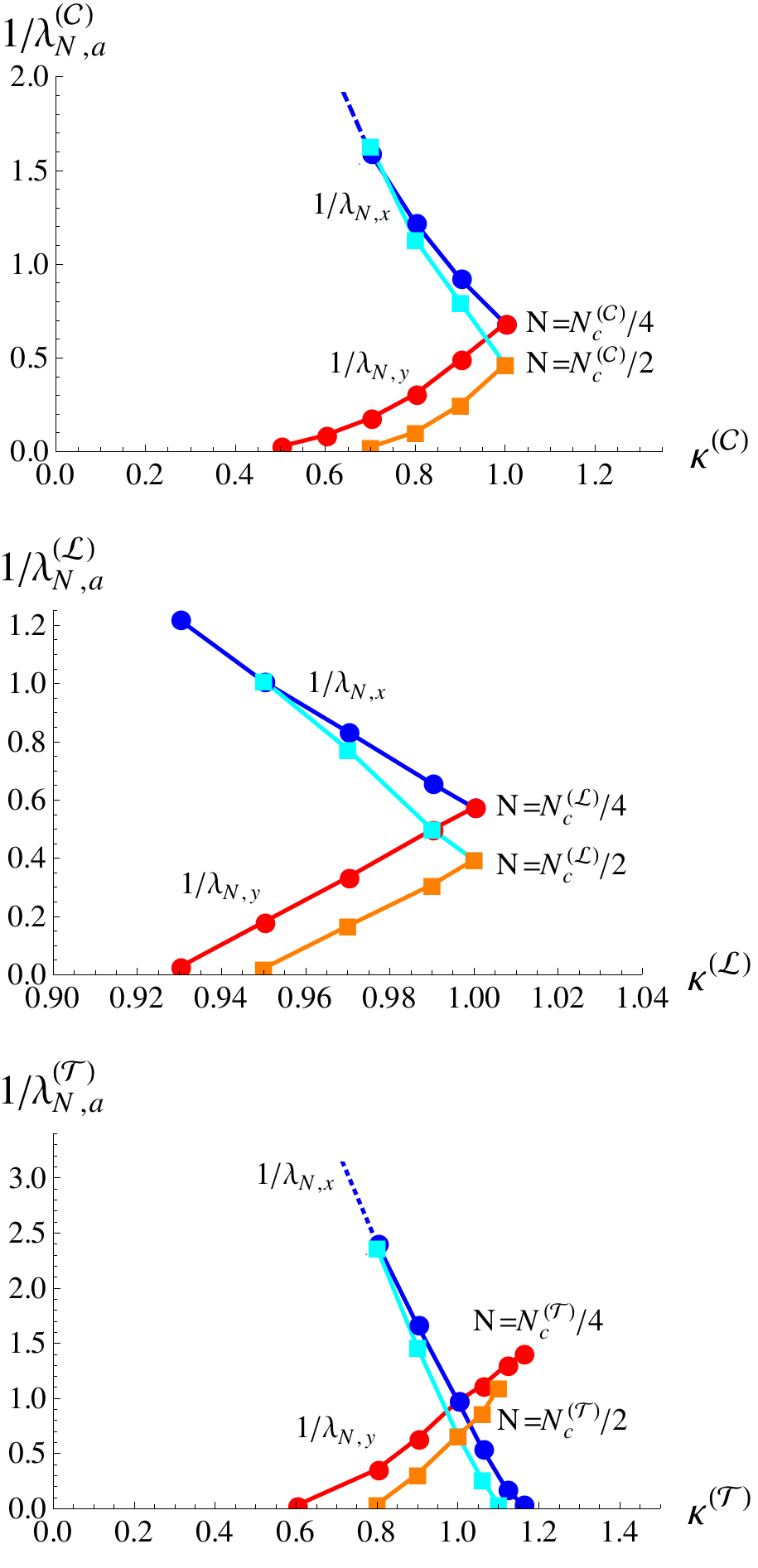}
\par\end{centering}
\caption{Inverse localisation length $1/\protect\lambda _{N,y}^{\left(
\Gamma \right) }$ (red and orange line) and $1/\protect\lambda %
_{N,x}^{\left( \Gamma \right) }$ (blue and cyan line) of the Hartree ground
state in the quantum waveguides $\Gamma \in \left\{ \mathcal{C},\mathcal{L},%
\mathcal{T}\right\} $ for different ratios $\protect\kappa ^{\left( \Gamma
\right) }=w_{y}^{\left( \Gamma \right) }/w_{x}^{\left( \Gamma \right) }$ of
lateral tube diameters. Curves shown refer to two choices of particle
numbers: $N=0.25N_{c}^{\left( \Gamma \right) }$ and $N=0.5N_{c}^{\left(
\Gamma \right) }$. Here $N_{c}^{\left( \Gamma \right) }$ denotes the
critical particle number of the respective waveguide assuming $\protect%
\kappa ^{\left( \Gamma \right) }=1$. Tube height is for all plots $%
w_{z}^{\left( \Gamma \right) }=2L$, lateral tube widths are $w_{x}^{\left( 
\mathcal{C}\right) }=2L$, $w_{x}^{\left( \mathcal{L}\right) }=L$, $%
w_{x}^{\left( \mathcal{T}\right) }=L$.$\ $In all plots $L$ denotes the unit
of length.}
\label{fig:LambdaKappaNN}
\end{figure}

In Fig.\ref{fig:LambdaKappaNN} we display the effect interactions have on
the localisation length $1/\lambda _{N,x}^{\left( \Gamma \right) }$ and $%
1/\lambda _{N,y}^{\left( \Gamma \right) }$ of the ground state of a BEC for
various particle numbers $N$ as a function of the ratio $\kappa ^{\left(
\Gamma \right) }=\frac{w_{y}^{\left( \Gamma \right) }}{w_{x}^{\left( \Gamma
\right) }}$ of lateral tube diameters (we assume $w_{y}^{\left( \mathcal{C}%
\right) }\leq w_{x}^{\left( \mathcal{C}\right) }$). The previously
established bounds for the localisation of a single particle are clearly
changed, see Fig.\ref{fig:LambdaKappaNN1}. According to our numerical
calculations a localised Hartree ground state exists (i) around the crossing
zone of the waveguide $\mathcal{C}\ $ only for $\kappa _{c}^{\left( \mathcal{%
C}\right) }\left( N\right) <\ \kappa ^{\left( \mathcal{C}\right) }$, (ii)
around the corner of the waveguide $\mathcal{L}$ only for $\kappa
_{c}^{\left( \mathcal{L}\right) }\left( N\right) <\kappa ^{\left( \mathcal{L}%
\right) }$ , and (iii) around the branching zone of the waveguide $\mathcal{T%
}$ only in the interval $\kappa _{c,1}^{\left( \mathcal{T}\right) }\left(
N\right) <\ \kappa ^{\left( \mathcal{T}\right) }<\kappa _{c,2}^{\left( 
\mathcal{T}\right) }\left( N\right) $. We find all the lower bounds $\kappa
_{c}^{\left( \mathcal{C}\right) }\left( N\right) $, $\kappa _{c}^{\left( 
\mathcal{L}\right) }\left( N\right) $ and $\kappa _{c,1}^{\left( \mathcal{T}%
\right) }\left( N\right) $ increase as $N$ increases, while the upper bound $%
\kappa _{c,2}^{\left( \mathcal{T}\right) }\left( N\right) $ decreases as $N$
increases.

The observed scaling laws for the localisation lengths $\lambda
_{N,a}^{\left( \Gamma \right) }$ (see Fig. \ref{fig:LambdaCLTPlot}) and the
non standard scaling law of the chemical potential $\mu _{N}^{\left( \Gamma
\right) }$ vs. particle number $N$ (see Fig. \ref{fig:muCLTPlot}), as
obtained from our full three-dimensional numerical calculations, can be
explained in terms of analytical results derived from a simple toy model
that we discuss in section \ref{Scaling Laws for Localised GP-Orbitals from
a One--Dimensional Toy Model.}.

\section{Scaling Laws for Localisation Length and Chemical Potential vs.
Particle Number $N$ from a Toy Model.}

\label{Scaling Laws for Localised GP-Orbitals from a One--Dimensional Toy
Model.}

When a single atom traverses the crossing zone $\mathcal{A}_{0}$ of size $2L$
of the waveguide $\mathcal{C}$ its \emph{transversal} kinetic energy
undergoes a sudden drop. As shown already in section \ref{Kinetic Energy
Induced Confinement at a Crossing.} for a single particle, see also \cite%
{lp0}, the influence of this sudden drop on the asymptotic decay of the
ground state can be modelled by an attractive delta-function potential $-%
\frac{2}{\lambda }\delta (x)$ , corresponding to a localisation length $%
\lambda >L$. Such a delta-function potential is equivalent to a jump
condition for the first derivative of the wave function taken at $x=0$:

\begin{equation}
\left[ \partial _{x}\phi (x)\ \right] _{x=0^{-}}^{x=0^{+}}\ =-\frac{2}{%
\lambda }\phi (0)  \label{GP 1-dim jump condition}
\end{equation}%
The Gross Pitaevskii equation (\ref{Gross-Pitaevskii I}) determining the $N$%
-particle Hartree ground state (\ref{ground state BEC}) of a BEC in terms of
the optimal GP-orbital $\psi ^{\left( \mathcal{C}\right) }\left( \mathbf{r}%
\right) $ can be projected at large distance $x$ to the crossing zone $%
\mathcal{A}_{0}$ to one dimension making the separation ansatz 
\[
\psi ^{\left( \mathcal{C}\right) }\left( \mathbf{r}\right) \rightarrow \psi
_{\perp }^{\left( \mathcal{C}\right) }\left( \mathbf{r}_{\perp }\right) \phi
(x) 
\]%
The function $\phi _{N}(x)$ in this case solves a one-dimensional non linear
Schr\"{o}dinger equation (scaled units):

\begin{equation}
\left[ -\partial _{x}^{2}+\varepsilon _{xt}-\mu _{N}-\frac{2}{\lambda _{N}}%
\delta (x)+\left( N-1\right) g_{s}\left\vert \phi (x)\right\vert ^{2}\right]
\phi (x)=0  \label{1D non linear Schroedinger Eq}
\end{equation}%
Here $g_{s}$ describes the strength of the effective repulsive two-body
interaction potential (as projected to one-dimension), and $\mu _{N}$ is the
chemical potential ensuring the usual normalization constraint of the
GP-orbital.

To solve this differential equation we make an ansatz for $\phi (x)$
depending on three parameters, the amplitude $A_{N}$ , the localisation
length $\lambda _{N}>0$ and a shift parameter $s_{N}>0$ : 
\begin{equation}
\phi _{N}\left( x\right) =\frac{A_{N}}{\sinh \left( \frac{|x|+s_{N}}{\lambda
_{N}}\right) }  \label{1D ansatz GP}
\end{equation}%
This ansatz solves (\ref{1D non linear Schroedinger Eq}) and the boundary
condition (\ref{GP 1-dim jump condition}), provided 
\begin{eqnarray}
\mu _{N} &=&\varepsilon _{xt}-\frac{1}{\lambda _{N}^{2}}
\label{1D parameters ansatz GP} \\
\lambda _{N} &=&\frac{\lambda }{1-\frac{\left( N-1\right) g_{s}\lambda }{4}}%
>0  \nonumber \\
\exp \left( -\frac{s_{N}}{\lambda _{N}}\right) &=&\sqrt{\frac{\frac{\lambda
_{N}}{\lambda }-1}{\frac{\lambda _{N}}{\lambda }+1}}  \nonumber \\
A_{N} &=&\frac{1}{\sqrt{2\lambda _{N}\left( \frac{\lambda _{N}}{\lambda }%
-1\right) }}  \nonumber
\end{eqnarray}%
The amplitude $A_{N}$ is fixed by the usual wavefunction normalization.
Apparently, for $N\rightarrow 1$ there holds $\lambda _{N}\rightarrow
\lambda +0$. In this limit, $A_{N}$ and the parameter ratio $\frac{s_{N}}{%
\lambda _{N}}$ both display a singularity, so that the expression obtained
for $\phi _{N=1}\left( x\right) \ $coincides with the wavefunction (\ref%
{bound state delta-function potential}) of a single particle.

As the particle number $N$ (per cross section area $w_{y}\times w_{z}$)
increases it is seen from Eq.(\ref{1D parameters ansatz GP}) that the
localisation length $\lambda _{N}$ increases, and it diverges as $N$
approaches a critical particel number $N_{c}$ given by%
\begin{equation}
N_{c}=1+\frac{4}{\lambda g_{s}}
\end{equation}%
This critical particle number $N_{c}$, depending on the localisation length $%
\lambda $ for \emph{one} particle and the effective interaction strength $%
g_{s}$ for \emph{two} particles, determines the maximal capacity of a
localised BEC ground state build from the respective optimal GP-orbitals to
bind Bose-atoms around a crossing or branching zone of quantum waveguides.
For $N>N_{c}$ a localised solution for the ground state orbital ceases to
exist. As the described localisation-delocalisation quantum transition is
sharp, it should be possible to determine in an experiment that critical
particle number $N_{c}\ $rather precisely.

Elimination of the interaction constant $g_{s}$ in terms of the observable
critical particle number $N_{c}$ leads to the following scaling law of the
localisation length $\lambda _{N}$ of the optimal GP-orbital: 
\begin{equation}
\lambda _{N}=\frac{\lambda }{1-\frac{N-1}{N_{c}-1}}  \label{eq:1D lambda_N}
\end{equation}%
In the range $1\leq N<N_{c}$ we obtain then for the chemical potential $\mu
_{N}$ the expression (scaled units)

\begin{equation}
\mu _{N}=\varepsilon _{xt}-\frac{1}{\lambda ^{2}}\left( 1-\frac{N-1}{N_{c}-1}%
\right) ^{2}  \label{eq:1Dmu}
\end{equation}%
At the critical particle number $N=N_{c}$ the chemical potential assumes the
value $\mu _{N=N_{c}}=\varepsilon _{xt}$.

Overall we find, that the dependence on particle number $N$ of the inverse
localisation length $1/\lambda _{N}^{\left( \mathcal{C}\right) }$ , and the
dependence on particle number $N$ of the function$\sqrt{\varepsilon
_{xt}^{\left( \mathcal{C}\right) }-\mu _{N}^{\left( \mathcal{C}\right) }}$,
as numerically calculated solving the full three-dimensional GP-equation and
displayed in Fig.\ref{fig:LambdaCLTPlot} and in Fig.\ref{fig:muCLTPlot} (the
green curves correspond to \emph{equal} lateral tube diameters $%
w_{x}^{\left( \mathcal{C}\right) }=w_{y}^{\left( \mathcal{C}\right) }$) both
agree very well with the analytical scaling laws (\ref{eq:1D lambda_N}) and (%
\ref{eq:1Dmu}). Indeed Eq.(\ref{eq:1Dmu}) fully coincides with the scaling
law Eq.(\ref{scaling law chemical potential}) reported in section \ref%
{ Heat Kernel Method for Numerical Solution of Gross-Pitaevskii
Equation.}.

The chemical potential $\mu _{N}$ of a system of $N$ interacting Bose atoms
is connected to the ground state energy $E\left( N\right) $ by 
\begin{equation}
\mu _{N}=E\left( N\right) -E\left( N-1\right)
\label{definition chemical potential}
\end{equation}%
Solving this difference equation for $E\left( N\right) $ assuming $N\geq 2$
gives 
\begin{equation}
E\left( N\right) =N\cdot \mu _{N=1}+\frac{1}{\lambda ^{2}}\frac{N\left(
N-1\right) }{N_{c}-1}\left( 1-\frac{\frac{N}{3}-\frac{1}{6}}{N_{c}-1}\right)
\label{groundstate energy}
\end{equation}%
$\allowbreak $The energy $E\left( N\right) $ should be observable as the
release energy of the system, say switching off the lasers creating the
'walls' of an hollow optical waveguide. In the non-interacting (ideal) Bose
gas there holds $N_{c}\rightarrow \infty $ , so that one finds the expected
result $E^{\left( 0\right) }\left( N\right) =N\cdot \left( \varepsilon _{xt}-%
\frac{1}{\lambda ^{2}}\right) $.

It is instructive to express for the $N$-particle BEC ground state (\ref%
{ground state BEC}) the expectation values of the kinetic energy (\ref%
{kinetic energy in BEC groundstate}) and the interaction energy (\ref%
{interaction energy in BEC groundstate}) in terms of the chemical potential $%
\mu _{N}$ and the total energy $E\left( N\right) $. Making use of the
general relations%
\begin{eqnarray}
E_{kin}(N)+E_{int}(N) &=&E\left( N\right) \\
E_{kin}(N)+2E_{int}(N) &=&N\mu _{N}\   \nonumber
\end{eqnarray}%
one obtains for $N>>1$ :%
\begin{equation}
\frac{E_{int}(N)}{E_{kin}(N)}=\frac{1}{1+3E\left( 1\right) \lambda ^{2}}%
+O\left( \frac{1}{N}\right)
\label{interaction&kinectic energy of groundstate}
\end{equation}%
In our three-dimensional numerical calculations reported in the previous
section for the crossing waveguide $\mathcal{C}$ we found this ratio assumes
for all tube size parameters $w_{z}^{\left( \mathcal{C}\right) }>0$ a value
substantially smaller than unity. This finding is confirmed quantitatively
by our $1D$-toy model inserting the ground state energy (\ref{groundstate
energy}) into the expressions (\ref{interaction&kinectic energy of
groundstate}). For a large particle number $N\gg 1$, which according to Fig.
(\ref{fig:NcPlot}) corresponds to a choice $w_{z}^{\left( \mathcal{C}\right)
}\gg L$, one finds $E_{int}(N_{c})\leq 0.15\times E_{kin}(N_{c})$ , in good
agreement with the results displayed in Fig.\ref{fig: ratio E_int/E_kin}.
Thus it is evident that for the localised Hartree ground state around the
crossing zone of the quantum waveguide $\mathcal{C}$ the Thomas-Fermi
approximation does not apply. This is in sharp contrast to the $N$-particle
ground state of a BEC that forms in a harmonic trap \cite{Pethick&Smith},
where for $N\gg 1$ the interaction energy is large compared to the kinetic
energy, so that the density profile of such a BEC is well reproduced by the
Thomas-Fermi approximation.

\section{Binary Mixture of Cold Bose Atoms in $\mathcal{C}$.}

\label{Binary Mixture of Cold Bose Atoms inside a Quantum Waveguide}

The previously described non standard trapping mechanism for cold particles
moving around the crossing zone of a waveguide $\mathcal{C}$ is kinetic
energy driven. It is then interesting to study a \emph{binary} BEC
consisting of two \emph{different} species of Bose atoms, say with mass $%
m_{A}>m_{B}$. The associated two-particle contact interaction parameters of
the atoms (using obvious notation for the respective $s$-wave scattering
lengths) we denote as%
\begin{eqnarray}
g_{AA} &=&\frac{4\pi \hslash ^{2}a_{A\ }}{m_{A}}
\label{binary mixture interaction strengths} \\
g_{BB} &=&\frac{4\pi \hslash ^{2}a_{B\ }}{m_{B}}  \nonumber \\
g_{AB} &=&2\pi \hslash ^{2}a_{AB\ }\left( \frac{1}{m_{A}}+\frac{1}{m_{B}}%
\right)  \nonumber
\end{eqnarray}%
, see for example \cite{Pethick&Smith}. Let then $N_{A}$ be the number of
Bose atoms of type $A$ , and $N_{B}=N-N_{A}$ be the number of Bose atoms of
type $B$. Within mean field theory the ground state of such a binary BEC is
then a generalization of the Hartree ground state describing a single atom
species Bose condensate: 
\begin{equation}
\Psi _{G}(\mathbf{r}_{1},\mathbf{r}_{2},...,\mathbf{r}_{N})=\prod%
\limits_{j=1}^{N_{A}}\psi _{A}(\mathbf{r}_{j})\,\prod\limits_{j=N_{A}+1\
}^{N_{A}+N_{B}}\psi _{B}(\mathbf{r}_{j})
\label{binary Bose gas Hartree ground state}
\end{equation}%
The task is then to find the optimal Hartree orbitals $\psi _{A}(\mathbf{r})$
and $\psi _{B}(\mathbf{r})$, that minimize the expectation value of the
Hamiltonian of the interacting Bose gas mixture in that ground state,
subject to the constraint that the number of particles, $N_{A}$ and $N_{B}$
respectively, are both conserved. This constraint engenders for $\psi _{A}(%
\mathbf{r})$ and $\psi _{B}(\mathbf{r})$ the normalization conditions 
\begin{equation}
\int_{\mathcal{C}}d^{3}r|\psi _{A}(\mathbf{r})|^{2}=1=\int_{\mathcal{C}%
}d^{3}r|\psi _{B}(\mathbf{r})|^{2}
\label{binary mixture normalization constraints optimal orbitals}
\end{equation}%
It is \emph{not} required that the optimal orbitals $\psi _{A}(\mathbf{r})$
and $\psi _{B}(\mathbf{r})$ are orthogonal.

Introducing Lagrange parameters $\mu _{A}$ and $\mu _{A}$ for these
normalization constraints (\ref{binary mixture normalization constraints
optimal orbitals}), the respective optimal orbitals are solutions to the
following $2\times 2$-system of coupled Hartree equations

\begin{widetext}

\begin{eqnarray}
\left[ H_{A,kin}+\left( N_{A}-1\right) g_{AA}\ |\psi _{A}(\mathbf{r}%
)|^{2}+N_{B}g_{AB}|\psi _{B}(\mathbf{r})|^{2}\right] \psi _{A}(\mathbf{r})
&=&\mu _{A}\psi _{A}(\mathbf{r})
\label{binary mixture coupled Hartree equations} \\
\left[ H_{B,kin}+N_{A}g_{AB}|\psi _{A}(\mathbf{r})|^{2}+\left(
N_{B}-1\right) g_{BB}\ |\psi _{B}(\mathbf{r})|^{2}\right] \psi _{B}(\mathbf{r%
}) &=&\mu _{B}\psi _{B}(\mathbf{r})  \notag
\end{eqnarray}%
\end{widetext}
Here $H_{A,kin}$ and $H_{B,kin}$ denote the kinetic energy operators
associated with a single $A$- or $B$-atom, respectively: 
\begin{equation}
\frac{m_{A}}{m_{B}}H_{A,kin}=-\frac{\hbar ^{2}}{2m_{B}}\nabla ^{2}=H_{B,kin}
\end{equation}%
It follows, that lighter atoms moving along the armes $\mathcal{A}%
_{j}\subset \mathcal{C}$ have a higher excitation threshold (\ref{excitation
threshold C}) than the heavier ones: 
\begin{equation}
\frac{\varepsilon _{xt,B}}{\varepsilon _{xt,A}}=\frac{m_{A}}{m_{B}}
\label{mass dependence of excitation threshold}
\end{equation}

To solve the coupled equations (\ref{binary mixture coupled Hartree
equations}) we consider (like in the afore mentioned case of an interacting
Bose gas consisting of only one atom species) a suitable auxiliary diffusion
process. Introducing $2\times 2$-matrix notation we write%
\begin{equation}
-\frac{\partial }{\partial \tau }\psi (\mathbf{r},\tau )=\left[
H_{kin}+U_{\psi }(\mathbf{r},\tau )\right] \psi (\mathbf{r},\tau )
\label{heat equation for GP-orbitals of mixture}
\end{equation}%
where%
\begin{equation}
H_{kin}\psi (\mathbf{r},\tau )=\left[ 
\begin{array}{cc}
H_{A,kin} & 0 \\ 
0 & \frac{m_{A}}{m_{B}}H_{A,kin}%
\end{array}%
\right] \left[ 
\begin{array}{c}
\psi _{A}(\mathbf{r},\tau ) \\ 
\psi _{B}(\mathbf{r},\tau )%
\end{array}%
\right]  \label{kinetic energy operator mixture}
\end{equation}%
Because the amplitude of the auxiliary wave functions $\psi _{A}(\mathbf{r}%
,\tau )$ and $\psi _{B}(\mathbf{r},\tau )$ decay exponentially with
diffusion time $\tau $ as the diffusion process progresses the interaction
term neeeds explicit normalization:

\begin{widetext}

\begin{eqnarray}
&& \\
U_{\psi }(\mathbf{r},\tau ) &=&\left[ 
\begin{array}{cc}
\ \frac{\left( N_{A}-1\right) g_{AA}\left\vert \psi _{A}(\mathbf{r},\tau
)\right\vert ^{2}}{\int_{\mathcal{C}}d^{3}r^{\prime }\left\vert \psi _{A}(%
\mathbf{r}^{\prime },\tau )\right\vert ^{2}}, & \frac{N_{B}g_{AB}\ \psi _{A}(%
\mathbf{r,\tau })\psi _{B}^{\dag }(\mathbf{r,\tau })}{\sqrt{\int_{\mathcal{C}%
}d^{3}r^{\prime }\left\vert \psi _{A}(\mathbf{r}^{\prime },\tau )\right\vert
^{2}}\sqrt{\int_{\mathcal{A}}d^{3}r^{\prime }\left\vert \psi _{B}(\mathbf{r}%
^{\prime },\tau )\right\vert ^{2}}} \\ 
\frac{N_{A}g_{AB}\ \psi _{B}(\mathbf{r,\tau })\psi _{A}^{\dag }(\mathbf{%
r,\tau })}{\sqrt{\int_{\mathcal{C}}d^{3}r^{\prime }\left\vert \psi _{A}(%
\mathbf{r}^{\prime },\tau )\right\vert ^{2}}\sqrt{\int_{\mathcal{C}%
}d^{3}r^{\prime }\left\vert \psi _{B}(\mathbf{r}^{\prime },\tau )\right\vert
^{2}}} & ,\ \frac{\left( N_{B}-1\right) g_{BB}\left\vert \psi _{B}(\mathbf{r}%
,\tau )\right\vert ^{2}}{\int_{\mathcal{C}}d^{3}r^{\prime }\left\vert \psi
_{B}(\mathbf{r}^{\prime },\tau )\right\vert ^{2}}%
\end{array}%
\right]  \notag
\end{eqnarray}%

\end{widetext}

Apparently, for large diffusion time $\tau $ then $U_{\psi }(\mathbf{r},\tau
)$ becomes independent on $\tau $. The seeked optimal Hartree orbitals are
given by 
\begin{eqnarray}
\psi _{A}(\mathbf{r}) &=&\lim_{\tau \rightarrow \infty }\frac{\psi _{A}(%
\mathbf{r},\tau )}{\sqrt{\int_{\mathcal{C}}d^{3}r^{\prime }\ \left\vert \psi
_{A}(\mathbf{r}^{\prime },\tau )\right\vert ^{2}}} \\
\psi _{B}(\mathbf{r}) &=&\lim_{\tau \rightarrow \infty }\frac{\psi _{B}(%
\mathbf{r},\tau )}{\sqrt{\int_{\mathcal{C}}d^{3}r^{\prime }\ \left\vert \psi
_{B}(\mathbf{r}^{\prime },\tau )\right\vert ^{2}}}  \nonumber
\end{eqnarray}%
In practice, the (normalized!) Hartree orbitals are calculated as numerical
solutions to the system of diffusion equations (\ref{heat equation for
GP-orbitals of mixture}) extending the afore mentioned splitting scheme (\ref%
{splitting scheme GP-equation}) to the case of a two component spinor, as
indicated in (\ref{kinetic energy operator mixture}).

One obtains directly from (\ref{binary mixture coupled Hartree equations}),
by taking a scalar product with $\psi _{A}(\mathbf{r})$ in the first line,
and with $\psi _{B}(\mathbf{r})$ in the second line, explicit expressions
for the Lagrange parameters $\mu _{A}$ and $\mu _{B}$ depending on the
interaction strengths $g_{AA}$ , $g_{AB}$ , $g_{BB}$ and the particle
numbers $N_{A}$ and $N_{B}$: 
\begin{eqnarray}
\mu _{A} &=&\int_{\mathcal{C}}d^{3}r\left[ 
\begin{array}{c}
\psi _{A}^{\dag }(\mathbf{r})H_{A,kin}\psi _{A}(\mathbf{r}) \\ 
+\left( N_{A}-1\right) g_{AA}|\psi _{A}(\mathbf{r})|^{4} \\ 
+N_{B}g_{AB}|\psi _{A}(\mathbf{r})|^{2}|\psi _{B}(\mathbf{r})|^{2}%
\end{array}%
\right]  \label{binary Bose mixture Lagrange parameters mu_A and mu_B} \\
\mu _{B} &=&\int_{\mathcal{C}}d^{3}r\left[ 
\begin{array}{c}
\psi _{B}^{\dag }(\mathbf{r})H_{B,kin}\psi _{B}(\mathbf{r}) \\ 
+\left( N_{B}-1\right) g_{BB}|\psi _{B}(\mathbf{r})|^{4} \\ 
+N_{A}g_{AB}|\psi _{A}(\mathbf{r})|^{2}|\psi _{B}(\mathbf{r})|^{2}%
\end{array}%
\right]  \nonumber
\end{eqnarray}%
We readily confirm the identity%
\begin{equation}
N_{A}\mu _{A}+N_{B}\mu _{B}=E_{kin}\left( N_{A},N_{B}\right) +2E_{int}\left(
N_{A},N_{B}\right)
\end{equation}%
, where 
\begin{eqnarray}
E_{kin}\left( N_{A},N_{B}\right) &=&\int_{\mathcal{C}}d^{3}r\ \left[ 
\begin{array}{c}
N_{A}\psi _{A}^{\dag }(\mathbf{r})H_{A,kin}\psi _{A}(\mathbf{r}) \\ 
+N_{B}\psi _{B}^{\dag }(\mathbf{r})H_{B,kin}\psi _{B}(\mathbf{r})%
\end{array}%
\right] \\
E_{int}\left( N_{A},N_{B}\right) &=&\int_{\mathcal{C}}d^{3}r\left[ 
\begin{array}{c}
\frac{N_{A}\left( N_{A}-1\right) }{2}g_{AA}|\psi _{A}(\mathbf{r})|^{4} \\ 
+\frac{N_{B}\left( N_{B}-1\right) }{2}g_{BB}|\psi _{B}(\mathbf{r})|^{4} \\ 
+N_{A}N_{B}g_{AB}|\psi _{A}(\mathbf{r})|^{2}|\psi _{B}(\mathbf{r})|^{2}%
\end{array}%
\right]  \nonumber
\end{eqnarray}%
denotes the kinetic energy, respectively the interaction energy in the $N$%
-particle binary BEC ground state (\ref{binary Bose gas Hartree ground state}%
).

With the total energy the system has, 
\begin{equation}
E\left( N_{A},N_{B}\right) =E_{kin}\left( N_{A},N_{B}\right) +E_{int}\left(
N_{A},N_{B}\right)
\end{equation}%
, and with $\mu _{A}$ and $\mu _{B}$ as stated in (\ref{binary Bose mixture
Lagrange parameters mu_A and mu_B}), there follows as an identity 
\begin{eqnarray}
\mu _{A} &=&E\left( N_{A},N_{B}\right) -E\left( N_{A}-1,N_{B}\right) \\
\mu _{B} &=&E\left( N_{A},N_{B}\right) -E\left( N_{A},N_{B}-1\right) 
\nonumber
\end{eqnarray}%
Because atoms species $A$ and $B$ are distinguishable, there exist two
different chemical potentials in a binary mixture.

The localisation lengths $\lambda _{A}$ and $\lambda _{B}$ for the two atom
species $A$ and $B$, respectively, follow from the asymptotic decay of the
respective Hartree orbitals $\psi _{A}(\mathbf{r})$ and $\psi _{B}(\mathbf{r}%
)$, see (\ref{localisation length GP orbital}). These localisation lengths
depend not only on the choice of interaction strength parameters (\ref%
{binary mixture interaction strengths}), but also on the lateral tube
diameters of the joining waveguides, and of course on the mixing ratio $%
\frac{N_{A}}{N_{B}}$ of particle numbers $N_{A}$ and $N_{B}$.

\begin{figure}[tbph]
\begin{centering}
\includegraphics[scale=0.9]{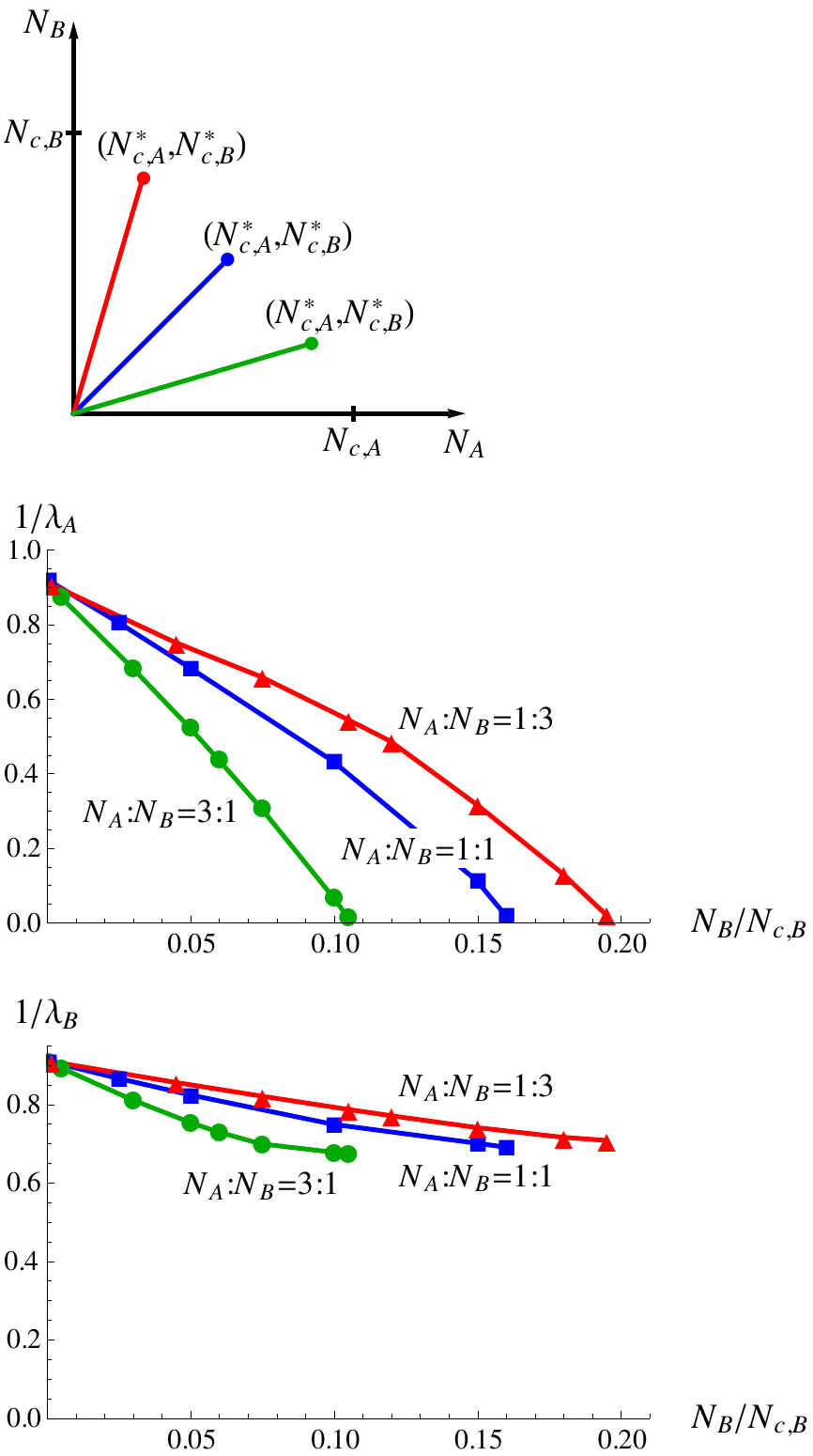}
\par\end{centering}
\caption{Demixing quantum transition of a binary cold BEC consisting of $A$-
and $B$-atoms confined around the crossing of $\mathcal{C}$ assuming equal
tube diameters $w_{x}^{\left( \mathcal{C}\right) }=w_{y}^{\left( \mathcal{C}%
\right) }=w_{z}^{\left( \mathcal{C}\right) }=2L$. Plots show the respective
inverse localisation lengths $1/\protect\lambda _{A}$ and $1/\protect\lambda %
_{B}$ vs. particle number $N=N_{A}+N_{B}=(\frac{N_{A}}{N_{B}}+1)N_{B}$ for
three different mixing ratios: red line $N_{A}:N_{B}=1:3$, blue line $%
N_{A}:N_{B}=1:1$, green line $N_{A}:N_{B}=3:1$. Mass ratio $\frac{m_{A}}{%
m_{B}}=\frac{87}{23}$ and interaction parameters $%
g_{AA}:g_{AB}:g_{BB}=1:1.7:2$ describe a binary BEC mixture consisting of $%
^{23}Na$- and $^{87}Rb$-atoms.}
\label{fig:LambdaAB}
\end{figure}

\begin{figure}[tbph]
\begin{centering}
\includegraphics[scale=0.9]{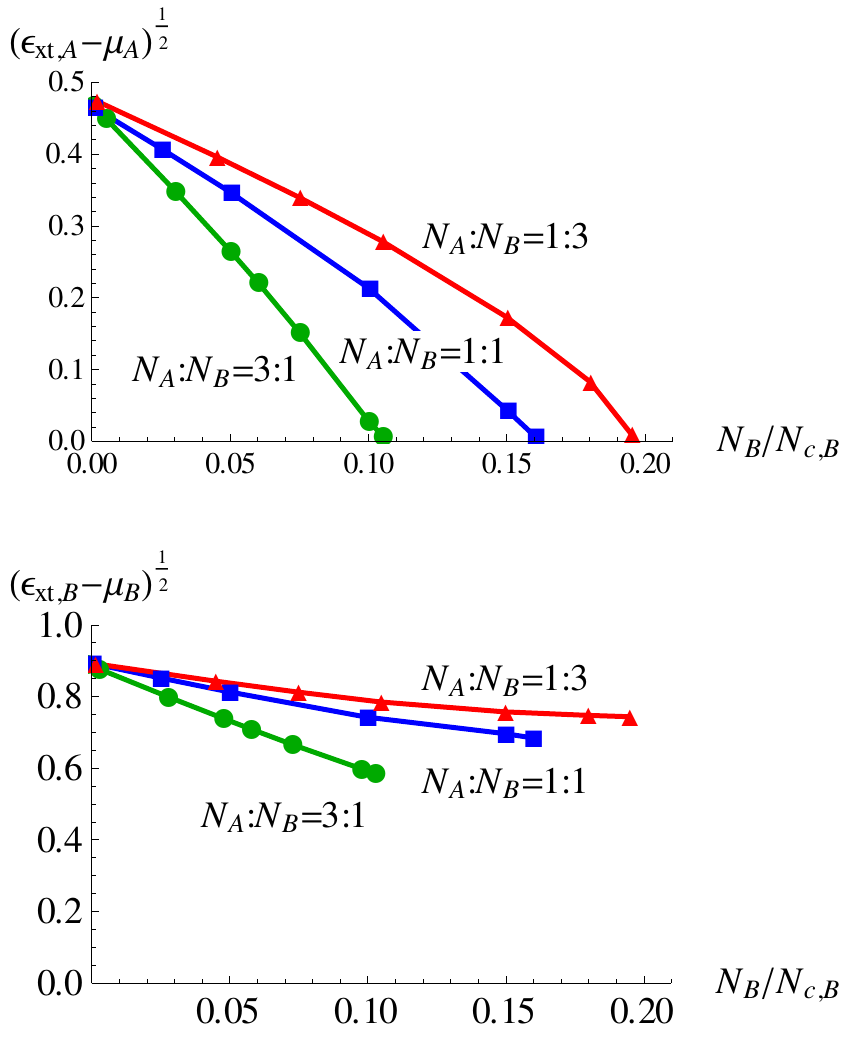}
\par\end{centering}
\caption{Chemical potentials $\protect\mu _{A}$ and $\protect\mu _{B}$ of a
binary BEC mixture vs. particle number $N=N_{A}+N_{B}=(\frac{N_{A}}{N_{B}}%
+1)N_{B}$ assuming a fixed mixing ratio $N_{A}:N_{B}$. Here $\protect%
\epsilon _{xt,B}=\protect\varepsilon _{L}\times \frac{\protect\pi ^{2}}{2}$
and $\protect\varepsilon _{xt,A}=\frac{m_{B}}{m_{A}}\protect\varepsilon %
_{xt,B}$ denote the respective excitation thresholds for atoms with mass $%
m_{A}$ and $m_{B}$. Unit of length is $L$, unit of energy $\protect%
\varepsilon _{L}=\frac{\hslash ^{2}}{2m_{B}\ L^{2}}$. Mass and interaction
parameters like in Fig.\protect\ref{fig:LambdaAB}.}
\label{fig:muSqrtRbNa}
\end{figure}

\begin{figure}[tbph]
\begin{centering}
\includegraphics[scale=0.9]{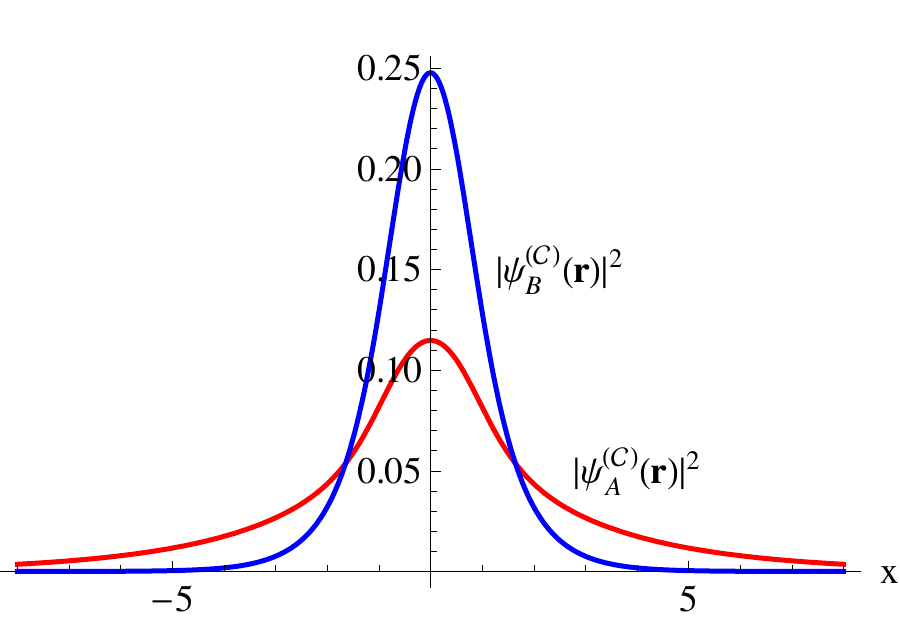}
\par\end{centering}
\caption{The profiles $|\protect\psi _{A}^{(\mathcal{C})}(\mathbf{r})|^{2}$
and $|\protect\psi _{B}^{(\mathcal{C})}(\mathbf{r})|^{2}$ along the axis $%
\mathbf{e}_{x}$ of the ground state GP-orbitals in a binary BEC mixture
consisting of $A$- and $B$-atoms, assuming a mixing ratio $N_{A}:N_{B}=1:10$
and a filling fraction $N_{B}/N_{c,B}^{\star }=0.9$. Mass and interaction
parameters like in Fig.\protect\ref{fig:LambdaAB}. }
\label{fig:ParticleDensityOneTen}
\end{figure}

We discuss now the results obtained for a cross shaped waveguide geometry $%
\mathcal{C}$ with equal tube sizes $w_{x}^{\left( \mathcal{C}\right)
}=w_{y}^{\left( \mathcal{C}\right) }=w_{z}^{\left( \mathcal{C}\right) }=2L$.
Choosing $m=m_{B}$ as unit of mass, $L$ as unit of length and $\varepsilon
_{L}=\frac{\hslash ^{2}}{2mL^{2}}$ as unit of energy, the respective
excitation thresholds for atom species $A$ and $B$ are $\varepsilon
_{xt,B}=\varepsilon _{L}\times \frac{\pi ^{2}}{2}$ and $\varepsilon _{xt,A}=%
\frac{m_{B}}{m_{A}}\varepsilon _{xt,B}$. As an example we study the trapping
of a dilute binary cold Bose gas consisting of $^{23}Na$- and $^{87}Rb$%
-atoms (in this case $\frac{m_{A}}{m_{B}}=\frac{87}{23}\simeq \allowbreak
3.\,\allowbreak 78$ ). The interaction strength parameters (\ref{binary
mixture interaction strengths}) for this sytem we take from \cite{Xiong}, $%
g_{AA}:g_{AB}:g_{BB}=1:1.7:2$. Calculating the optimal Hartree orbitals $%
\psi _{A}(\mathbf{r})$ and $\psi _{B}(\mathbf{r})$ with this set of
interaction parameters we find, see Fig.\ref{fig:ParticleDensityOneTen} and
in particular Fig.\ref{fig:LambdaAB}, that the orbitals associated with the
heavier $A$-atoms display a longer localisation length than those of the
lighter $B$-atoms, as the total particle number $N=N_{A}+N_{B}=(1+\frac{N_{A}%
}{N_{B}})\times N_{B}$ is increased at a fixed \emph{mixing} ratio $\frac{%
N_{A}}{N_{B}}$. When a pair of critical particle numbers $\left(
N_{c,A}^{\star },N_{c,B}^{\star }\right) $ is reached in this process, see
Fig.\ref{fig:LambdaAB}, there happens a sudden \emph{demixing} \emph{quantum
transition}. The heavier $A$-atoms delocalise, so that the condensate that
then remains localised around the crossing of $\mathcal{C}$ is a pure single
atom BEC consisting only of the lighter $B$-atoms. Correspondingly, the
chemical potential $\mu _{A}$ approaches the excitation threshold $%
\varepsilon _{xt,A}$ of the $A$-atoms as $N_{B}\rightarrow N_{c,B}^{\star }$
from below, see Fig.\ref{fig:muSqrtRbNa}. The critical particle number $%
N_{c,B}^{\star }$ characterizing this demixing quantum transition decreases
as the mixing ratio $\frac{N_{A}}{N_{B}}$ is increased. In Fig.\ref%
{fig:LambdaAB} $N_{c,B}$ denotes (for the waveguide $\mathcal{C}$ under
consideration) the maximum particle number $N_{B}$ that can be trapped in
the \emph{pure} Hartree ground state consisting only of $B$-atoms. The
localisation length $\lambda _{B}$ and the chemical potential $\mu _{B}$ of
the GP-orbital $\psi _{B}(\mathbf{r})$ undergo, depending on the mixing
ratio $\frac{N_{A}}{N_{B}}$ and on the mass ratio $\frac{m_{A}}{m_{B}}$ , at
a particular particle number $N_{B}=N_{c,B}^{\star }$ a jump. Both
quantities, the localisation length $\lambda _{B}$ and the chemical
potential $\mu _{B}$ , assume then in the remaining interval $N_{c,B}^{\star
}<$ $N_{B}<N_{c,B}$ values corresponding to a localised single atom species
Hartree ground state, see Fig.\ref{fig:LambdaCLTPlot}, Fig. \ref%
{fig:muCLTPlot}. Like in the single atom species case, see Fig. \ref%
{fig:muCLTPlot}, the scaling of the chemical potentials $\mu _{A}$ and $\mu
_{B}$ vs. particle number $N$ at fixed mixing ratio $\frac{N_{A}}{N_{B}}$,
see Fig. \ref{fig:muSqrtRbNa}, differs noticeably from the scaling of the
chemical potential for standard conservative atom trap potentials \cite%
{Pethick&Smith}.

It should be pointed out that for binary mixtures of Bose atoms confined in
standard \emph{conservative} atom trap potentials, well known stability
criteria \cite{Pethick&Smith} describe possible coexistence and also
segregation of phases dependent on the interaction parameters $g_{AA}$, $%
g_{AB}$, $g_{BB}$. However, such criteria are not directly applicable to the
above described sudden demixing transition, because around the branching or
crossing zone of a QW the kinetic energy of a localised binary BEC dominates
by far the interaction energy, so that the Thomas-Fermi approximation is
false. For instance, if one adds a small number $N_{A}\ll N_{B}$ of $A$%
-atoms to a cloud of $B$-atoms confined in a standard \emph{conservative}
atom trap potential, the $A$-atoms either reside at the surface formed by
the $B$-atom cloud, or are positioned deep inside of the $B$-atom cloud ,
depending on the interaction strengths \ref{binary mixture interaction
strengths}. Our calculations of atom density profiles $n_{A}\left( \mathbf{r}%
\right) =N_{A}\left\vert \psi _{A}\left( \mathbf{r}\right) \right\vert ^{2}$
and $n_{B}\left( \mathbf{r}\right) =N_{B}\left\vert \psi _{B}\left( \mathbf{r%
}\right) \right\vert ^{2}$ , see for example Fig.\ref%
{fig:ParticleDensityOneTen}, indicate for a wide range of interaction
parameters, that this scenario does not apply for cold Bose atoms trapped
around the branching zone or crossing of a QW.

\section{Conclusions}

We have studied (within the range of validity of mean field theory)
localised matter wave ground states of cold Bose atoms for different
prototypes of quantum waveguides with broken translational symmetry: i) a
waveguide system $\mathcal{C}$ akin to the shape of a swiss cross, ii) a
waveguide $\mathcal{L}$ in the guise of a cranked $L$ with a rectangular
corner, iii) a $T$-shaped waveguide $\mathcal{T}$ consisting of three
branching arms, see Fig.\ref{fig:waveguides}. The associated trapping
mechanism is non standard, because the force confining the particles around
the branching zone or crossing of waveguides cannot be derived from a
potential.

Based on an analytic expression, that approximates for small propagation
times $\Delta \tau $ the quantum propagator of a single particle at
imaginary time, we solved numerically the three-dimensional Gross-Pitaevskii
equation inside those quantum waveguides using a suitable splitting scheme,
and found depending on the choice of the ratio $\kappa ^{\left( \Gamma
\right) }$ of lateral tube widths, for fixed particle number $N$, various
localised Hartree ground states describing non standard trapping of cold
interacting Bose atoms. The kernel representing the imaginary time quantum
propagator implemented into the algorithm obeys by construction the
Dirichlet boundary conditions at the walls $\partial \Gamma $ of the
associated waveguides $\Gamma \in \left\{ \mathcal{C},\mathcal{L},\mathcal{T}%
\right\} $ exactly.

Observing, that the transversal kinetic energy of a particle undergoes a
rapid drop, when it traverses along a straight line the branching zone of
the respective arms inside the waveguides $\Gamma $, we suggested an
explanation for the existence of a localised ground state in section \ref%
{Kinetic Energy Induced Confinement at a Crossing.}. We also discussed the
non existence of localised states in the waveguides $\Gamma $ for too small,
respectively too large, lateral tube widths ratios $\kappa ^{\left( \Gamma
\right) }$, see Fig.\ref{fig:LambdaKappaNN1} and Fig.\ref{fig:LambdaKappaNN}%
. Analytical scaling laws obtained in section \ref{Scaling Laws for
Localised GP-Orbitals from a One--Dimensional Toy Model.} for the dependence
on $N$ of the localisation length $\lambda _{N}$ and the chemical potential $%
\mu _{N}$ agree very well with the results of the three-dimensional
numerical calculations. We found that the kinetic energy of a BEC confined
by this non standard trapping mechanism is by a factor seven(!) larger than
the interaction energy, see Fig.\ref{fig: ratio E_int/E_kin}, so that the
density profile of a BEC trapped around the branching or crossing zone of
waveguides, see for example Fig.\ref{fig:PsiProfileC} and Fig.\ref%
{fig:ParticleDensityOneTen}, cannot be described by the Thomas-Fermi
approximation.

For the case of a binary mixture of two different Bose atom species $A$ and $%
B$ we observed non standard trapping of both atom species for subcritical
particle numbers $N_{A}$ and $N_{B}$ around the branching or crossing zone
of quantum waveguides. A sudden demixing quantum transition takes place at a
critical particle number $N^{\star }=$ $N_{c,A}^{\star }+N_{c,B}^{\star }$
as the total particle number $N=N_{A}+N_{B}$ is increased at fixed mixing
ratio $N_{A}/N_{B}$, see Fig. \ref{fig:LambdaAB}. Depending on the mass
ratio $m_{A}/m_{B}$ the heavier atom species delocalises first for a \emph{%
wide} range of interaction parameters. We found that in this case the
dominant energy is not the interaction energy, but the kinetic energy of the
atoms. This feature could perhaps be used to seperate isotopes.

Finally we mention, that the choice of a hard wall boundary condition (\ref%
{boundary value condition GP-orbital}) at the walls $\partial \Gamma $ of
our waveguides $\Gamma \in \left\{ \mathcal{C},\mathcal{L},\mathcal{T}%
\right\} $ serves in our calculations just as a convenient model. Choosing
more general Robin boundary conditions (with a positive slip length), or
replacing the walls of the tubes by a steep harmonic potential (which should
be more appropriate to describe confinement generated by optical dipole
forces) in no way changes qualitatively any of the above described
localisation phenomena of cold matter waves around the branching or crossing
zones of quantum waveguides.

\begin{acknowledgments}
We thank J\'{o}szef Fort\'{a}gh for useful discussions.
\end{acknowledgments}

\clearpage

\appendix 
\section{}
\label{Appendix A}

The Magnus expansion theorem states for the product of the exponential of
two linear operators $\hat{A}$ and $\hat{B}$ \cite{Wilcox} 
\begin{equation}
e^{\hat{A}}\circ e^{\hat{B}}=e^{\hat{A}+\hat{B}+\frac{1}{2}\left[ \hat{A},%
\hat{B}\right] +\frac{1}{12}\left[ \hat{A}-\hat{B},\left[ \hat{A},\hat{B}%
\right] \right] +...}
\end{equation}

, where $\left[ \hat{A},\hat{B}\right] =\hat{A}\circ \hat{B}-\hat{B}\circ 
\hat{A}$. By explicit calculation, it can then be shown introducing a small
parameter $\tau $: 
\begin{eqnarray}
\widehat{S}(\tau ) &=&e^{-\frac{\tau }{2}\hat{A}}\circ e^{-\tau \hat{B}%
}\circ e^{-\frac{\tau }{2}\hat{A}} \\
&=&e^{-\tau \left( \hat{A}+\hat{B}\right) +\frac{\tau ^{3}}{24}\left[ \hat{A}%
+2\hat{B},\left[ \hat{A},\hat{B}\right] \right] +O\left( \tau ^{5}\right)
...}  \nonumber
\end{eqnarray}

All even powers of $\ \tau $ in the exponent cancel as can be seen from the
identity $\widehat{S}(\tau )\widehat{S}(-\tau )=\hat{1}$. There follows with
real parameters $\lambda _{1}$, $\lambda _{2}>0$ : 
\begin{equation}
\begin{array}{c}
e^{-\lambda_{1}\tau\hat{B}}\circ e^{-\frac{\tau}{2}\hat{A}}\circ
e^{-\lambda_{2}\tau\hat{B}}\circ e^{-\frac{\tau}{2}\hat{A}}\circ
e^{-\lambda_{1}\tau\hat{B}} \\ 
\\ 
=\exp\left\{ 
\begin{array}{c}
-\left(2\lambda_{1}+\lambda_{2}\right)\tau\hat{B}\ -\tau\hat{A} \\ 
-\frac{\tau^{3}}{24}\left(4\lambda_{1}-\lambda_{2}\right)\left[\hat{A},\left[%
\hat{A},\hat{B}\right]_{-}\right]_{-} \\ 
-\frac{\tau^{3}}{24}\left(\left(\lambda_{1}+\lambda_{2}\right)4\lambda_{1}-2%
\lambda_{2}^{2}\right)\left[\hat{B},\left[\hat{B,}\hat{A}\right]_{-}\right]%
_{-} \\ 
+o\left(\tau^{5}\right)%
\end{array}%
\right\}%
\end{array}
\label{eq:Magnus}
\end{equation}

Let us assume $||\hat{B}||\ll ||\hat{A}||$. In order that equation (\ref%
{eq:Magnus}) represents an accurate approximation to the original time
development operator $e^{-\tau \left( \hat{A}+\hat{B}\right) } $ for small $%
\tau$ we require now $\lambda _{1}=\frac{1}{6}$ , $\lambda _{2}=\frac{2}{3}$%
. We consequently obtain that the accuracy of the approximation%
\begin{equation}
e^{-\tau \left( \hat{A}+\hat{B}\right) }\simeq e^{-\frac{\tau }{6}\hat{B}%
}\circ e^{-\frac{\tau }{2}\hat{A}}\circ e^{-\frac{2\tau }{3}\hat{B}}\circ
e^{-\frac{\tau }{2}\hat{A}}\circ e^{-\frac{\tau }{6}\hat{B}}
\end{equation}

is of order $O\left( \tau ^{2}||\hat{B}||^{2}+\tau ^{4}||\hat{A}%
||^{4}\right) $. This property provides the basis of the splitting scheme as
stated in (\ref{splitting scheme GP-equation}).

\section{}
\label{Appendix B}

It is convenient to write $w_{a}^{\left( \mathcal{C}\right) }=2L_{a}$ for
the respective tube diameters $w_{a}^{\left( \mathcal{C}\right) }$ of the
arms $\mathcal{A}_{j}\subset \mathcal{C}$, see Fig.\ref{fig:waveguides}.
One-dimensional heat kernels obeying to homogeneous Dirichlet boundary
conditions at the end points of the intervals $\left[ L_{a},\infty \right] $
, $\left[ -\infty ,-L_{a}\right] $ and $\left[ -L_{a},L_{a}\right] $ may be
found by the standard mirror method of Sommerfeld:

\begin{eqnarray}
u,u^{\prime } &\in &\mathbb{R} \\
k(u-u^{\prime },\tau ) &=&\frac{1}{\sqrt{4\pi \tau }}\exp \left[ -\frac{%
(u-u^{\prime })^{2}}{4\tau }\right]  \nonumber \\
k_{[L_{a},\infty ]}^{(D)}(u,u^{\prime };\tau ) &=&k(u-u^{\prime };\tau
)-k(u+u^{\prime }-2L_{a};\tau )  \nonumber \\
k_{[-\infty ,-L_{a}]}^{(D)}(u,u^{\prime };\tau ) &=&k(u-u^{\prime };\tau
)-k(u+u^{\prime }+2L_{a};\tau )  \nonumber \\
k_{[-L_{a},L_{a}]}^{(D)}(u,u^{\prime };\tau ) &=&\sum_{n\in \mathbb{Z}}\left[
k(u-u^{\prime }+4nL_{a};\tau )\right.  \nonumber \\
&&\left. -k(u+u^{\prime }+(4n+2)L_{a};\tau )\right]  \nonumber
\end{eqnarray}%
\newline
We show in \cite{Supplementary Material}, that the short time expansion of
the three-dimensional imaginary time quantum propagator $K(\mathbf{r},%
\mathbf{r}^{\prime };\Delta \tau )=\left\langle \mathbf{r}\right\vert
e^{-\Delta \tau H_{kin}}\left\vert \mathbf{r}^{\prime }\right\rangle $
obeying to Dirichlet boundary conditions at the walls $\partial \mathcal{C}$
of a cross shaped waveguide $\mathcal{C}$, assumes for a small diffusion
time $\Delta \tau >0$ the following explicit guise: 
\begin{widetext}
\begin{equation}
\left[ K(\mathbf{r},\mathbf{r}^{\prime };\Delta \tau )\right] _{\mathbf{r}%
\in \mathcal{A}_{j},\mathbf{r}^{\prime }\in \mathcal{A}_{l}}=\mathcal{K}_{%
\mathcal{A}_{j},\mathcal{A}_{l}}\left( \mathbf{r},\mathbf{r}^{\prime
};\Delta \tau \right) =\mathcal{K}_{\mathcal{A}_{j},\mathcal{A}_{l}}^{\left(
\perp \right) }\left( \mathbf{r}_{\perp },\mathbf{r}_{\perp }^{\prime
};\Delta \tau \right) k_{[-L_{z},L_{z}]}^{\left( D\right) }(z,z^{\prime
};\Delta \tau )  \label{exact heat kernel through Delta_tau}
\end{equation}%
\begin{equation*}
\mathcal{K}_{\mathcal{A}_{0},\mathcal{A}_{0}}^{\left( \perp \right) }\left( 
\mathbf{r}_{\perp },\mathbf{r}_{\perp }^{\prime };\Delta \tau \right) =\left[
\begin{array}{c}
k_{[-L_{x},L_{x}]}^{\left( D\right) }(x,x^{\prime };\Delta \tau
)k_{[-L_{y},L_{y}]}^{\left( D\right) }(y,y^{\prime };\Delta \tau ) \\ 
\\ 
+\left[ C_{\mathcal{A}_{3},\mathcal{A}_{0}}(x,x^{\prime };\Delta \tau )-C_{%
\mathcal{A}_{1},\mathcal{A}_{0}}(x,x^{\prime };\Delta \tau )\right]
k_{[-L_{y},L_{y}]}^{\left( D\right) }(y,y^{\prime };\Delta \tau ) \\ 
\\ 
+k_{[-L_{x},L_{x}]}^{\left( D\right) }(x,x^{\prime };\Delta \tau )\left[ C_{%
\mathcal{A}_{4},\mathcal{A}_{0}}(y,y^{\prime };\Delta \tau )-C_{\mathcal{A}%
_{2},\mathcal{A}_{0}}(y,y^{\prime };\Delta \tau )\right]%
\end{array}%
\right]
\end{equation*}%
\begin{eqnarray}
\mathcal{K}_{\mathcal{A}_{0},\mathcal{A}_{1}}^{\left( \perp \right) }\left( 
\mathbf{r}_{\perp },\mathbf{r}_{\perp }^{\prime };\Delta \tau \right) &=&C_{%
\mathcal{A}_{1},\mathcal{A}_{1}}(x,x^{\prime };\Delta \tau
)k_{[-L_{y},L_{y}]}^{\left( D\right) }(y,y^{\prime };\Delta \tau )  \notag \\
\mathcal{K}_{\mathcal{A}_{0},\mathcal{A}_{2}}^{\left( \perp \right) }\left( 
\mathbf{r}_{\perp },\mathbf{r}_{\perp }^{\prime };\Delta \tau \right)
&=&k_{[-L_{x},L_{x}]}^{\left( D\right) }(x,x^{\prime };\Delta \tau )C_{%
\mathcal{A}_{2},\mathcal{A}_{2}}(y,y^{\prime };\Delta \tau )  \notag \\
\mathcal{K}_{\mathcal{A}_{0},\mathcal{A}_{3}}^{\left( \perp \right) }\left( 
\mathbf{r}_{\perp },\mathbf{r}_{\perp }^{\prime };\Delta \tau \right) &=&-C_{%
\mathcal{A}_{3},\mathcal{A}_{3}}(x,x^{\prime };\Delta \tau
)k_{[-L_{y},L_{y}]}^{\left( D\right) }(y,y^{\prime };\Delta \tau )  \notag \\
\mathcal{K}_{\mathcal{A}_{0},\mathcal{A}_{4}}^{\left( \perp \right) }\left( 
\mathbf{r}_{\perp },\mathbf{r}_{\perp }^{\prime };\Delta \tau \right)
&=&-k_{[-L_{x},L_{x}]}^{\left( D\right) }(x,x^{\prime };\Delta \tau )C_{%
\mathcal{A}_{4},\mathcal{A}_{4}}(y,y^{\prime };\Delta \tau )  \notag
\end{eqnarray}%
\begin{eqnarray*}
\mathcal{K}_{\mathcal{A}_{1},\mathcal{A}_{0}}^{\left( \perp \right) }\left( 
\mathbf{r}_{\perp },\mathbf{r}_{\perp }^{\prime };\Delta \tau \right) &=&%
\left[ C_{\mathcal{A}_{3},\mathcal{A}_{0}}(x,x^{\prime };\Delta \tau )-C_{%
\mathcal{A}_{1},\mathcal{A}_{0}}(x,x^{\prime };\Delta \tau )\right]
k_{[-L_{y},L_{y}]}^{\left( D\right) }(y,y^{\prime };\Delta \tau ) \\
\mathcal{K}_{\mathcal{A}_{1},\mathcal{A}_{1}}^{\left( \perp \right) }\left( 
\mathbf{r}_{\perp },\mathbf{r}_{\perp }^{\prime };\Delta \tau \right) &=&%
\left[ k_{[L_{x},\infty ]}^{\left( D\right) }(x,x^{\prime };\Delta \tau )+C_{%
\mathcal{A}_{1},\mathcal{A}_{1}}(x,x^{\prime };\Delta \tau )\right]
k_{[-L_{y},L_{y}]}^{\left( D\right) }(y,y^{\prime };\Delta \tau ) \\
\mathcal{K}_{\mathcal{A}_{1},\mathcal{A}_{2}}^{\left( \perp \right) }\left( 
\mathbf{r}_{\perp },\mathbf{r}_{\perp }^{\prime };\Delta \tau \right) &=&0=%
\mathcal{K}_{\mathcal{A}_{1},\mathcal{A}_{4}}^{\left( \perp \right) }\left( 
\mathbf{r}_{\perp },\mathbf{r}_{\perp }^{\prime };\Delta \tau \right) \\
\mathcal{K}_{\mathcal{A}_{1},\mathcal{A}_{3}}^{\left( \perp \right) }\left( 
\mathbf{r}_{\perp },\mathbf{r}_{\perp }^{\prime };\Delta \tau \right) &=&-C_{%
\mathcal{A}_{3},\mathcal{A}_{3}}(x,x^{\prime };\Delta \tau
)k_{[-L_{y},L_{y}]}^{\left( D\right) }(y,y^{\prime };\Delta \tau )
\end{eqnarray*}%
\begin{eqnarray*}
\mathcal{K}_{\mathcal{A}_{2},\mathcal{A}_{0}}^{\left( \perp \right) }\left( 
\mathbf{r}_{\perp },\mathbf{r}_{\perp }^{\prime };\Delta \tau \right)
&=&k_{[-L_{x},L_{x}]}^{\left( D\right) }(x,x^{\prime };\Delta \tau )\left[
C_{\mathcal{A}_{4},\mathcal{A}_{0}}(y,y^{\prime };\Delta \tau )-C_{\mathcal{A%
}_{2},\mathcal{A}_{0}}(y,y^{\prime };\Delta \tau )\right] \\
\mathcal{K}_{\mathcal{A}_{2},\mathcal{A}_{1}}^{\left( \perp \right) }\left( 
\mathbf{r}_{\perp },\mathbf{r}_{\perp }^{\prime };\Delta \tau \right) &=&0=%
\mathcal{K}_{\mathcal{A}_{2},\mathcal{A}_{3}}^{\left( \perp \right) }\left( 
\mathbf{r}_{\perp },\mathbf{r}_{\perp }^{\prime };\Delta \tau \right) \\
\mathcal{K}_{\mathcal{A}_{2},\mathcal{A}_{2}}^{\left( \perp \right) }\left( 
\mathbf{r}_{\perp },\mathbf{r}_{\perp }^{\prime };\Delta \tau \right)
&=&k_{[-L_{x},L_{x}]}^{\left( D\right) }(x,x^{\prime };\Delta \tau )\left[
k_{[L_{y},\infty ]}^{\left( D\right) }(y,y^{\prime };\Delta \tau )+C_{%
\mathcal{A}_{2},\mathcal{A}_{2}}(y,y^{\prime };\Delta \tau )\right] \\
\mathcal{K}_{\mathcal{A}_{2},\mathcal{A}_{4}}^{\left( \perp \right) }\left( 
\mathbf{r}_{\perp },\mathbf{r}_{\perp }^{\prime };\Delta \tau \right)
&=&-k_{[-L_{x},L_{x}]}^{\left( D\right) }(x,x^{\prime };\Delta \tau )C_{%
\mathcal{A}_{4},\mathcal{A}_{4}}(y,y^{\prime };\Delta \tau )
\end{eqnarray*}%
\begin{eqnarray*}
\mathcal{K}_{\mathcal{A}_{3},\mathcal{A}_{0}}^{\left( \perp \right) }\left( 
\mathbf{r}_{\perp },\mathbf{r}_{\perp }^{\prime };\Delta \tau \right) &=&%
\left[ C_{\mathcal{A}_{3},\mathcal{A}_{0}}(x,x^{\prime };\Delta \tau )-C_{%
\mathcal{A}_{1},\mathcal{A}_{0}}(x,x^{\prime };\Delta \tau )\right]
k_{[-L_{y},L_{y}]}^{\left( D\right) }(y,y^{\prime };\Delta \tau ) \\
\mathcal{K}_{\mathcal{A}_{3},\mathcal{A}_{1}}^{\left( \perp \right) }\left( 
\mathbf{r}_{\perp },\mathbf{r}_{\perp }^{\prime };\Delta \tau \right) &=&C_{%
\mathcal{A}_{1},\mathcal{A}_{1}}(x,x^{\prime };\Delta \tau
)k_{[-L_{y},L_{y}]}^{\left( D\right) }(y,y^{\prime };\Delta \tau ) \\
\mathcal{K}_{\mathcal{A}_{3},\mathcal{A}_{2}}^{\left( \perp \right) }\left( 
\mathbf{r}_{\perp },\mathbf{r}_{\perp }^{\prime };\Delta \tau \right) &=&0=%
\mathcal{K}_{\mathcal{A}_{3},\mathcal{A}_{4}}^{\left( \perp \right) }\left( 
\mathbf{r}_{\perp },\mathbf{r}_{\perp }^{\prime };\Delta \tau \right) \\
\mathcal{K}_{\mathcal{A}_{3},\mathcal{A}_{3}}^{\left( \perp \right) }\left( 
\mathbf{r}_{\perp },\mathbf{r}_{\perp }^{\prime };\Delta \tau \right) &=&%
\left[ k_{[-\infty ,-L_{x}]}^{\left( D\right) }(x,x^{\prime };\Delta \tau
)-C_{\mathcal{A}_{3},\mathcal{A}_{3}}(x,x^{\prime };\Delta \tau )\right]
k_{[-L_{y},L_{y}]}^{\left( D\right) }(y,y^{\prime };\Delta \tau )
\end{eqnarray*}%
\begin{eqnarray*}
\mathcal{K}_{\mathcal{A}_{4},\mathcal{A}_{0}}^{\left( \perp \right) }\left( 
\mathbf{r}_{\perp },\mathbf{r}_{\perp }^{\prime };\Delta \tau \right)
&=&k_{[-L_{x},L_{x}]}^{\left( D\right) }(x,x^{\prime };\Delta \tau )\left[
C_{\mathcal{A}_{4},\mathcal{A}_{0}}(y,y^{\prime };\Delta \tau )-C_{\mathcal{A%
}_{2},\mathcal{A}_{0}}(y,y^{\prime };\Delta \tau )\right] \\
\mathcal{K}_{\mathcal{A}_{4},\mathcal{A}_{1}}^{\left( \perp \right) }\left( 
\mathbf{r}_{\perp },\mathbf{r}_{\perp }^{\prime };\Delta \tau \right) &=&0=%
\mathcal{K}_{\mathcal{A}_{4},\mathcal{A}_{3}}^{\left( \perp \right) }\left( 
\mathbf{r}_{\perp },\mathbf{r}_{\perp }^{\prime };\Delta \tau \right) \\
\mathcal{K}_{\mathcal{A}_{4},\mathcal{A}_{2}}^{\left( \perp \right) }\left( 
\mathbf{r}_{\perp },\mathbf{r}_{\perp }^{\prime };\Delta \tau \right)
&=&k_{[-L_{x},L_{x}]}^{\left( D\right) }(x,x^{\prime };\Delta \tau )C_{%
\mathcal{A}_{2},\mathcal{A}_{2}}(y,y^{\prime };\Delta \tau ) \\
\mathcal{K}_{\mathcal{A}_{4},\mathcal{A}_{4}}^{\left( \perp \right) }\left( 
\mathbf{r}_{\perp },\mathbf{r}_{\perp }^{\prime };\Delta \tau \right)
&=&k_{[-L_{x},L_{x}]}^{\left( D\right) }(x,x^{\prime };\Delta \tau )\left[
k_{[-\infty ,-L_{y}]}^{\left( D\right) }(y,y^{\prime };\Delta \tau )-C_{%
\mathcal{A}_{4},\mathcal{A}_{4}}(y,y^{\prime };\Delta \tau )\right]
\end{eqnarray*}

\begin{eqnarray}
C_{\mathcal{A}_{1},\mathcal{A}_{0}}(x,x^{\prime };\tau ) &=&\sum_{n=-\infty
}^{\infty }\textnormal{sgn}\left[ \ x^{\prime }-\left( 4n+1\right) L_{x}%
\right] k(\left\vert x-L_{x}\right\vert +\ \left\vert x^{\prime }-\left(
4n+1\right) L_{x}\right\vert ;\tau )  \label{C-Integrale III}\\
C_{\mathcal{A}_{2},\mathcal{A}_{0}}(y,y^{\prime };\tau ) &=&\sum_{n=-\infty
}^{\infty }\textnormal{sgn}\left[ \ y^{\prime }-\left( 4n+1\right) L_{y}%
\right] k(\left\vert y-L_{y}\right\vert +\ \left\vert y^{\prime }-\left(
4n+1\right) L_{y}\right\vert ;\tau )  \notag \\
C_{\mathcal{A}_{1},\mathcal{A}_{1}}(x,x^{\prime };\tau ) &=&\textnormal{sgn}%
\left( x^{\prime }-L_{x}\right) k(\left\vert x-L_{x}\right\vert +\left\vert
x^{\prime }-L_{x}\right\vert ;\tau )  \notag \\
C_{\mathcal{A}_{2},\mathcal{A}_{2}}(y,y^{\prime };\tau ) &=&\textnormal{sgn}%
\left( y^{\prime }-L_{y}\right) k(\left\vert y-L_{y}\right\vert +\left\vert
y^{\prime }-L_{y}\right\vert ;\tau )  \notag \\
C_{\mathcal{A}_{3},\mathcal{A}_{0}}(x,x^{\prime };\tau ) &=&\sum_{n=-\infty
}^{\infty }\textnormal{sgn}\left[ x^{\prime }+\left( 4n+1\right) L_{x}\right]
k(\left\vert x+L_{x}\right\vert +\left\vert x^{\prime }+\left( 4n+1\right)
L_{x}\right\vert ;\tau )  \notag \\
C_{\mathcal{A}_{4},\mathcal{A}_{0}}(y,y^{\prime };\tau ) &=&\sum_{n=-\infty
}^{\infty }\textnormal{sgn}\left[ y^{\prime }+\left( 4n+1\right) L_{y}\right]
k(\left\vert y+L_{y}\right\vert +\left\vert y^{\prime }+\left( 4n+1\right)
L_{y}\right\vert ;\tau )  \notag \\
C_{\mathcal{A}_{3},\mathcal{A}_{3}}(x,x^{\prime };\tau ) &=&\textnormal{sgn}%
\left( x^{\prime }+L_{x}\right) k(\left\vert x+L_{x}\right\vert +\left\vert
x^{\prime }+L_{x}\right\vert ;\tau )  \notag \\
C_{\mathcal{A}_{4},\mathcal{A}_{4}}(y,y^{\prime };\tau ) &=&\textnormal{sgn}%
\left( y^{\prime }+L_{y}\right) k(\left\vert y+L_{y}\right\vert +\left\vert
y^{\prime }+L_{y}\right\vert ;\tau )  \notag \\
&& \notag \\
&& \notag \\
&& \notag \\
&& \notag \\
&& \notag \\
&& \notag \\
&& \notag
\end{eqnarray}%
\end{widetext}


\end{document}


\title{Supplementary material to : Cold Bose Atoms Around the Crossing of a
Quantum Waveguide.}
\author{A. Markowsky and N. Schopohl }
\email{corresponding author: nils.schopohl@uni-tuebingen.de}
\affiliation{ Eberhard Karls-Universit\"at T\"ubingen, Institut f\"ur Theoretische Physik, Auf
der Morgenstelle 14, D-72076 T\"ubingen, Germany\\
}
\date{\today}

\begin{abstract}
In this supplementary materials section, we provide (i) additional
information on the details of the determination of the imaginary time
quantum propagator of a single particle moving inside a cross shaped
waveguide and obeying to Dirichlet boundary conditions at the walls, (ii) we
give detailed information on the numerical approximation method employed
based on three-dimensional barycentric interpolation on a Chebyshev tensor
grid.
\end{abstract}

\maketitle

\section{Supplementary Note}

In what follows we present supplemental details not included in the original
article\cite{M+Sch}. First we give a full account of the analytical
calculations carried out to determine the imaginary time quantum propagator
associated with a single particle moving around the crossing of two quantum
waveguides. In the second part we explain the numerical methods employed, in
particular barycentric interpolation on a Chebyshev grid.

\subsection{Imaginary Time Quantum Propagator}

The mathematical analogy between a diffusion equation and the Schr\"{o}%
dinger equation is well established, for example \cite{Ramond},\cite{Berger}%
, so that in quantum field theory and mathematical physics the imaginary time
quantum propagator is often shortly referred to as the \emph{heat kernel}.
This object plays an important role in analyzing the spectrum of operators
like 
\begin{equation}
H_{kin}=-\frac{\hslash ^{2}}{2m}\nabla ^{2}
\end{equation}%
acting on functions $\psi ^{\left( \mathcal{C}\right) }(\mathbf{r})$ with a
support $\mathcal{C}$ , so that $H_{kin}$ is not separable in $\mathcal{C}$.

To find the ground state $\psi _{0}^{\left( \mathcal{C}\right) }(\mathbf{r})$
of the posed Schr\"{o}dinger eigenvalue problem for the cross shaped domain $%
\mathcal{C}$ (see see Fig.1 and equation (6) in \cite{M+Sch}) we consider an
auxiliary function $\psi (\mathbf{r};\tau )$ that is defined by a diffusion
process with a diffusion time $\tau $: 
\begin{eqnarray}
-\frac{\partial }{\partial \tau }\psi (\mathbf{r};\tau ) &=&H_{kin}\psi (%
\mathbf{r};\tau )  \label{Diffusion Eq.} \\
\lim_{\tau \rightarrow 0^{+}}\psi (\mathbf{r};\tau ) &\equiv &\psi ^{\left(
in\right) }(\mathbf{r})  \nonumber \\
\psi (\mathbf{r};\tau )|_{\mathbf{r\in \partial }\mathcal{C}} &=&0
\end{eqnarray}

A formal solution to such a diffusion equation problem \cite{Ramond} is
provided by the \emph{heat kernel } : 
\begin{eqnarray}
K(\mathbf{r},\mathbf{r}^{\prime };\tau ) &=&\left\langle \mathbf{r}|e^{-\tau
H_{kin}}|\mathbf{r}^{\prime }\right\rangle  \label{formal time evolution} \\
\psi (\mathbf{r};\tau ) &=&\int_{\mathcal{C}}d^{3}r^{\prime }K(\mathbf{r},%
\mathbf{r}^{\prime };\text{$\tau $})\psi ^{\left( in\right) }(\mathbf{r}%
^{\prime })  \nonumber
\end{eqnarray}%
It obeys inside the waveguide $\mathcal{C}$ to the homogeneous equation of
motion 
\begin{equation}
\left( \partial _{\tau }+H_{kin}\right) K(\mathbf{r},\mathbf{r}^{\prime };%
\text{$\tau $})=0  \label{heat kernel  equation of motion  I}
\end{equation}%
At intial time $\tau =0$ the kernel $K(\mathbf{r},\mathbf{r}^{\prime };\tau $%
$)$ obeys for a fixed source position $\mathbf{r}^{\prime }\in \mathcal{C}$
to the initial value condition%
\begin{equation}
\lim_{\tau \rightarrow 0^{+}}K(\mathbf{r},\mathbf{r}^{\prime };\tau )=\delta
^{(3)}(\mathbf{r}-\mathbf{r}^{\prime })
\label{heat kernel  initial value  I}
\end{equation}%
In addition, $K(\mathbf{r},\mathbf{r}^{\prime };\tau $$)$ fullfills at the
walls $\partial \mathcal{C}$ standard Dirichlet boundary conditions: 
\begin{equation}
K(\mathbf{r},\mathbf{r}^{\prime };\tau )|_{\mathbf{r}\in \partial \mathcal{C}%
}=0  \label{Hard wall boundary value condition II}
\end{equation}%
The latter requirement represents the difficult part of the problem, because
the operator $H_{kin}$ is not separable in the waveguide geometry $\mathcal{C%
}$.

In practice, all waveguide tubes $\mathcal{A}_{1}$,$...$, $\mathcal{A}_{4}$
comprising the arms of $\mathcal{C}$ are of a finite extension $\widetilde{L}
$ along their respective tube axes. Choosing this length $\widetilde{L}\ $%
sufficiently large (but finite), the perturbation of a localised state by
the finiteness of the extension of the waveguide system should be hardly
descernible \cite{dng}. Keeping $\widetilde{L}$ $<\infty $, the Hamiltonian $%
H_{kin}$, which now operates on wave functions with support in the truncated
subdomain, has a discrete point spectrum. In this case the connection
between the auxiliary function $\psi (\mathbf{r};\tau )$ and the seeked
ground state mode $\psi _{0}^{\left( \mathcal{C}\right) }(\mathbf{r})$ of
the Schr\"{o}dinger eigenvalue problem (equation (6) in \cite{M+Sch}) is
revealed considering a complete countable orthonormal system of
eigenfunctions $\psi _{n}^{\left( \mathcal{C}\right) }\left( \mathbf{r}%
\right) $ of the operator $H_{kin}$ with eigenvalues $E_{0}^{\left( \mathcal{%
C}\right) }<E_{1}^{\left( \mathcal{C}\right) }<E_{2}^{\left( \mathcal{C}%
\right) }....$.

All the modes $\psi _{n}^{\left( \mathcal{C}\right) }\left( \mathbf{r}%
\right) $ obeying to Dirichlet boundary value conditions (\ref{Hard wall
boundary value condition II}) at the walls $\partial \mathcal{C}$ of the
domain $\mathcal{C}$ , there holds for the kernel $K(\mathbf{r},\mathbf{r}%
^{\prime };\tau )$ the spectral representation (\cite{Ramond}) 
\begin{equation}
K(\mathbf{r},\mathbf{r}^{\prime };\tau )=\sum_{n=0}^{\infty }e^{-\tau
E_{n}^{\left( \mathcal{C}\right) }}\psi _{n}^{\left( \mathcal{C}\right)
}\left( \mathbf{r}\right) \left[ \psi _{n}^{\left( \mathcal{C}\right)
}\left( \mathbf{r}^{\prime }\right) \right] ^{\dag }  \label{heat kernel I}
\end{equation}%
Inserting (\ref{heat kernel I}) into (\ref{formal time evolution}) there
holds for $E_{1}^{\left( \mathcal{C}\right) }-E_{0}^{\left( \mathcal{C}%
\right) }\gg \frac{1}{\tau }$ asymptotically

\begin{equation}
\psi (\mathbf{r};\tau )\varpropto e^{-\tau E_{0}^{\left( \mathcal{C}\right)
}}\psi _{0}^{\left( \mathcal{C}\right) }(\mathbf{r})\ \left\langle \psi
_{0}^{\left( \mathcal{C}\right) },\psi ^{\left( in\right) }\right\rangle
\end{equation}%
The normalized ground state $\psi _{0}^{\left( \mathcal{C}\right) }(\mathbf{r%
})$ with energy eigenvalue $\ E_{0}^{\left( \mathcal{C}\right) }$ then
corresponds to the limit $\tau \rightarrow \infty $ of this diffusion
process:%
\begin{equation}
\psi _{0}^{\left( \mathcal{C}\right) }(\mathbf{r})=\lim_{\tau \rightarrow
\infty }\frac{\psi (\mathbf{r};\tau )}{\sqrt{\int_{\mathcal{C}%
}d^{3}r^{\prime }\ \left\vert \psi (\mathbf{r}^{\prime };\tau )\right\vert
^{2}}}  \label{normaliized ground state wavefunction}
\end{equation}%
Stopping the diffusion process at a large (but finite) diffusion time $\tau $
thus provides a suitable variational wave function for the ground state $%
\psi _{0}^{\left( \mathcal{C}\right) }(\mathbf{r})$, provided the cut off
length $\widetilde{L}\ $was chosen sufficiently large.

In the numerical calculations we find it convenient to determine the
associated eigenvalue $E_{0}^{\left( \mathcal{C}\right) }$ of the eigenmode $%
\psi _{0}^{\left( \mathcal{C}\right) }(\mathbf{r})$ directly from the
propagator, thus avoiding evaluation of second order spatial derivatives:

\begin{eqnarray}
&& \\
E_{0}^{\left( \mathcal{C}\right) } &=&\left\langle \psi _{0}^{\left( 
\mathcal{C}\right) }|H_{kin}|\psi _{0}^{\left( \mathcal{C}\right)
}\right\rangle  \nonumber \\
&=&\lim_{\tau \rightarrow 0}\frac{1-\left\langle \psi _{0}^{\left( \mathcal{C%
}\right) }|e^{-\tau H_{kin}}|\psi _{0}^{\left( \mathcal{C}\right)
}\right\rangle }{\tau }  \nonumber
\end{eqnarray}

An important consistency property of heat kernels is the semigroup identity,
implying that a kernel $K(\mathbf{r},\mathbf{r}^{\prime };\tau _{n})$ to be
evaluated at a later time $\tau _{n}=\tau _{n-1}+\Delta \tau $ may be
reconstructed recursively from earlier times $\tau _{n-1\ }$

\begin{equation}
K(\mathbf{r},\mathbf{r}^{\prime };\tau _{n})=\int_{\mathcal{C}}d^{3}\mathbf{r%
}_{1}K(\mathbf{r},\mathbf{r}_{1};\Delta \tau )K(\mathbf{r}_{1},\mathbf{r}%
^{\prime };\tau _{n-1})  \label{Chapman-Kolmogorov I}
\end{equation}%
So for the reconstruction of $K(\mathbf{r},\mathbf{r}^{\prime };\tau _{n})$
at larger values $\tau _{n}\ $it suffices to know an accurate approximation
for the short time asymptotics of the kernel $K(\mathbf{r},\mathbf{r}%
^{\prime };\Delta \tau )$.

Because all five subdomains $\mathcal{A}_{j}\subset \mathcal{C}$ have a
rectangular shape, see Fig.1 in \cite{M+Sch}, the kernel $K(\mathbf{r},%
\mathbf{r}^{\prime };\Delta \tau )$ of the kinetic energy operator obeying
to (\ref{heat kernel initial value I}) and (\ref{Hard wall boundary value
condition II}) can be obtained introducing tensorised one-dimensional
auxiliary kernels obeying either to homogeneous Dirichlet or to homogeneous
Neumann boundary conditions at the boundaries of the respective subdomains $%
\mathcal{A}_{j}$. Common to those one-dimensional kernels is, that they all
are constructed from the fundamental solution 
\begin{eqnarray}
u,u^{\prime } &\in &\mathbb{R}  \label{heat kernel III} \\
k(u-u^{\prime };\tau ) &=&\frac{1}{\sqrt{4\pi \ \tau }}\exp \left[ -\frac{%
\left( u-u^{\prime }\right) ^{2}}{4\tau }\right]   \nonumber
\end{eqnarray}%
This function solves the homogeneous one-dimensional diffusion equation
along the real axis $\mathbb{R}$ , 
\begin{equation}
\left( \frac{\partial }{\partial \tau }-\frac{\partial ^{2}}{\partial u^{2}}%
\right) k(u-u^{\prime };\tau )=0
\label{standard 1D heat equation initial value problem}
\end{equation}%
, and obeys at time $\tau =0$ to the intial value condition 
\begin{equation}
\lim_{\tau \rightarrow 0^{+}}k(u-u^{\prime };\tau )=\delta \left(
u-u^{\prime }\right) 
\end{equation}%
Upon analytic continuation  $\tau \rightarrow it$   the fundamental solution
(\ref{heat kernel III}) coincides with the well known Feynman propagator of
a freely moving particle.

We now write $w_{a}^{\left( \mathcal{C}\right) }=2L_{a}$ for the respective
tube diameters $w_{a}^{\left( \mathcal{C}\right) }$ of the arms $\mathcal{A}%
_{j}\subset \mathcal{C}$, see Fig.1 in \cite{M+Sch}. One-dimensional kernels
obeying to homogenous Dirichlet boundary conditions at the end points of the
intervals $\left[ L_{a},\infty \right] $ , $\left[ -\infty ,-L_{a}\right] $
and $\left[ -L_{a},L_{a}\right] $ may then be found by the mirror method of
Sommerfeld:

\begin{eqnarray}
L &\in &\left\{ L_{x},L_{y},L_{z}\right\}  \label{heat kernel  IVa} \\
k_{[L,\infty ]}^{(D)}(u,u^{\prime };\tau ) &=&k(u-u^{\prime };\tau
)-k(u+u^{\prime }-2L;\tau )  \nonumber \\
k_{[-\infty ,-L]}^{(D)}(u,u^{\prime };\tau ) &=&k(u-u^{\prime };\tau
)-k(u+u^{\prime }+2L;\tau )  \nonumber \\
k_{[-L,L]}^{(D)}(u,u^{\prime };\tau ) &=&\sum_{n\in \mathbb{Z}}\left[ 
\begin{array}{c}
k(u-u^{\prime }+4nL;\tau ) \\ 
-k(u+u^{\prime }+(4n+2)L;\tau )%
\end{array}%
\right]  \nonumber
\end{eqnarray}%
\newline
Correspondingly, one-dimensional kernels obeying to homogeneous Neumann
boundary conditions at the end points of the respective intervals are given
by

\begin{eqnarray}
&&  \label{heat kernel  IV b} \\
k_{[L,\infty ]}^{(N)}(u,u^{\prime };\tau ) &=&k(u-u^{\prime };\tau
)+k(u+u^{\prime }-2L;\tau )  \nonumber  \label{heat kernel  IV b} \\
k_{[-\infty ,-L]}^{(N)}(u,u^{\prime };\tau ) &=&k(u-u^{\prime };\tau
)+k(u+u^{\prime }+2L;\tau )  \nonumber \\
k_{[-L,L]}^{(N)}(u,u^{\prime };\tau ) &=&\sum_{n\in \mathbb{Z}}\left[ 
\begin{array}{c}
k(u-u^{\prime }+4nL;\tau ) \\ 
+k(u+u^{\prime }+(4n+2)L;\tau )%
\end{array}%
\right]  \nonumber
\end{eqnarray}%
\newline
For example, the Dirichlet kernel $k_{\left[ -L_{a},L_{a}\right]
}^{(D)}\left( u,u^{\prime };\tau \right) $ permits a periodic continuation
outside the interval $\left[ -L_{a},L_{a}\right] $ :%
\begin{equation}
k_{\left[ -L_{a},L_{a}\right] }^{(D)}\left( u+4L_{a},u^{\prime };\tau
\right) =k_{\left[ -L_{a},L_{a}\right] }^{(D)}\left( u,u^{\prime };\tau
\right)
\end{equation}%
So it can also be represented as a trigonometric series: 
\begin{eqnarray}
&&  \label{eq:trigseries} \\
k_{\left[ -L_{a},L_{a}\right] }^{(D)}\left( u,u^{\prime };\tau \right)
&=&\sum_{n\in \mathbb{Z}}^{\infty }e^{-n^{2}\frac{\pi ^{2}}{4L_{a}^{2}}\tau
}\left( e^{in\pi \frac{u-u^{\prime }}{2L_{a}}}-e^{in\pi \frac{u+u^{\prime
}-2L_{a}}{2L_{a}}}\right)  \nonumber
\end{eqnarray}%
While this trigonometric series converges fast for large values of $\tau $,
the dual representation (\ref{heat kernel IVa}) converges fast for small $%
\tau $. For the recursive construction of the solution to the
Gross-Pitaevskii equation, see Eq.(35) in \cite{M+Sch}, only small values $%
\tau $ are of interest calculating the kernel $K(\mathbf{r},\mathbf{r}%
^{\prime },\Delta \tau )$. The representation (\ref{heat kernel IVa}) is
therefore more suitable for our purposes.

A wave packet approaching the crossing zone $\mathcal{A}_{0}$ may there
undergo scattering. It is then natural to represent the solution $\psi (%
\mathbf{r};\tau )$ to (\ref{Diffusion Eq.}) inside each subdomain $\mathcal{A%
}_{j}\subset \mathcal{C}\ $in the following manner:

\begin{eqnarray}
\psi _{\mathcal{A}_{j}}(\mathbf{r};\tau ) &=&\left[ \psi (\mathbf{r};\tau )%
\right] _{\mathbf{r}\in \mathcal{A}_{j}}  \label{subdomain functions II} \\
\psi _{\mathcal{A}_{0}}(\mathbf{r};\tau ) &=&\psi _{\mathcal{A}_{0}}^{\left(
D\right) }(\mathbf{r};\tau )+\left[ 
\begin{array}{c}
\phi _{\mathcal{A}_{1}}(\mathbf{r};\tau )+\phi _{\mathcal{A}_{2}}(\mathbf{r}%
;\tau ) \\ 
+\phi _{\mathcal{A}_{3}}(\mathbf{r};\tau )+\phi _{\mathcal{A}_{4}}(\mathbf{r}%
;\tau )%
\end{array}%
\right]  \nonumber
\end{eqnarray}%
\begin{eqnarray}
\psi _{\mathcal{A}_{1}}(\mathbf{r};\tau ) &=&\psi _{\mathcal{A}_{1}}^{\left(
D\right) }(\mathbf{r};\tau )\ +\phi _{\mathcal{A}_{1}}(\mathbf{r};\tau )+\
\phi _{\mathcal{A}_{3}}(\mathbf{r};\tau )  \nonumber \\
\psi _{\mathcal{A}_{2}}(\mathbf{r};\tau ) &=&\psi _{\mathcal{A}_{2}}^{\left(
D\right) }(\mathbf{r};\tau )\ +\phi _{\mathcal{A}_{2}}(\mathbf{r};\tau )+\
\phi _{\mathcal{A}_{4}}(\mathbf{r};\tau )  \nonumber
\end{eqnarray}%
\begin{eqnarray}
\psi _{\mathcal{A}_{3}}(\mathbf{r};\tau ) &=&\psi _{\mathcal{A}_{3}}^{\left(
D\right) }(\mathbf{r};\tau )\ +\phi _{\mathcal{A}_{1}}(\mathbf{r};\tau )+\
\phi _{\mathcal{A}_{3}}(\mathbf{r};\tau )  \nonumber \\
\psi _{\mathcal{A}_{4}}(\mathbf{r};\tau ) &=&\psi _{\mathcal{A}_{4}}^{\left(
D\right) }(\mathbf{r};\tau )\ +\phi _{\mathcal{A}_{2}}(\mathbf{r};\tau )+\
\phi _{\mathcal{A}_{4}}(\mathbf{r};\tau )  \nonumber
\end{eqnarray}%
The subdomain functions $\psi _{\mathcal{A}_{j}}^{\left( D\right) }(\mathbf{r%
};\tau )$ solve (\ref{Diffusion Eq.}) subject to \emph{inhomogeneous}
initial value conditions:

\begin{equation}
\lim_{\tau \rightarrow 0^{+}}\psi _{\mathcal{A}_{j}}^{\left( D\right) }(%
\mathbf{r};\tau )=\left[ \psi ^{\left( in\right) }(\mathbf{r})\right] _{%
\mathbf{r\in }\mathcal{A}_{j}}
\end{equation}%
Furthermore the functions $\psi _{\mathcal{A}_{j}}^{\left( D\right) }(%
\mathbf{r};\tau )$ obey (by construction) at the side, top and bottom walls
of the respective subdomains $\mathcal{A}_{j}$ to homogeneous Dirichlet
boundary value conditions%
\begin{equation}
\psi _{\mathcal{A}_{j}}^{\left( D\right) }(\mathbf{r};\tau )|_{\mathbf{r\in
\partial }\mathcal{A}_{j}}=0
\end{equation}%
In terms of the afore mentioned tensor product of one-dimensional Dirichlet
kernels the subdomain functions $\psi _{\mathcal{A}_{j}}^{\left( D\right) }(%
\mathbf{r};\tau )$ are given by%
\begin{widetext}
\begin{eqnarray}
&&  \label{auxiliarx functions psi_Dirichlet_A_j} \\
\psi _{\mathcal{A}_{0}}^{\left( D\right) }\left( \mathbf{r};\tau \right)
&=&\int_{-L_{x}}^{L_{x}}dx^{\prime }k_{[-L_{x},L_{x}]}^{\left( D\right)
}(x,x^{\prime };\tau )\int_{-L_{y}}^{L_{y}}dy^{\prime
}k_{[-L_{y},L_{y}]}^{\left( D\right) }(y,y^{\prime };\tau
)\int_{-L_{z}}^{L_{z}}dz^{\prime }k_{[-L_{z},L_{z}]}^{\left( D\right)
}(z,z^{\prime };\tau )\psi _{\mathcal{A}_{0}}^{\left( in\right) }(\mathbf{r}%
^{\prime })  \notag \\
\psi _{\mathcal{A}_{1}}^{\left( D\right) }\left( \mathbf{r};\tau \right)
&=&\int_{L_{x}}^{\infty }dx^{\prime }k_{[L_{x},\infty ]}^{\left( D\right)
}(x,x^{\prime };\tau )\int_{-L_{y}}^{L_{y}}dy^{\prime
}k_{[-L_{y},L_{y}]}^{\left( D\right) }(y,y^{\prime };\tau
)\int_{-L_{z}}^{L_{z}}dz^{\prime }k_{[-L_{z},L_{z}]}^{\left( D\right)
}(z,z^{\prime };\tau )\psi _{\mathcal{A}_{1}}^{\left( in\right) }(\mathbf{r}%
^{\prime })  \notag \\
\psi _{\mathcal{A}_{2}}^{\left( D\right) }\left( \mathbf{r};\tau \right)
&=&\int_{-L_{x}}^{L_{x}}dx^{\prime }k_{[-L_{x},L_{x}]}^{\left( D\right)
}(x,x^{\prime };\tau )\int_{L_{y}}^{\infty }dy^{\prime }k_{[L_{y},\infty
]}^{\left( D\right) }(y,y^{\prime };\tau )\int_{-L_{z}}^{L_{z}}dz^{\prime
}k_{[-L_{z},L_{z}]}^{\left( D\right) }(z,z^{\prime };\tau )\psi _{\mathcal{A}%
_{2}}^{\left( in\right) }(\mathbf{r}^{\prime })  \notag \\
\psi _{\mathcal{A}_{3}}^{\left( D\right) }\left( \mathbf{r};\tau \right)
&=&\int_{-\infty }^{-L_{x}}dx^{\prime }k_{[-\infty ,-L_{x}]}^{\left(
D\right) }(x,x^{\prime };\tau )\int_{-L_{y}}^{L_{y}}dy^{\prime
}k_{[-L_{y},L_{y}]}^{\left( D\right) }(y,y^{\prime };\tau
)\int_{-L_{z}}^{L_{z}}dz^{\prime }k_{[-L_{z},L_{z}]}^{\left( D\right)
}(z,z^{\prime };\tau )\psi _{\mathcal{A}_{3}}^{\left( in\right) }(\mathbf{r}%
^{\prime })  \notag \\
\psi _{\mathcal{A}_{4}}^{\left( D\right) }\left( \mathbf{r};\tau \right)
&=&\int_{-L_{x}}^{L_{x}}dx^{\prime }k_{[-L_{x},L_{x}]}^{\left( D\right)
}(x,x^{\prime };\tau )\int_{-\infty }^{-L_{y}}dy^{\prime }k_{[-\infty
,-L_{y}]}^{\left( D\right) }(y,y^{\prime };\tau
)\int_{-L_{z}}^{L_{z}}dz^{\prime }k_{[-L_{z},L_{z}]}^{\left( D\right)
}(z,z^{\prime };\tau )\psi _{\mathcal{A}_{4}}^{\left( in\right) }(\mathbf{r}%
^{\prime })  \notag
\end{eqnarray}%
\end{widetext}In contrast, the functions $\phi _{\mathcal{A}_{j}}(\mathbf{r}%
;\tau )$ on the right hand side of (\ref{subdomain functions II}) solve
inside the crossing zone $\mathcal{A}_{0}$ , and also inside the respective
arms $\mathcal{A}_{j}$ for $j\in \left\{ 1,2,3,4\right\} $, the diffusion
equation (\ref{Diffusion Eq.}) subject to \emph{homogeneous} initial value
conditions 
\begin{equation}
\lim_{\tau \rightarrow 0^{+}}\phi _{\mathcal{A}_{j}}(\mathbf{r};\tau )=0
\end{equation}%
Like the subdomain functions $\psi _{\mathcal{A}_{j}}^{\left( D\right) }(%
\mathbf{r};\tau )$ also the functions $\phi _{\mathcal{A}_{j}}(\mathbf{r}%
;\tau )$ obey at the walls of the respective arms $\mathcal{A}_{j}\subset 
\mathcal{C}$ to homogeneous Dirichlet boundary conditions. But when crossing
the common boundary $\partial \mathcal{A}_{0,j}$ of the central box $%
\mathcal{A}_{0}$ with one of the neighbouring arms $\mathcal{A}_{j}$, the
normal derivatives of the functions $\phi _{\mathcal{A}_{j}}(\mathbf{r};\tau
)$ undergo a jump. For example, at the (inner) interface $\partial \mathcal{A%
}_{0,1}$ between $\mathcal{A}_{0}$ and $\mathcal{A}_{1}$, there holds for $%
\mathbf{r\in }\partial \mathcal{A}_{0,1}$ 
\begin{eqnarray}
&&  \label{phi  functions Neumann@boundary A0} \\
\frac{\partial }{\partial x}\phi _{\mathcal{A}_{1}}(\mathbf{r};\tau
)|_{x=L-0} &=&f_{1}(\mathbf{r};\tau )=-\frac{\partial }{\partial x}\phi _{%
\mathcal{A}_{1}}(\mathbf{r};\tau )|_{x=L+0}  \nonumber
\end{eqnarray}%
, with analogous jump conditions at the other interfaces $\partial \mathcal{A%
}_{0,j}$ between $\mathcal{A}_{0}$ and the other arms $\mathcal{A}_{j}$.The
four functions $f_{j}\left( \mathbf{r};\tau \right) $ with $\mathbf{r}\in
\partial \mathcal{A}_{0,j}$ are determined selfconsistently by the
requirement, that the first derivative of the full solution $\psi (\mathbf{r}%
;\tau )$, see (\ref{subdomain functions II}), should be continuous crossing
the respective interfaces $\partial \mathcal{A}_{0,j}$.

Consider, for instance, a convolution of the Neumann kernel (\ref{heat
kernel IV b}) with a function $f\left( \tau \right) $: 
\begin{equation}
\phi \left( u;\tau \right) =-\int_{0}^{\tau }d\tau ^{\prime }k_{\left[
L,\infty \right] }^{\left( N\right) }\left( u,L;\tau -\tau ^{\prime }\right)
f\left( \tau ^{\prime }\right)  \label{convolution Neumann heat kernel}
\end{equation}%
Apparently, $\phi \left( u;\tau \right) $ solves for $x\neq L$ and $\tau >0$
the homogeneous one-dimensional diffusion equation subject to the initial
value condition $\phi \left( u;0\right) =0$. What's more, carefully
evaluating the derivative $\frac{\partial }{\partial \tau }\phi \left(
u;\tau \right) $ with regard to the upper limit of the integral in (\ref%
{convolution Neumann heat kernel}), one obtains 
\begin{eqnarray}
&& \\
\left( \frac{\partial }{\partial \tau }-\frac{\partial ^{2}}{\partial u^{2}}%
\right) \phi \left( u;\tau \right) &=&-2f\left( \tau \right) \delta \left(
u-L\right)  \nonumber \\
\lim_{u\rightarrow L\pm 0^{+}}\frac{\partial \phi \left( u;\tau \right) }{%
\partial u} &=&\pm f\left( \tau \right)  \nonumber
\end{eqnarray}%
This engenders that we may represent the afore mentioned auxiliary functions 
$\phi _{\mathcal{A}_{j}}(\mathbf{r};\tau )$ in terms of yet unknown boundary
functions $f_{j}(\mathbf{r};\tau )$ as convolution integrals with the
respective Neumann kernels:%
\begin{widetext}

\begin{eqnarray}
&&  \label{auxiliary functions phi_A_j} \\
\phi _{\mathcal{A}_{1}}(\mathbf{r};\tau ) &=&-\int_{0}^{\tau }d\tau ^{\prime
}k_{[L_{x},\infty ]}^{\left( N\right) }(x,L_{x};\tau -\tau ^{\prime }) 
 \int_{-L_{y}}^{L_{y}}dy^{\prime }k_{[-L_{y},L_{y}]}^{\left(
D\right) }(y,y^{\prime };\tau -\tau ^{\prime
})\int_{-L_{z}}^{L_{z}}dz^{\prime }k_{[-L_{z},L_{z}]}^{\left( D\right)
}(z,z^{\prime };\tau -\tau ^{\prime })f_{\mathcal{A}_{1}}(y^{\prime
},z^{\prime };\tau ^{\prime })  \notag
\end{eqnarray}%
\begin{eqnarray*}
\phi _{\mathcal{A}_{2}}(\mathbf{r};\tau ) &=&-\int_{0}^{\tau }d\tau ^{\prime
}k_{[L_{y},\infty ]}^{\left( N\right) }(y,L_{y};\tau -\tau ^{\prime }) 
\int_{-L_{x}}^{L_{x}}dx^{\prime }k_{[-L_{x},L_{x}]}^{\left(
D\right) }(x,x^{\prime };\tau -\tau ^{\prime
})\int_{-L_{z}}^{L_{z}}dz^{\prime }k_{[-L_{z},L_{z}]}^{\left( D\right)
}(z,z^{\prime };\tau -\tau ^{\prime })f_{\mathcal{A}_{2}}(x^{\prime
},z^{\prime };\tau ^{\prime })
\end{eqnarray*}%
\begin{eqnarray*}
\phi _{\mathcal{A}_{3}}(\mathbf{r};\tau ) &=&\int_{0}^{\tau }d\tau ^{\prime
}k_{[-\infty ,-L_{x}]}^{\left( N\right) }(x,-L_{x};\tau -\tau ^{\prime }) 
\int_{-L_{y}}^{L_{y}}dy^{\prime }k_{[-L_{y},L_{y}]}^{\left(
D\right) }(y,y^{\prime };\tau -\tau ^{\prime
})\int_{-L_{z}}^{L_{z}}dz^{\prime }k_{[-L_{z},L_{z}]}^{\left( D\right)
}(z,z^{\prime };\tau -\tau ^{\prime })f_{\mathcal{A}_{3}}(y^{\prime
},z^{\prime };\tau ^{\prime })
\end{eqnarray*}%
\begin{eqnarray*}
\phi _{\mathcal{A}_{4}}(\mathbf{r};\tau ) &=&\int_{0}^{\tau }d\tau ^{\prime
}k_{[-\infty ,-L_{y}]}^{\left( N\right) }\left( y,-L_{y};\tau -\tau ^{\prime
}\ \right) \int_{-L_{x}}^{L_{x}}dx^{\prime }k_{[-L_{x},L_{x}]}^{\left(
D\right) }(x,x^{\prime };\tau -\tau ^{\prime
})\int_{-L_{z}}^{L_{z}}dz^{\prime }k_{[-L_{z},L_{z}]}^{\left( D\right)
}(z,z^{\prime };\tau -\tau ^{\prime })f_{\mathcal{A}_{4}}(x^{\prime
},z^{\prime };\tau ^{\prime })
\end{eqnarray*}%
\end{widetext}

It follows from what has been said, that each linear combination $\psi _{%
\mathcal{A}_{j}}(\mathbf{r};\tau )$ as defined in (\ref{subdomain functions
II}) solves inside the respective domains $\mathcal{A}_{j}$ the
three-dimensional diffusion equation (\ref{Diffusion Eq.}) subject to the
posed initial- and boundary value conditions. By construction $\psi (\mathbf{%
r};\tau )$ is continuous at the four interfaces $\partial \mathcal{A}_{0,j}$
separating the center domain $\mathcal{A}_{0}$ from the arms $\mathcal{A}%
_{j} $:

\begin{eqnarray}
&&\lim_{\eta \rightarrow 0^{+}}\psi _{\mathcal{A}_{0}}(\mathbf{r}-\eta 
\mathbf{n}_{\partial \mathcal{A}_{0,j}};\tau )|_{\mathbf{r}\in \partial 
\mathcal{A}_{0,j}} \\
&=&\lim_{\eta \rightarrow 0^{+}}\psi _{\mathcal{A}_{j}}(\mathbf{r}+\eta 
\mathbf{n}_{\partial \mathcal{A}_{0,j}};\tau )|_{\mathbf{r}\in \partial 
\mathcal{A}_{0,j}}  \nonumber
\end{eqnarray}%
Here the surface normal vector $\mathbf{n}_{\partial \mathcal{A}_{0,j}}$
points outward from the interfacial plane $\partial \mathcal{A}_{0,j}$ into
the respective arm $\mathcal{A}_{j}$. What remains is the determination of
the boundary functions $f_{j}(\mathbf{r};\tau )$ for $\mathbf{r\in }\partial 
\mathcal{A}_{0,j}$ from the requirement of the continuity of the normal
derivatives $\mathbf{n}_{\partial \mathcal{A}_{0,j}}\cdot \mathbf{\nabla }%
\psi _{\mathcal{A}_{j}}$ crossing the common boundaries $\partial \mathcal{A}%
_{0,j}$ :

\begin{eqnarray}
&&\lim_{\eta \rightarrow 0^{+}}\mathbf{n}_{\partial \mathcal{A}_{0,j}}\cdot 
\mathbf{\nabla }\psi _{\mathcal{A}_{0}}(\mathbf{r}-\eta \mathbf{n}_{\partial 
\mathcal{A}_{0,j}};\tau )|_{\mathbf{r}\in \partial \mathcal{A}_{0,j}} \\
&=&\lim_{\eta \rightarrow 0^{+}}\mathbf{n}_{\partial \mathcal{A}_{0,j}}\cdot 
\mathbf{\nabla }\psi _{\mathcal{A}_{j}}(\mathbf{r}+\eta \mathbf{n}_{\partial 
\mathcal{A}_{0,j}};\tau )|_{\mathbf{r}\in \partial \mathcal{A}_{0,j}} 
\nonumber
\end{eqnarray}

Taking into account the afore mentioned jump conditions (\ref{phi functions
Neumann@boundary A0}) for the respective normal derivatives of the functions 
$\phi _{\mathcal{A}_{j}}(\mathbf{r};\tau )$ at the surfaces $\partial 
\mathcal{A}_{0,j}$, see (\ref{phi functions Neumann@boundary A0}), one
obtains then a system of coupled linear Volterra integral equations (of the
second kind) enabling the selfconsistent determination of the boundary
functions $f_{j}(\mathbf{r};\tau )$ for $\mathbf{r}\mathbf{\in }\partial 
\mathcal{A}_{0,j}$ and $j\in \left\{ 1,2,3,4\right\} $:

\begin{eqnarray}
&&  \label{Volterra} \\
{\small f}_{j}{\small (\mathbf{r};\tau )} &{\small =F}&_{j}{\small (\mathbf{r%
};\tau )+}\sum_{{\small l=1}}^{{\small 4}}\int_{\partial \mathcal{A}_{0,l}}%
{\small d}^{2}{\small \mathbf{r}}^{\prime }\int_{0}^{\tau }{\small d\tau }%
^{\prime }\mathcal{T}_{j,l}\left( {\small \mathbf{r}},{\small \mathbf{r}}%
^{\prime };{\small \tau -\tau }^{\prime }\right) {\small f}_{l}{\small (%
\mathbf{r}}^{\prime }{\small ;\tau }^{\prime }{\small )}  \nonumber
\end{eqnarray}%
The above described procedure leads to the following explicit expressions
for the inhomogeneous terms $F_{j}(\mathbf{r};\tau )$ and the pieces $%
\mathcal{T}_{j,l}\left( \mathbf{r},\mathbf{r}^{\prime };\tau -\tau ^{\prime
}\right) $ of the integral kernel:

\begin{equation}
\mathcal{T}_{j,l}=\left[%
\begin{array}{cccc}
0 & \mathcal{T}_{1,2} & 0 & \mathcal{T}_{1,4} \\ 
\mathcal{T}_{2,1} & 0 & \mathcal{T}_{2,3} & 0 \\ 
0 & \mathcal{T}_{3,2} & 0 & \mathcal{T}_{3,4} \\ 
\mathcal{T}_{4,1} & 0 & \mathcal{T}_{4,3} & 0%
\end{array}%
\right]_{j,l}
\end{equation}

\begin{widetext}

\begin{eqnarray}
&&\\
\mathcal{T}_{1,2}(u,u^{\prime},z,z^{\prime};\tau)
&=&-\frac{1}{2}k_{[L_{y},\infty]}^{(N)}(u,L_{y};\tau)\left[\partial_{x}k_{[-L_{x},L_{x}]}^{(D)}(x,u^{\prime};\tau)\right]_
{x=L_{x}}k_{[-L_{z},L_{z}]}^{(D)}(z,z^{\prime};\tau)\nonumber\\
\mathcal{T}_{1,2}(u,u^{\prime},z,z^{\prime};\tau)&=&\mathcal{T}_{2,1}(u,u^{\prime},z,z^{\prime};\tau)\nonumber\\
\mathcal{T}_{1,4}(u,u^{\prime},z,z^{\prime};\tau)
&=&\frac{1}{2}k_{[-\infty,-L_{y}]}^{(N)}(u,-L_{y};\tau)\left[\partial_{x}k_{[-L_{x},L_{x}]}^{(D)}(x,u^{\prime};\tau)\right]_
{x=L_{x}}k_{[-L_{z},L_{z}]}^{(D)}(z,z^{\prime};\tau)\nonumber\\
\mathcal{T}_{1,4}(u,u^{\prime},z,z^{\prime};\tau)&=&\mathcal{T}_{2,3}(u,u^{\prime},z,z^{\prime};\tau)\nonumber\\
\mathcal{T}_{3,2}(u,u^{\prime},z,z^{\prime};\tau)
&=&-\frac{1}{2}k_{[L_{y},\infty]}^{(N)}(u,L_{y};\tau)\left[\partial_{x}k_{[-L_{x},L_{x}]}^{(D)}(x,u^{\prime};\tau)\right]_
{x=-L_{x}}k_{[-L_{z},L_{z}]}^{(D)}(z,z^{\prime};\tau)\nonumber\\
\mathcal{T}_{3,2}(u,u^{\prime},z,z^{\prime};\tau)&=&\mathcal{T}_{4,1}(u,u^{\prime},z,z^{\prime};\tau)\nonumber\\
\mathcal{T}_{3,4}(u,u^{\prime},z,z^{\prime};\tau)
&=&\frac{1}{2}k_{[-\infty,-L_{y}]}^{(N)}(u,-L_{y};\tau)\left[\partial_{x}k_{[-L_{x},L_{x}]}^{(D)}(x,u^{\prime};\tau)\right]_
{x=-L_{x}}k_{[-L_{z},L_{z}]}^{(D)}(z,z^{\prime};\tau)\nonumber\\
\mathcal{T}_{3,4}(u,u^{\prime},z,z^{\prime};\tau)&=&\mathcal{T}_{4,3}(u,u^{\prime},z,z^{\prime};\tau)\nonumber
\end{eqnarray}

\begin{eqnarray}
\label{F equation}\\
F_{1}(u,z;\tau)=\frac{1}{2}\left[\begin{array}{c}
\int_{-L_{x}}^{L_{x}}dv^{\prime}\int_{-L_{y}}^{L_{y}}du^{\prime}\left[\partial_{v}k_{[-L_{x},L_{x}]}^{(D)}(v,v^{\prime};\tau)\right]_{v=L_{x}}
k_{[-L_{y},L_{y}]}^{(D)}(u,u^{\prime};\tau)\int_{-L_{z}}^{L_{z}}dz^{\prime}k_{[-L_{z},L_{z}]}^{(D)}(z,z^{\prime};\tau)\,
\psi_{\mathcal{A}_{0}}^{(in)}(v^{\prime},u^{\prime},z^{\prime})\\
\\
-\int_{L_{x}}^{\infty}dv^{\prime}\int_{-L_{y}}^{L_{y}}du^{\prime}\left[\partial_{v}k_{[L_{x},\infty]}^{(D)}(v,v^{\prime};\tau)\right]_{v=L_{x}}
k_{[-L_{y},L_{y}]}^{(D)}(u,u^{\prime};\tau)
\int_{-L_{z}}^{L_{z}}dz^{\prime}k_{[-L_{z},L_{z}]}^{(D)}(z,z^{\prime};\tau)\,\psi_{\mathcal{A}_{1}}^{(in)}(v^{\prime},u^{\prime},z^{\prime})
\end{array}\right]\nonumber
\end{eqnarray}

\begin{equation*}
F_{2}(u,z;\tau)=\frac{1}{2}\left[\begin{array}{c}
\int_{-L_{x}}^{L_{x}}du^{\prime}\int_{-L_{y}}^{L_{y}}dv^{\prime}k_{[-L_{x},L_{x}]}^{(D)}(u,u^{\prime};\tau)
\left[\partial_{v}k_{[-L_{y},L_{y}]}^{(D)}(v,v^{\prime};\tau)\right]_{v=L_{y}}
\int_{-L_{z}}^{L_{z}}dz^{\prime}k_{[-L_{z},L_{z}]}^{(D)}(z,z^{\prime};\tau)\,\psi_{\mathcal{A}_{0}}^{(in)}(u^{\prime},v^{\prime},z^{\prime})\\
\\
-\int_{-L_{x}}^{L_{x}}du^{\prime}\int_{L_{y}}^{\infty}dv^{\prime}k_{[-L_{x},L_{x}]}^{(D)}(u,u^{\prime};\tau)
\left[\partial_{v}k_{[L_{y},\infty]}^{(D)}(v,v^{\prime};\tau)\right]_{v=L_{y}}
\int_{-L_{z}}^{L_{z}}dz^{\prime}k_{[-L_{z},L_{z}]}^{(D)}(z,z^{\prime};\tau)\,\psi_{\mathcal{A}_{2}}^{(in)}(u^{\prime},v^{\prime},z^{\prime})
\end{array}\right]
\end{equation*}
 
\begin{equation*}
F_{3}(u,z;\tau)=\frac{1}{2}\left[\begin{array}{c}
\int_{-L_{x}}^{L_{x}}dv^{\prime}\int_{-L_{y}}^{L_{y}}du^{\prime}k_{[-L_{y},L_{y}]}^{(D)}(u,u^{\prime};\tau)
\left[\partial_{v}k_{[-L_{x},L_{x}]}^{(D)}(v,v^{\prime};\tau)\right]_{v=-L_{x}}
\int_{-L_{z}}^{L_{z}}dz^{\prime}k_{[-L_{z},L_{z}]}^{(D)}(z,z^{\prime};\tau)\,\psi_{\mathcal{A}_{0}}^{(in)}(v^{\prime},u^{\prime},z^{\prime})\\
\\
-\int_{-\infty}^{-L_{x}}dv^{\prime}\int_{-L_{y}}^{L_{y}}du^{\prime}
\left[\partial_{v}k_{[-\infty,-L_{x}]}^{(D)}(v,v^{\prime};\tau)\right]_{v=-L_{x}}k_{[-L_{y},L_{y}]}^{(D)}(u,u^{\prime};\tau)
\int_{-L_{z}}^{L_{z}}dz'k_{[-L_{z},L_{z}]}^{(D)}(z,z^{\prime};\tau)\,\psi_{\mathcal{A}_{3}}^{(in)}(v^{\prime},u^{\prime},z^{\prime})
\end{array}\right]
\end{equation*}

\begin{equation*}
F_{4}(u,z;\tau)=\frac{1}{2}\left[\begin{array}{c}
\int_{-L_{x}}^{L_{x}}du^{\prime}\int_{-L_{y}}^{L_{y}}dv^{\prime}k_{[-L_{x},L_{x}]}^{(D)}(u,u^{\prime};\tau)
\left[\partial_{v}k_{[-L_{y},L_{y}]}^{(D)}(v,v^{\prime};\tau)\right]_{v=-L_{y}}
\int_{-L_{z}}^{L_{z}}dz'k_{[-L_{z},L_{z}]}^{(D)}(z,z^{\prime};\tau)\,\psi_{\mathcal{A}_{0}}^{(in)}(u^{\prime},v^{\prime},z^{\prime})\\
\\
-\int_{-L_{x}}^{L_{x}}du^{\prime}\int_{-\infty}^{-L_{y}}dv^{\prime}k_{[-L_{x},L_{x}]}^{(D)}(u,u^{\prime};\tau)
\left[\partial_{v}k_{[-\infty,-L_{y}]}^{(D)}(v,v^{\prime};\tau)\right]_{v=-L_{y}}
\int_{-L_{z}}^{L_{z}}dz'k_{[-L_{z},L_{z}]}^{(D)}(z,z^{\prime};\tau)\,\psi_{\mathcal{A}_{4}}^{(in)}(u^{\prime},v^{\prime},z^{\prime})
\end{array}\right]
\end{equation*}

\end{widetext}

\subparagraph{Short-Time Asymptotics}

For a sufficiently small time parameter $\tau $ the solution of (\ref%
{Volterra}) may be represented (symbolically) by the Neumann series: 
\begin{eqnarray}
f &=&\left( 1-\mathcal{T}\right) ^{-1}\circ F  \label{Neumann series} \\
&=&F+\mathcal{T}\circ F+\mathcal{T}\circ \mathcal{T}\circ F+\mathcal{T}\circ 
\mathcal{T}\circ \mathcal{T}\circ F+....  \nonumber
\end{eqnarray}%
The zero order term $F$ describes transmission from left to right, and up to
down, respectively. First order scattering processes involving contributions
from neighbouring arms $\mathcal{A}_{j}$ are provided by the first order
term $\mathcal{T}\circ F$. The higher order terms $\mathcal{T}\circ \mathcal{%
T}\circ F+\mathcal{T}\circ \mathcal{T}\circ \mathcal{T}\circ F+....$
describe multiple scattering events.

The functions $\phi _{\mathcal{A}_{j}}\left( \mathbf{r};\Delta \tau \right) $
defined in (\ref{auxiliary functions phi_A_j}) tend to zero for $\Delta \tau
\rightarrow 0$. For a short time $\tau ^{\prime }$ , so that $0\leq \tau
^{\prime }\leq \Delta \tau $, it then suffices to keep only the lowest order
term in the series (\ref{Neumann series}) and to approximate under the
integral $\int_{0}^{\Delta \tau }d\tau ^{\prime }...$ in the defining
equation (\ref{auxiliary functions phi_A_j}) 
\begin{equation}
f_{j}(\mathbf{r};\tau )=F_{j}(\mathbf{r};\tau )+O\left( small\right)
\label{approximation Neumann series I}
\end{equation}

There follows, taking into account (\ref{Chapman-Kolmogorov I}) and
inserting the results (\ref{auxiliarx functions psi_Dirichlet_A_j}), and (\ref%
{auxiliary functions phi_A_j}) together with (\ref{F equation}), into (\ref{subdomain functions II}), the
recursion%
\begin{eqnarray}
\tau _{n+1} &=&\tau _{n}+\Delta \tau \text{ for }n=0,1,2,3,..
\label{time evolution of wavefunction@small tau I} \\
\psi _{\mathcal{A}_{j}}\left( \mathbf{r};\tau _{n+1}\right)
&=&\sum_{l=0}^{4}\int_{\mathcal{A}_{l}}d^{3}r^{\prime }\mathcal{K}_{\mathcal{%
A}_{j},\mathcal{A}_{l}}\left( \mathbf{r},\mathbf{r}^{\prime };\Delta \tau
\right) \psi _{\mathcal{A}_{l}}\left( \mathbf{r}^{\prime };\tau _{n}\right) 
\nonumber
\end{eqnarray}%
, see Eq.(12) in \cite{M+Sch}. Here, for $\mathbf{r\in }\mathcal{A}_{j}$ and 
$\mathbf{r}^{\prime }\mathbf{\in }\mathcal{A}_{l}$ , the kernel functions $%
\mathcal{K}_{\mathcal{A}_{j},\mathcal{A}_{l}}\left( \mathbf{r},\mathbf{r}%
^{\prime };\Delta \tau \right) =\left[ K(\mathbf{r},\mathbf{r}^{\prime
};\Delta \tau )\right] _{\mathbf{r}\in \mathcal{A}_{j},\mathbf{r}^{\prime
}\in \mathcal{A}_{l}}$ represent the various pieces of the \emph{short-time}
expansion of the kernel $K(\mathbf{r},\mathbf{r}^{\prime };\Delta \tau )$ ,
see appendix B in \cite{M+Sch} .

\subsection{Barycentric Interpolation using Chebyshev Grid}

In order to evaluate the integrals in (\ref{time evolution of
wavefunction@small tau I}) we approximate the functions $\psi _{\mathcal{A}%
_{l}}\left( \mathbf{r}^{\prime };\tau _{n}\right) $ by an interpolation
polynomial. To discuss the principal idea we first consider interpolation of
a Lipschitz continuous function $f(x)$ in one dimension on the interval $%
x\in \left[ L_{1},L_{2}\right] $. On an \emph{equidistant} grid with $m+1$
grid points, say $x_{j}=L_{2}-\frac{2j}{m}\left( L_{2}-L_{1}\right) $ for $%
j\in \left\{ 0,1,...m\right\} $, it is well known that a high order
interpolation polynomial $p_{m}(x)$ of degree $m\gg 1$ ceases to approximate
the function $f\left( x\right) $ well, because such an interpolation
polynomial is exponentially ill conditioned near to the boundaries of the
interval $\left[ L_{1},L_{2}\right] $, the so called Runge phenomenon\cite%
{Trefethen II}. For the approximation of the function $f\left( x\right) $ on
the interval $x\in \left[ L_{1},L_{2}\right] $ by a high order interpolation
polynomial $p_{m}(x)$, accurately and uniformly, a \emph{non equidistant}
grid that clusters around the endpoints $L_{1}$, $L_{2}$ is mandatory.
Instead of approximating the function $f\left( x\right) $ by a linear
combination of Chebyshev polynomials $T_{n}(x)$ modern numerical
approximation methods \cite{Trefethen II} recommend barycentric
interpolation of $f(x)$ on a \emph{Chebyshev grid }, say with $m+1$ grid
points%
\begin{eqnarray}
&&  \label{Chebyshev grid} \\
x_{j} &=&\frac{L_{2}+L_{1}}{2}+\frac{L_{2}-L_{1}}{2}\cos \left( \frac{\pi }{m%
}j\right) \text{ for }j\in \left\{ 0,1,2,...,m\right\}  \nonumber
\end{eqnarray}%
The associated barycentric interpolation polynomial $p_{m}(x)$ of degree $m$
for a function $f\left( x\right) $ is given by: 
\begin{eqnarray}
x &\in &\left[ L_{1},L_{2}\right] \\
p_{m}(x) &=&\sum_{j=0}^{m}w_{m}(x;x_{j})f(x_{j})  \nonumber
\end{eqnarray}%
The cardinal functions 
\begin{eqnarray}
w_{m}(x;x_{j}) &=&\frac{\frac{c_{j}}{x-x_{j}}}{\sum_{l=0}^{m}\frac{c_{j}}{%
x-x_{l}}} \\
c_{0} &=&\frac{1}{2}  \nonumber \\
c_{j} &=&\left( -1\right) ^{j},\,\,1\leq j\leq m-1  \nonumber \\
c_{m} &=&\frac{(-1)^{m}}{2}  \nonumber
\end{eqnarray}%
associated with this interpolation scheme on a Chebyshev grid (\ref%
{Chebyshev grid}) have the property%
\begin{eqnarray}
j,k &\in &\left\{ 0,1,2,...,m\right\} \\
\lim_{x\rightarrow x_{k}}w_{m}(x,x_{j}) &=&\delta _{j,k}  \nonumber \\
p_{m}(x_{j}) &=&f(x_{j})  \nonumber
\end{eqnarray}

The three-dimensional integrals over the box shaped domains $\mathcal{A}_{l}$
for $l\in \left\{ 0,1,2,3,4\right\} $ can be evaluated using tensor products
of one-dimensional Chebyshev grids. For example, in the central box $%
\mathcal{A}_{0}=\left[ -L_{x},L_{x}\right] \times \left[ -L_{y},L_{y}\right]
\times \left[ -L_{z},L_{z}\right] $ our choice of gridpoints is according to
(\ref{Chebyshev grid}): 
\begin{eqnarray}
j_{a} &\in &\left\{ 0,...,m_{a}\right\} \\
\mathbf{r}^{(j_{x},j_{y},j_{z})} &=&(x_{j_{x}},y_{j_{y}},z_{j_{z}})\in 
\mathcal{A}_{0}  \nonumber \\
x_{j_{x}} &=&L_{x}\cos \left( \frac{\pi }{m_{x}}j_{x}\right)  \nonumber \\
y_{j_{y}} &=&L_{y}\cos \left( \frac{\pi }{m_{y}}j_{y}\right)  \nonumber \\
z_{j_{z}} &=&L_{z}\cos \left( \frac{\pi }{m_{z}}j_{z}\right)  \nonumber
\end{eqnarray}

In our calculations of the ground state orbitals $\psi _{0,\gamma }\left( 
\mathbf{r}\right) $ as discussed in \cite{M+Sch}, we replaced the four
infinitely extended arms $\mathcal{A}_{l}\ $ by arms $\widetilde{\mathcal{A}}%
_{l}$ of a finite length $\widetilde{L}$. This is admitted \cite{dng},
provided $\widetilde{L}$ is chosen much larger than the respective
localisation lengths $\lambda _{N,\gamma }$. The interval $\left[
L_{x},\infty \right] $ is then replaced by $\ \left[ L_{x},\widetilde{L}%
\right] $. The associated three-dimensional Chebyshev interpolation
polynomial $p_{\mathcal{A}_{1}}(\mathbf{r},\tau )$ approximating the
function $\psi _{\mathcal{A}_{1}}(\mathbf{r},\tau )$ inside the domain $%
\mathcal{A}_{1}$ is 
\begin{eqnarray}
&& \\
p_{\mathcal{A}_{1}}(\mathbf{r},\tau ) &=&\sum_{\substack{ {\small 0\leq j}%
_{a}{\small \leq m}_{a}  \\ {\small a\in }\left\{ x,y,z\right\} }}%
w_{m_{x},m_{y},m_{z}}\left( \mathbf{r};\mathbf{r}^{(j_{x},j_{y},j_{z})}%
\right) \psi _{\mathcal{A}_{1}}(\mathbf{r}^{(j_{x},j_{y},j_{z})},\tau ) 
\nonumber
\end{eqnarray}%
\[
w_{m_{x},m_{y},m_{z}}\left( \mathbf{r};\mathbf{r}^{(j_{x},j_{y},j_{z})}%
\right) \equiv
w_{m_{x}}(x;x_{j_{x}})w_{m_{y}}(y;y_{j_{y}})w_{m_{z}}(z;z_{j_{z}}) 
\]%
Similar expressions are obtained for the interpolation polynomials $p_{%
\mathcal{A}_{l}}(\mathbf{r},\tau )$ approximating the other functions $\psi
_{\mathcal{A}_{l}}(\mathbf{r},\tau )$. Replacing thus in (\ref{time
evolution of wavefunction@small tau I}) the functions $\psi _{\mathcal{A}%
_{l}}(\mathbf{r}^{\prime },\tau _{n})$ by their interpolation polynomials $%
p_{\mathcal{A}_{l^{\prime }}}(\mathbf{r}^{\prime },\tau _{n})$ we obtain at
the grid points $\mathbf{r}^{(j_{x},j_{y},j_{z})}\in \mathcal{A}_{j}$ inside
each of the five box shaped subdomains $\mathcal{A}_{j}$ comprising the
cross shaped waveguide geometry $\mathcal{C}$ the following recursion:

\begin{widetext}

\begin{eqnarray}
&& \\
\psi _{\mathcal{A}_{j}}(\mathbf{r}^{(j_{x},j_{y},j_{z})},\tau _{n+1})
&=&\sum_{0\leq l\leq 4}\sum_{\substack{ {\small 0\leq j}_{a}^{\prime }%
{\small \leq m}_{a} \\ {\small a\in }\left\{ x,y,z\right\} }}\kappa _{%
\mathcal{A}_{j},\mathcal{A}_{l}}\left( \mathbf{r}^{(j_{x},j_{y},j_{z})},%
\mathbf{r}^{(j_{x}^{\prime },j_{y}^{\prime },j_{z}^{\prime })};\Delta \tau
\right) \psi _{\mathcal{A}_{l}}(\mathbf{r}^{(j_{x}^{\prime },j_{y}^{\prime
},j_{z}^{\prime })},\tau _{n})  \nonumber \\
\kappa _{\mathcal{A}_{j},\mathcal{A}_{l}}\left( \mathbf{r}%
^{(j_{x},j_{y},j_{z})},\mathbf{r}^{(j_{x}^{\prime },j_{y}^{\prime
},j_{z}^{\prime })};\Delta \tau \right)  &\equiv &\int_{\mathcal{A}%
_{l}}d^{3}r^{\prime }\mathcal{K}_{\mathcal{A}_{j},\mathcal{A}_{l}}\left( 
\mathbf{r}^{(j_{x},j_{y},j_{z})},\mathbf{r}^{\prime };\Delta \tau \right)
w_{m_{x},m_{y},m_{z}}\left( \mathbf{r}^{\prime };\mathbf{r}^{(j_{x}^{\prime
},j_{y}^{\prime },j_{z}^{\prime })}\right)   \nonumber
\end{eqnarray}

\end{widetext}

An advantage of the described algorithm is that for a fixed time step $%
\Delta \tau $ the matrix elements $\kappa _{\mathcal{A}_{j},\mathcal{A}_{l}}(%
\mathbf{r}^{(j_{x},j_{y},j_{z,})},\mathbf{r}^{(j_{x}^{\prime
},\,j_{y}^{\prime },\,j_{z}^{\prime })},\Delta \tau )$ need to be calculated
(using a high accuracy numerical integration rule) just once! The update
rule (\ref{time evolution of wavefunction@small tau I}) is thus reduced to
quick matrix-vector multiplications. It forms the core of the computational
algorithm employed in \cite{M+Sch} in the Eqs. (12) and (35).